\documentclass[11pt,oneside,letterpaper]{article}
\pdfoutput=1
\usepackage{amssymb}
\usepackage{amsmath}
\usepackage[dvips]{graphicx}
\usepackage{setspace}
\usepackage{amsfonts}
\usepackage{fancyhdr}
\usepackage{xcolor}
\usepackage{graphicx}
\usepackage{rotating}
\usepackage{comment}
\usepackage{color}
\usepackage{cite}
\usepackage{braket}

\usepackage{tikz}

\usepackage{subfigure}
\usetikzlibrary{decorations.pathmorphing}

\definecolor{darkgreen}{rgb}{0,0.5,0}
\definecolor{darkblue}{rgb}{0,0,0.6}
\definecolor{purple}{rgb}{0.4,.2,0.7}
\newcommand{\p}{\partial}

\newcommand{\f}{\frac}

\newcommand{\be}{\begin{equation}}
\newcommand{\ee}{\end{equation}}

\usepackage[colorlinks=true,citecolor=darkgreen,linkcolor=black,urlcolor=purple]{hyperref}

\usepackage{pdfsync}

\makeatletter
\newcommand*{\defeq}{\mathrel{\rlap{%
                     \raisebox{0.3ex}{$\m@th\cdot$}}%
                     \raisebox{-0.3ex}{$\m@th\cdot$}}%
                     =} 
\makeatother

\def\be{\begin{eqnarray}}
\def\ee{\end{eqnarray}}

\newcommand{\bea}{\begin{eqnarray}}
\newcommand{\eea}{\end{eqnarray}}

\newcommand{\beg}{\begin{equation} \begin{gathered}}
\newcommand{\eeg}{\end{gathered} \end{equation}}

\let\G=\Gamma
\let\l=\left
\let\r=\right

\let\f=\frac

\let\d=\partial
\def\be{\begin{equation}}
\def\ee{\end{equation}}
\def\ba{\begin{array}}
\def\ea{\end{array}}

\def\eps{\epsilon}

\def \F{{}_2 F_1}
\def \G{\Gamma}
\def \ddt{{d\over 2}}
\def \half{\frac{1}{2}}

\def\ba#1\ea{\begin{align}#1\end{align}}
\def\bs#1\es{\begin{split}#1\end{split}}

\renewcommand{\p}{\partial}

\allowdisplaybreaks  

\numberwithin{equation}{section}

\def \be {\begin{equation}}
\def \ee {\end{equation}}
\def \half {{1\over 2}}

\def\comma{\,,}
\def\period{\,.}

\def\O{{\cal{O}}}

\begin{document}
\begin{flushright}
{\tt CERN-TH-2021-118}
\end{flushright}
\onehalfspacing
\begin{center}

~
\vskip5mm

{\LARGE  {
Analyticity and Unitarity for Cosmological Correlators
}}

\vskip10mm

Lorenzo Di Pietro$^{1,2}$, \ Victor Gorbenko$^{3}$,\ \ Shota Komatsu,$^{4}$ 

\vskip15mm

{\it $^{1}$ Dipartimento di Fisica, Universit\`a di Trieste,
Strada Costiera 11, I-34151 Trieste, Italy}\\
{\it $^{2}$ INFN, Sezione di Trieste, Via Valerio 2, I-34127 Trieste, Italy}\\
{\it $^{3}$ SITP, Stanford University, Palo Alto, California, USA } \\
{\it $^{4}$ Department of Theoretical Physics, CERN, 1211 Meyrin, Switzerland}
\vskip5mm

\vskip5mm

\end{center}

\vspace{4mm}

\begin{abstract}
\noindent

We study the fundamentals of quantum field theory on a rigid de Sitter space. We show that the perturbative expansion of late-time correlation functions to all orders can be equivalently generated by a non-unitary Lagrangian on a Euclidean AdS geometry. This finding simplifies dramatically perturbative computations, as well as allows us to establish basic properties of these correlators, which comprise a Euclidean CFT.   
We use this to infer the analytic structure of the spectral density that captures the conformal partial wave expansion of a late-time four-point function, to derive an OPE expansion, and to constrain the operator spectrum. Generically, dimensions and OPE coefficients do not obey the usual CFT notion of unitarity. Instead, unitarity of the de Sitter theory manifests itself as the positivity of the spectral density. This statement does not rely on the use of Euclidean AdS Lagrangians and holds non-perturbatively. We illustrate and check these properties by explicit calculations in a scalar theory by computing first tree-level, and then full one-loop-resummed exchange diagrams. An exchanged particle appears as a resonant feature in the spectral density which can be potentially useful in experimental searches.

 \end{abstract}

\pagebreak
\pagestyle{plain}
\setcounter{tocdepth}{2}
{}
\vfill
{\footnotesize \tableofcontents}

\newpage


\section{Introduction}
De Sitter (dS) space is the most symmetric and in this sense the simplest cosmological spacetime. According to a vast observational evidence the cosmological evolution of our own universe has at least two epochs which appear to have spacetime metric very similar to that of dS. These two epochs are the current accelerated expansion and the earliest period in the history of the universe to which we have a direct experimental access -- the period of inflation. Despite its basic status, dS space is significantly less studied as compared to its mostly symmetric cousins, anti-de Sitter (AdS) and Minkowski spaces. The reason lying in several conceptual and technical difficulties that it exhibits, some of which we are going to touch upon below. In this paper we take up a relatively modest goal: to study the properties of correlation functions in quantum field theory on a rigid dS space. Unfortunately, at present rather little is known about them. Some of the questions we ask in this paper, for example, what  imprint an exchanged heavy particle leaves on the observed quantities, were analyzed in \cite{Arkani-Hamed:2015bza}. Although it has a very clear answer when flat space scattering is concerned, in cosmology a complete understanding of such elementary issues is still lacking. 
Note that, even at tree-level, the closed-form expression for the  exchange diagram was obtained only very recently~\cite{Sleight:2020obc}, and a systematic treatment of infrared divergences of light particles were understood not long ago either~\cite{Gorbenko:2019rza}.  

Our main focus is on the ``boundary'' correlation functions of the fields, evaluated at  the future infinity of dS space. Even though formally not observables, these objects are closely related to inflationary correlation functions measured by an observer located in the future from the reheating surface and who sees simultaneously a very large number of Hubble patches.

The motivation for this endeavor is three-fold. First, the calculation of such correlation functions is an integral part of the computation of primordial non-gaussianities in inflation. Second, a mastery of perturbative techniques may assist us in the quest for a fundamental description of gravitational theories, in the spirit of AdS/CFT, which is so far illusionary in the cosmological setting. 

Third, quantum field theory on a rigid dS offers yet another, perhaps much less known, route to cosmology in quantum gravity. Some large $N$ quantum field theories on a rigid dS background are known to be holographically dual to a gravitational spacetime that includes a FRW-like geometry with a crunching singularity, see the Appendix of \cite{Maldacena:2010un}, as well as \cite{Hertog:2004rz,Hertog:2005hu,Turok:2007ry,Craps:2007ch,Barbon:2011ta,Barbon:2013nta,Kumar:2015jxa} for concrete examples in string theory. This connection to cosmology is less direct than what was discussed above, but it would allow us to study quantum gravity in a cosmological setup using tools of quantum field theory on a rigid dS, which we develop in this paper. 

The punchline of this work is the following:
\begin{itemize}
\item{We develop a systematic framework to reduce any perturbative calculation in dS to that in Euclidean AdS (EAdS), where plentiful computational techniques have been developed. We achieve this by writing down a Lagrangian in EAdS which directly computes boundary correlation functions in dS.
Our result systematizes and extends various observations made in the literature on the connection between Feynman diagrams in dS and EAdS.
}
\item{Based on the EAdS Lagrangian we derived, we discuss the analytic structure of boundary correlation functions in dS with a particular focus on the operator product expansion (OPE). We then discuss a possibility of interpreting the OPE in terms of quasi-normal modes in a static patch of dS.
}
\item{We study the implication of bulk unitarity on the boundary correlators, and relate it to certain positivity conditions. The derivation does not rely on perturbation theory and is therefore valid at a {\it non-perturbative} level.}
\end{itemize}

In this paper we do not consider gravity, 
or a breaking of dS isometries inevitably present in inflation. Much of our discussion can be generalized to include either or both, even though many additional conceptual and technical complications arise. We thus find it important to first understand as clearly as possible the simplest possible case and defer the generalizations to the future.

The paper is organized in the following way. After reviewing the basics of dS, we make several additional introductory remarks about the choice of observables in cosmology in section~\ref{sec:wavefunction}. We also emphasize the distinction between different notions of ``late-time CFTs'' and discuss their relationships. We then go through the three bullet points above in sections~\ref{sec:EAdS},  \ref{sec:Analyticity} and \ref{sec:Unitarity} in a generic setting. In section \ref{sec:pert} we perform explicit tree-level and resummed-one-loop calculations of scalar four-point functions and examine the implications of analyticity and unitarity in this example. In particular, in figure~\ref{fig:resonance} we demonstrate that weakly coupled heavy particles appear as narrow resonances in our parametrization, which makes it potentially useful in phenomenological applications. Several open problems are presented in section~\ref{sec:Conclusions}.

\paragraph{Note added:} During the course of this work, we learned of the then ongoing work \cite{Hogervorst:2021uvp}, in which the authors propose a non-perturbative conformal bootstrap program for late-time correlation functions on a rigid dS. The paper offers a complementary view on some of the issues we consider here, the main overlap with our work being the discussion of dS unitarity and the positivity of the spectral function which we discuss in section \ref{sec:Unitarity}. We thank the authors of \cite{Hogervorst:2021uvp} for fruitful exchanges and discussions. 
\section{De Sitter, wave functions and correlators}
\label{sec:wavefunction}
In this section, we first review the basic facts about the dS spacetime in subsection \ref{subsec:basics}, such as the coordinate systems, the isometries and the unitary representations. We then discuss the notion of wave functions and correlation functions of fields in dS in subsection \ref{subsec:corr}. In particular, we make clear distinction between two  different notions of ``late-time CFTs''; one that describes the wave function and the other that describes the correlation function. The materials reviewed in these two subsections are well-known, but we decided to include them partly to fix the notations and conventions and also to avoid possible confusions in the subsequent sections.
Finally in section \ref{sec:alternate}, we suggest a possible interpretation of the correlator CFT as two wave-function CFTs, each corresponding to a different boundary condition, coupled together by (almost) marginal double-trace operators.
\subsection{De Sitter: basics}\label{subsec:basics}
Here we quickly summarize the basic facts about dS, with the aim of fixing the notations and conventions for later sections. More detailed and pedagogical reviews can be found in \cite{Spradlin:2001pw,Anninos:2012qw}.
\paragraph{De Sitter and three coordinates.} The simplest way to define the $(d+1)$-dimensional dS space is to realize it as a hypersurface inside $\mathbb{R}^{d+1,1}$ satisfying
\be\label{eq:embeddingdS}
\eta_{AB}X^{A}X^{B}=L^2\comma
\ee
with
\be
\eta_{AB}={\rm diag}(-1,\underbrace{1,\ldots, 1}_{d+1})\period
\ee
Here $L$ is the Hubble radius which controls the size of dS. In the rest of this paper, we will set it to $1$ for simplicity.

There are three coordinate systems (slicings) of dS often used in the literature. The first one is the {\it global coordinates}, which correspond to parametrizing $X^{A}$ as
\be
X^{A}=\left(\sinh\tau, \cosh \tau\, n_{1},\ldots  ,\cosh \tau\, n_{(d+1)}\right)~, \qquad \sum_{j=1}^{d+1}(n_j)^2=1\period
\ee
The metric in these coordinates is
\be\label{eq:metricGL}
ds^2=-d\tau^2+\cosh^2 \tau d\Omega_{d}^2\comma
\ee
where $\tau\in (-\infty,\infty)$ and $d\Omega_d^2$ is the metric for the $d$-dimensional sphere.
This coordinate system covers the entire hypersurface defined by \eqref{eq:embeddingdS}, hence the name. This is an analog of the global coordinates in AdS, which also covers the entire Lorentzian AdS.

The second one is the {\it Poincar\'e coordinates}, defined by
\be
X^{A}=\left(\frac{1-\eta^2+|x|^2}{2\eta},\frac{x^{\mu}}{\eta}, \frac{1+\eta^2-|x|^2}{2\eta}\right)\comma
\ee
with $\eta\in (-\infty, 0)$. In these coordinates, the metric of dS takes the form
\be\label{dsPoinc}
ds^2=\frac{-d\eta^2+dx^{\mu}dx_{\mu}}{\eta^2}\period
\ee
These coordinates cover a ``half'' of the global dS as can be seen from the Penrose diagram in figure \ref{fig:Penrose}. This is an analog of the Poincar\'e patch of AdS.

The last one is the {\it static patch}. This is in a sense the most physical coordinate system since it covers precisely the region accesible to a single observer inside dS, see figure \ref{fig:patches}. The metric in this coordinate system reads\footnote{A more standard parametrization of the static patch is given by 
\be
ds^2=-\left(1-r^2\right)dt^2+\left(1-r^2\right)^{-1}dr^2+r^2 d\Omega_{d-1}^2\period
\ee
It is related to \eqref{eq:metricST} by $r=\sin\theta$. 
}
\be\label{eq:metricST}
ds^2=-\cos^2\theta dt^2 +d\theta^2+\sin^2\theta d\Omega_{d-1}^2\comma
\ee
where $\theta\in(0,\pi/2)$ and $d\Omega_{d-1}^2$ is the metric for the $(d-1)$-dimensional sphere. It corresponds to the following parametrization of the embedding coordinates:
\be
X^{A}=\left(\cos\theta\sinh  t,\cos\theta\cosh t,\, \sin\theta\,  \tilde{n}_1,\ldots , \,\sin\theta\,  \tilde{n}_{d}\right)~,\qquad \sum_{j=1}^{d}(\tilde{n}_j)^2=1\period
\ee
In AdS, the closest analog of this is the Rindler AdS, which also covers a half of the Poincar\'e patch. Much like in the Rindler AdS (or its analog in flat space---the Rindler space), physics inside the static patch appears thermal.

\paragraph{Embedding and invariant distances.}As mentioned in the introduction, the main focus of this paper is the study of correlation functions on the late-time surface, which is a co-dimension 1 surface defined by $\eta\to 0^{-}$ in the Poincar\'e coordinates. A point on the surface can be parametrized by a ``boundary version'' of the embedding coordinates, which is given by a projective null cone 
\be
\eta_{AB}P^{A}P^{B}=0\comma\qquad P\sim \lambda P\period
\ee
The relation to the Poincar\'e coordinates is given by
\be\label{eq:bdyemb}
P^{A}=\left(\frac{1+|y|^2}{2},y^{\mu}, \frac{1-|y|^2}{2}\right)\period
\ee
Using $X$'s and $P$'s, we can construct invariants under the dS isometry. The two quantities that are important are the {\it two-point invariant} $s(X_1,X_2)$ and the {\it bulk-boundary distance} $X\cdot P$:
\begin{align}
s(X_1,X_2)&\equiv X_1\cdot X_2=\frac{\eta_1^2+\eta_2^2-|x_{12}|^2}{2\eta_1\eta_2}\comma\label{sdef}\\
X\cdot P&=\frac{\eta^2-|x-y|^2}{2\eta}\period\label{eq:BBdis}
\end{align}
Note that the two-point invariant is related to the {\it chordal distance} between the two points $\zeta (X_1,X_2)$ by the following relation:
\be\label{eq:chordal}
\zeta(X_1,X_2)=2\left(1-s(X_1,X_2)\right)=\frac{-\eta_{12}^2+|x_{12}|^2}{\eta_1\eta_2}\period
\ee
\paragraph{Comparison with AdS.} Embedding coordinates have analogs in Euclidean AdS space. Here we summarize them and show their relation to the (EAdS) Poincar\'e coordinates\footnote{The boundary of EAdS can also be expressed in terms of the embedding coordinates, but since they take exactly the same form as \eqref{eq:bdyemb}, we will not display them here.}:
\begin{align}
&\eta_{AB}X_{\rm AdS}^{A}X_{\rm AdS}^{B}=-1\comma\quad &&X_{\rm AdS}^{A}=\left(\frac{1+z^2+x^{\mu}x_{\mu}}{2z},\frac{x^{\mu}}{z},\frac{1-z^{2}-|x|^{2}}{2z}\right)\comma
\end{align}
where $(z,x^{\mu})$ are the Poincar\'e coordinates
\be
ds^2=\frac{dz^2+dx^{\mu}dx_{\mu}}{z^2}\period
\ee
The EAdS version of the two-point invariant is 
\be
\label{sAdSdef}
s_{\rm AdS}(X_1,X_2)=X_1\cdot X_2=-\left(\frac{z_1^2+z_2^2+|x_{12}|^2}{2z_1z_2}\right)\period
\ee
%
%
\paragraph{Isometries and unitary representations.}The isometry group of dS is isomorphic to the Euclidean conformal group $SO(d+1,1)$. This can be seen most explicitly in embedding coordinates in which the isometries are given by the following simple differential operators
\be\label{eq:differential}
\begin{aligned}
\mathcal{L}_{AB}&=X_{A}\frac{\partial}{\partial X^{B}}-X_{B}\frac{\partial}{\partial X^{A}}\\
&=P_{A}\frac{\partial}{\partial P^{B}}-P_{B}\frac{\partial}{\partial P^{A}}\period
\end{aligned}
\ee
Here the first line describes how the operator transforms a bulk point in dS while the second line describes how it transforms a point on the late-time surface.
Computing the commutation relations, one can easily check that it is isomorphic to $SO(d+1,1)$.

Much like in flat space, states in a unitary QFT in dS are classified by unitary irreducible representations of the isometries. This is a rich but well-studied subject, with literature both on the math side and on the physics side \cite{newton1950note,thomas1941unitary,Dobrev:1976vr,dobrev1977harmonic}. In particular, the three representations that play an important role in dS are the {\it principal series}, the {\it complementary series} and the {\it discrete series} (see also section 2 of \cite{Baumann:2017jvh} for a summary of the subject). 
To describe these representations, it is useful to use the language of conformal field theory and parametrize them in terms of a conformal dimension $\Delta$, which is related to the value of quadratic Casimir as
\be
\mathcal{C}_2=-\frac{1}{2}\sum_{A,B}\mathcal{L}_{AB}\mathcal{L}^{AB}=\Delta (\Delta-d)+J(J+d-2)\comma
\ee
where $J$ is the spin of the state.

In this language, the principal series corresponds to states with $\Delta=\frac{d}{2}+ i\nu$, where $\nu$ is a positive real number. When applied to a free scalar theory in dS, it corresponds to a ``heavy'' field, namely a field whose mass satisfies $m>\frac{d}{2}$
in units of the Hubble parameter. This can be checked from the relation between the conformal dimension and the mass in dS,
\be
m^2=\Delta(d-\Delta)\period
\ee
Being a quadratic equation, this has two solutions which we denote by $\Delta_{\pm}=\frac{d}{2}\pm i\nu$. As a representation, they are equivalent and this is why we can restrict to $\frac{d}{2}+i\nu$ with positive $\nu$,
\be\label{eq:massNu}
\nu=+\sqrt{m^2-\frac{d^2}{4}}\period
\ee However, we often keep them both in the computation since they show up as two different power-law decays in correlation functions.
The complementary series (with zero spin) corresponds to states with conformal dimension $0<\Delta<\frac{d}{2}$. In a free scalar theory in dS, it corresponds to a ``light'' field with mass
$\frac{d}{2}>m>0$.
Finally, there is a discrete series, which corresponds to states with integer or half-integer conformal dimension. In free theory, they exist only for a particle with spin, and correspond to so-called {\it partially massless fields}. Although they can leave imprints on the inflationary observables, in this paper we focus on scalar correlation functions and therefore will not study them. Summarizing the results for states with zero spin, we have
\begin{align}
\text{Principal series:}& \qquad \Delta=\frac{d}{2}+i\nu\comma\qquad &&m>\frac{d}{2}\comma\label{sum:pr}\\
\text{Complementary series:}& \qquad 0<\Delta<\frac{d}{2}\comma\qquad &&\frac{d}{2}>m>0\period\label{sum:com}
\end{align}

\paragraph{Remark on the notation.}In the rest of this paper, we often use $\Delta$ and $\nu$ interchangeably to parametrize the representation of states and the mass of particles. In particular, we find it convenient to use $\nu$ even for the light particles. When doing so, we need to decide which of $\Delta_{\pm}=\frac{d}{2}\pm i\nu$ satisfies \eqref{sum:com}. This is just a choice of the notation and throughout this paper, we choose $\Delta_{-}$ to satisfy the condition $\frac{d}{2}>\Delta_{-}>0$. Namely we have
\be\label{eq:nuLight}
\text{Complementary series:} \qquad\Delta_{-}=\frac{d}{2}-i\nu \qquad i\nu\in \left(0,\frac{d}{2}\right).
\ee 
\subsection{Cosmological correlators, wave functions and other observables in dS}\label{subsec:corr}
We now review the notion of wave functions and correlation functions of fields in dS, the relationship between the two and their basic properties. We also mention briefly the dS S-matrix and the static patch approach. For a researcher working in the field all these topics are well-known; however, we hope that this discussion can help to avoid potential confusions for those less familiar with the subject. Another goal of this section is to highlight the subtleties related with defining fundamental physical observables in dS space and more generally in cosmology. Since in this paper we are dealing with perturbative massive quantum field theories on a fixed background, defining observables is a relatively simple task; however, ideally one would like to do it in a way that can remain consistent once gauge fields, most importantly the metric, are also dynamical. In subsection~\ref{sec:alternate} we make a proposal for a procedure relating the wave functions and correlators, that may help to address some of these subtleties.

\begin{figure}[t] 
	\centering
		\includegraphics[width=1\linewidth,angle=0]{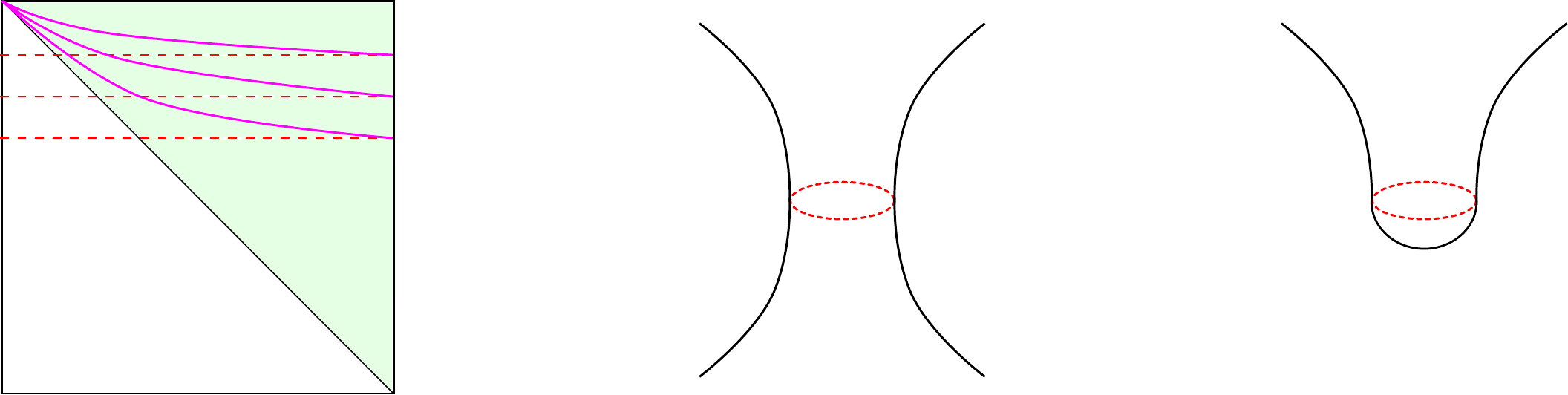}
	\caption{\small \textbf{Left:} Penrose diagram of dS, the expanding Poincar\'e patch is indicated in green. Constant global (red) and Poincar\'e (purple) slices are also indicated. \textbf{Middle:} global dS. \textbf{Right:} No-boundary state geometry in which contracting part of the global dS is analytically continued to become a half-sphere.}
	\label{fig:Penrose}
\end{figure}

We can try to learn a lesson from spacetimes that are asymptotically flat or AdS, in which case well-defined observables in theories with dynamical gravity are asymptotic, i.e. defined on the boundary of corresponding space. As can be seen on the Penrose diagram \ref{fig:Penrose}, the global dS space has two asymptotic boundaries -- future and past ones, both of which are space-like. This motivated several authors to consider the transition amplitude from the past to the future boundary, or the dS S-matrix \cite{Witten:2001kn,Strominger:2001pn}. The past boundary, however, is not exactly on the same footing as the future one. In particular, one can work with a certain Euclidean continuation of the geometry which has only one boundary in the future \cite{Hartle:1983ai}, see figure \ref{fig:Penrose} and a more detailed explanation in Appendix~\ref{App:global}. The no-boundary geometry has the benefit that it excludes the contracting part of dS, which hardly has cosmological relevance. From this point of view,  a natural object to study is the wave function of the universe in the no-boundary state \cite{Maldacena:2002vr}.
These two objects correspond to fixing the values of all fields on the boundaries of the corresponding geometries. The wave function can also be thought as a transition amplitude from the fixed in state. In what follows we will not discuss further general transition amplitudes, referring an interested reader to some recent works on the scattering approach \cite{
Marolf:2012kh,Cotler:2019dcj,Albrychiewicz:2020ruh}.

The late time wave function has many nice properties, which parallel those of the AdS partition function -- the central object in AdS/CFT correspondence. In particular, modulo certain local terms, it takes the following form:
\be\label{eq:wavephiexp}
\begin{aligned}
\Psi\l[\phi\r]=&\exp\l[\int dx_1 dx_2 \phi(x_1)\phi(x_2)G_2(x_1,x_2)\right.\\
&\qquad \left.+\int dx_1 dx_2 dx_3 \phi(x_1)\phi(x_2)\phi(x_3)G_3(x_1,x_2,x_3)+\ldots\r]\,,
\end{aligned}
\ee
where $\phi(x)$ stands for the late time value of some field propagating in dS (where $x$'s parametrize the future boundary). The function $G_n$ has all the properties of a correlation function in some Euclidean CFT, which we call CFT$_{\Psi}$. At least formally, we can express the wave function $\Psi$ as a generating function of CFT$_{\Psi}$,
\be\label{eq:CFTrepPsi}
\Psi [\phi]=\left<e^{\int dx \mathcal{O}(x)\phi(x)}\right>_{\text{CFT}_{\Psi}}\comma
\ee
where $\mathcal{O}(x)$ is some operator in CFT$_{\Psi}$ whose connected correlation functions give $G_n$.
These $G_n$'s have been a subject of many recent publications \cite{Arkani-Hamed:2017fdk,Hillman:2019wgh,Benincasa:2019vqr,Meltzer:2021zin,Pajer:2020wnj,Cespedes:2020xqq,Goodhew:2021oqg,Goodhew:2020hob,Jazayeri:2021fvk,Baumann:2021fxj} in which they are sometimes referred to as ``correlators''. We would like to stress that these are {\it{not}} what we call ``cosmological correlators''. In this paper, we will call $G_n$'s the wave function coefficients.

Indeed, the wave function itself is not an observable in any physical system. Instead, the objects of primary interest to phenomenologists and observational cosmologists are the correlation functions of the fields themselves\footnote{To be careful, correlation functions are also not directly measurable in a single copy of a system and to determine them experimentally one needs to prepare and measure the system many times, which is, unfortunately, impossible to do with an observable universe. What cosmologists use, however, is the approximate translational and rotational symmetry of the universe. Then by measuring correlation functions in many different locations and orientations one effectively repeats the measurement many times.}, $\langle\phi(x_1)\phi(x_2)\ldots\rangle$, calculated at late times. These are the objects we call cosmological correlators. They can be studied either in the global slicing \eqref{eq:metricGL} or in the Poincar\'e slicing \eqref{dsPoinc}. As can be seen on figure~\ref{fig:Penrose}, these slices practically coincide at late times. So at infinite future, there is no difference between the two.

In quantum mechanics, the knowledge of the wave function ensures the knowledge of all correlation functions, and this is also {\it{formally}} the case in dS:
\be
\label{corrfrompsi}
\l\langle \phi(x_1)\phi(x_2)\ldots\phi(x_n)\r\rangle = \int D\phi \Psi\l[\phi\r] \Psi^*\l[\phi\r] \phi(x_1)\phi(x_2)\ldots\phi(x_n)\,,
\ee
where $\int D\phi$ denotes the path integral over all dynamical fields of the theory at the boundary. In reality the knowledge of the wave function is not much different from the knowledge of an action or a Lagrangian of some theory -- much work is still needed to compute the correlation functions. Moreover, the corresponding action is highly non-local, as can be seen in \eqref{eq:wavephiexp}. Of course, at the lowest orders in perturbation theory it is straightforward to extract the correlators from the wave function coefficients; however, it is unclear what is the status of this procedure non-perturbatively. This is especially worrisome when the metric is dynamical, so taking the above path integral involves solving some Euclidean quantum gravity, which is by no means guaranteed to be possible.\footnote{For some special higher-spin theories this procedure can in fact be executed \cite{Anninos:2017eib}.}
Even though the use of eq. \eqref{corrfrompsi} is a viable strategy for the perturbative calculation of correlators (for an application of this approach see \cite{Ghosh:2014kba} ), in this paper we will find it more convenient to do it directly using the in-in formalism, bypassing the wave function computation. 

Let us anticipate some of the properties of these correlation functions. In our case of QFT on a rigid dS it is clear that the correlators of fields on the future boundary will be invariant under the dS isometry group $SO(1,d+1)$, which is also the Euclidean conformal group in $d$ dimensions. Thus, we expect the boundary correlators of a QFT in dS to form the correlators of some Euclidean CFT, which we will call CFT$_C$.\footnote{This relation for us is purely kinematical -- we do not imply any independent microscopic definition of CFT$_C$ at this point.} In the remaining part of this paper, we will confirm this expectation. For now, let us stress that the ``correlators'' CFT$_C$ is very different from the ``wave function'' CFT$_{\Psi}$. At weak coupling in the bulk both CFTs have an analog of a large-$N$ counting parameter and, as we will see, CFT$_C$, has roughly twice the number of single trace operators as compared to CFT$_{\Psi}$. Since in the calculation of correlators the boundary values of fields are not fixed, one may expect to have two operators corresponding to two modes of each fundamental field with dimensions related by the shadow transform: 
$\Delta_1+\Delta_2=d$. In fact, we will see that this relation is not exact and gets corrected in perturbation theory. 

Let us make a brief comment on how this picture gets altered when gravity is dynamical. The calculation of the wave function remains intact, at least in perturbation theory, and CFT$_{\Psi}$ gets endowed with the stress tensor. The change is much more dramatical on the correlator side: it is not even clear how to define local and gauge invariant observables from the boundary point of view. Since some form of ``boundary gravity'' remains a part of the calculation, one cannot simply parametrize the operator positions by coordinates in some system. We do not have any concrete proposal at this point; however, let us make one vague analogy: the relation between wave function and correlators in dS is somewhat reminiscent of the relation between S-matrix and inclusive cross-sections in flat space. In fact, also in Minkowski it is the cross section which is really observable and, moreover, a partial trace over soft gravitational modes is sometimes necessary to make it well-defined. It would be interesting to see if the techniques developed for dealing with these modes can helps us in defining proper observables also in cosmology. 

\begin{figure}[t] 
	\centering
		\includegraphics[width=.6\linewidth,angle=0]{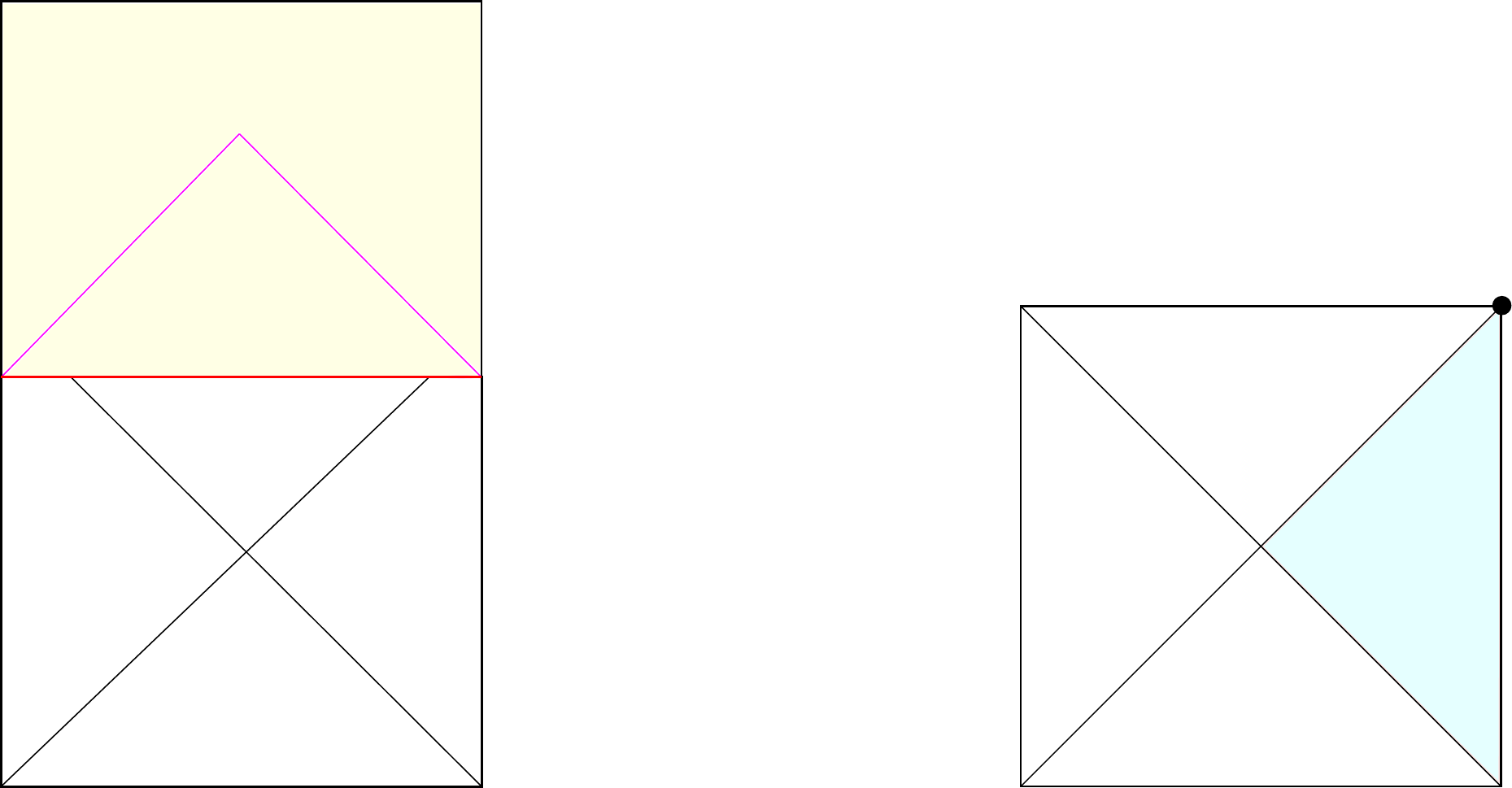}
	\caption{\small \textbf{Left:} Penrose diagram of dS, cutoff at the reheating surface (red) and glued to the non-expanding spacetime (yellow). The lightcone of an observer that sees the entire reheating surface is indicated in purple. \textbf{Right:} static patch indicated in blue. It coincides with the causal past of a point (the black dot) on the late-time surface, in the expanding part of dS.}
	\label{fig:patches}
\end{figure} 

Finally, let us come back to another issue related to asymptotic observables in dS. In fact, even in the absence of gravity, there isn't a single physical observer that could see the entire future infinity of dS space, as is obvious from the Penrose diagram. Thus one may object to the statement that correlators are useful physical objects to study. There are ways to deal with this problem. The first one is to imagine that dS spacetime is actually the limit of a very long inflation, and that after some late time the entire universe ``reheats'' into a flat space, after which there is an asymptotic flat space observer that has access to the entire reheating surface, see figure \ref{fig:patches}, left\footnote{For simplicity of drawing, we drew a Lorentzian cylinder instead of flat space in figure \ref{fig:patches}.}. Then our boundary correlators can be though of as the limit of correlators measured on the reheating surface, which is then taken to infinity. In practice, this is the limit taken in most phenomenological inflationary calculations since all observed modes are super-horizon. An alternative approach is to declare that the only fundamental physical observables are those accessible by a single causal observer in dS, that is restricted to a static patch (see figure \ref{fig:patches}). For a detailed discussion of this point of view see, e.g., \cite{Anninos:2011af,Anninos:2012qw}. The two approaches seem radically different; however, there is a hope that the two are related. In our study of boundary correlators in QFT on dS we will see features apparently corresponding to the physics of the static patch. We discuss this relation further in the subsection \ref{subsec:QNM} and in the conclusions section \ref{sec:Conclusions}.

\subsection{Wave function with alternate boundary conditions and double-trace deformation}
\label{sec:alternate}
Let us look at the relation \eqref{corrfrompsi} again. For the moment we will imagine that we have some independent knowledge of CFT$_\Psi$, possibly at the non-perturbative level. One can then view the relation \eqref{corrfrompsi} as generating a new theory by taking two CFTs, one corresponding to $\Psi$ and the other to $\Psi^*$, and coupling them through an integral over the common sources $\phi$. Even though an integral over sources is not a priori a well-defined procedure, as it requires knowledge of the CFT partition function for arbitrary values of the sources, for perturbative theories at leading order one can think of it as adding double-trace operators made of single trace operators from the two theories. Generically such operators will be strongly irrelevant. For example, if the metric is dynamical there will be an operator proportional to the product of the stress tensors of the two theories, whose scaling dimension is $2d$. It is thus hard to make sense of \eqref{corrfrompsi} as any kind of perturbative renormalization group (RG) flow. 

Instead we propose an alternative way to relate the two CFTs, in which the RG flow can stay perturbative.
The first step is to write $\Psi^{\ast}$ not in terms of $\phi(x)$ but in terms of its canonical conjugate momentum $\pi (x)$: From the dS point of view, this amounts to imposing a different boundary condition\footnote{The wave function in dS with an alternate boundary condition and its relation to the double-trace deformation was discussed recently in \cite{Isono:2020qew}. However, this is simply an analog of what was discussed in EAdS \cite{Witten:2001ua} and should not be confused with the double-trace deformation discussed in this section, which directly describes the correlator CFT.} at future infinity, which corresponds to  what is known as an {\it alternate boundary condition} in AdS\cite{Klebanov:1999tb}.\footnote{In AdS such boundary conditions can break unitarity, unless the fields masses belong to a particular mass range. As we discuss in details in section~\ref{sec:Unitarity} this is not a concern for us since unitarity has a different implementation in dS.} Semi-classically, the new wave function $\tilde{\Psi} [\pi]$ is related to the original wave function by the Fourier-transformation\footnote{In the discussion below, we omit factors of $i$'s and minuses to avoid cluttering of notations.}
\be\label{eq:CFTFT}
\Psi^{\ast}[\phi]\sim\int D\pi \, e^{\int dx\phi(x)\pi (x)}\tilde{\Psi}[\pi]\period
\ee
The second step is to write a counterpart of \eqref{eq:CFTrepPsi} for $\tilde{\Psi}[\pi]$,
\be\label{eq:momentumCFT}
\tilde{\Psi}[\pi]=\left<e^{\int dx \tilde{\mathcal{O}}(x)\pi (x)}\right>_{\text{CFT}_{\tilde{\Psi}}}\comma
\ee
where CFT$_{\tilde{\Psi}}$ is a CFT that would generate the correlation functions of $\pi (x)$:
\be
\begin{aligned}
\tilde \Psi\l[\pi\r]=&\exp\l[\int dx_1 dx_2 \pi(x_1)\pi(x_2)\tilde G_2(x_1,x_2)\right.\\
&\qquad\left.+\int dx_1 dx_2 dx_3 \pi(x_1)\pi(x_2)\pi(x_3)\tilde G_3(x_1,x_2,x_3)+\ldots\r]\,.
\end{aligned}
\ee

The last step is to plug  \eqref{eq:CFTrepPsi}, \eqref{eq:CFTFT} and \eqref{eq:momentumCFT} into \eqref{corrfrompsi} and integrate out the source fields $\phi (x)$ and $\pi (x)$. In the semi-classical limit, this is a Gaussian integration and we get
\be\label{eq:coupledCFT}
\langle\phi (x_1)\ldots \phi (x_n)\rangle\sim \left<\tilde{\mathcal{O}}(x_1)\ldots \tilde{\mathcal{O}}(x_n)\,\,e^{\int dx \mathcal{O}(x)\tilde{\mathcal{O}}(x)}\right>_{\text{CFT}_{\Psi}\times \text{CFT}_{\tilde{\Psi}}}\period
\ee
In the semi-classical limit (namely free fields in dS), the single-trace operator spectrum of CFT$_{\Psi}$ and CFT$_{\tilde{\Psi}}$ would be related by the shadow transform, $\Delta_{\mathcal{O}_i}+\Delta_{\tilde{\mathcal{O}_i}}=d$, and the double-trace operators $\mathcal{O}_i\tilde{\mathcal{O}}_i$ become marginal \cite{Witten:2001ua}. Therefore \eqref{eq:coupledCFT} can be viewed as a marginal double-trace deformation of two coupled CFTs.


 In an interacting, but weakly coupled theory instead, the operators $\mathcal{O}_i$ and $\tilde{\mathcal{O}}_i$ have dimensions that do not have to add up to $d$. Thus the operators $\mathcal{O}_i \tilde{\mathcal{O}}_i$ will be weakly relevant or irrelevant, at least not exactly marginal. Nevertheless, it is tempting to conjecture that CFT$_C$ can still be described by conformal perturbation theory around CFT$_{\Psi}\times \text{CFT}_{\tilde{\Psi}}$ by appropriately tuning double-trace
couplings to critical values:
\be
\label{3cfts}
\text{CFT}_C\stackrel{?}{=}\text{CFT}_{\Psi}\times \text{CFT}_{\tilde \Psi}+\sum_i c_i \mathcal{O}_i \tilde{\mathcal{O}}_i\,.
\ee
This conjecture can be clearly tested in the perturbative examples of QFTs in dS that we study in this paper, and we will leave this as an interesting future problem. Note that CFT$_C$ has to be real, while the individual CFTs (CFT$_{\Psi}$ and CFT$_{\tilde \Psi}$) as well as their direct product are complex according to the definition given in \cite{Gorbenko:2018ncu}.\footnote{Real non-unitary CFTs can have complex anomalous dimensions as long as they come in complex conjugate pairs.} Thus if \eqref{3cfts} is correct, reality has to be restored by the double-trace deformation.

Suppose for the moment that some relation similar to \eqref{3cfts} is correct. How can it help in studying the cosmological theories with dynamical gravity in the bulk? A glimpse of hope lies in the fact that, as opposed to the relation \eqref{corrfrompsi}, the new proposal \eqref{3cfts} no longer involves a path integral over the metrics. Of course the complication related to the presence of residual boundary gravity did not completely go away, but it got moved to the calculation of $\tilde\Psi$. Indeed, AdS theories with alternate boundary conditions for the metric correspond to CFTs coupled to ``induced'' boundary gravity \cite{Leigh:2003ez,Compere:2008us,Giombi:2013yva}. Such theories do not contain the usual Einstein-Hilbert term in the action and \cite{Compere:2008us}\ has demonstrated that they are power-counting renormalizable. Thus they may be UV complete and possibly defined non-perturbatively. The next question is whether we can make sense of the corresponding coupling between the stress tensor of  CFT$_{\Psi}$ and the corresponding field in the theory CFT$_{\tilde \Psi}$\footnote{In the gravitational case calling this theory a CFT is an abuse of notation, it is a CFT coupled to the special Euclidean gravity}. A good starting point to address these issues is to consider either a three-dimensional bulk, where the boundary gravity is tractable \cite{Giombi:2013yva,Cotler:2019nbi}, or a dS-like solution in two-dimensional JT gravity, for which the alternate boundary conditions were also studied recently \cite{Goel:2020yxl}. Another direction is to analyze abelian or non-abelian gauge theories in dS, for which some of the subtleties discussed here are also present. Leaving all these interesting puzzles for the future, in the next section we come back to the problems that constitute the main topic of this paper and that we understand incomparably better.

\section{EAdS Lagrangian for dS correlators}
\label{sec:EAdS}
\subsection{The in-in formalism and dS propagators}\label{subsec:prop}
Let us start with a brief review of the in-in formalism routinely used in inflationary calculations. An in-in correlation function of some set of fields located at a given time slice at time $t$ is defined by \cite{Maldacena:2002vr, Weinberg:2005vy, Seery:2007we, Adshead:2009cb, Senatore:2009cf}
\be
\label{inin}
\l\langle \Omega(t_0)\right\vert\bar T \left(e^{i\int_{t_0}^t dt' H}\right)\phi(x_1,t)\phi(x_2,t)\ldots \phi(x_n,t) T \left(e^{-i\int_{t_0}^t dt' H}\right)\left\vert\Omega(t_0) \r  \rangle\,.
\ee
Here $T$/$\bar{T}$ denote time/anti-time ordering, and the state $\left\vert\Omega(t_0) \r  \rangle$ is the generalization of the Bunch-Davies vacuum to the interacting theory, which implies that very high energy modes are in the Minkowski vacuum of the theory. To compute these correlation functions in perturbation theory it is convenient to use the interaction picture, which is defined by splitting the full hamiltonian in a free and an interacting part $H = H_0 + H_I$ and using $H_0$ to evolve the operators, and $H_I$ to evolve the states. Then \eqref{inin} can be rewritten as (see e.g. the derivation in \cite{Senatore:2016aui})
\begin{equation}\label{ininpert}
\frac{\l\langle 0 \right\vert\bar T \left(e^{i\int_{-\infty(1+i \epsilon)}^t dt' H_I^{\rm int}}\right)\, \phi^{\rm int}(x_1,t)\phi^{\rm int}(x_2,t)\ldots \phi^{\rm int}(x_n,t) \,T \left(e^{-i\int_{-\infty(1-i \epsilon)}^t dt' H_I^{\rm int}}\right)\left\vert 0 \r  \rangle}{\l\langle 0 \right\vert\bar T \left(e^{i\int_{-\infty(1+i \epsilon)}^t dt' H_I^{\rm int}}\right)\,T \left(e^{-i\int_{-\infty(1-i \epsilon)}^t dt' H_I^{\rm int}}\right)\left\vert 0 \r  \rangle}\,.
\end{equation}
Here the superscript `$\rm int$' denotes the interaction picture, $|0\rangle$ is the free Bunch-Davies vacuum, i.e. the Fock vacuum, and the role of the denominator is to cancel out disconnected vacuum bubble diagrams. The perturbative expansion is then simply generated by expanding the anti time-ordered/time-ordered exponentials on the left and on the right, and computing the resulting correlation function using Wick contractions. In the following it will be understood that we are using \eqref{ininpert} for the perturbative calculations and we will drop the superscript `$\rm int$'. The same formula applies for insertions of composite operators rather than fields.

By applying the Wick theorem to \eqref{ininpert} we encounter three types of propagators, which we denote $G^{ll}$, $G^{rr}$ and $G^{lr}$, depending on whether the Wick contraction appears within the left, right, or between the left and right Hamiltonians. When the contraction appears between two external fields one is free to use any propagator since for spacelike separations they coincide. For a scalar field of mass $m$ the propagators are
\be
\label{dSprops}
G^{ll}_\nu=W_\nu\l(s+i\eps\r)\,, \quad G^{rr}_\nu=W_\nu\l(s-i\eps\r)\,, \quad G^{lr}_\nu=W_\nu\l(s-i\eps\,{\text{sgn}}(t_l-t_r)\r)\,.
\ee
Here the parameter $\nu$ is related to the mass of the particle by \eqref{eq:massNu}
and the two-point invariant $s$ is defined in \eqref{sdef}.
$W_\nu$ is proportional to a hypergeometric function:
\be
\label{defW}
W_\nu(s)=\frac{\Gamma\l(\frac{d}{2}\pm i \nu\r)}{(4 \pi)^{\frac{d+1}{2}}\Gamma\l(\frac{d+1}{2}\r)}\F\l(\ddt+i\nu,\ddt-i\nu,\frac{d+1}{2},\frac{1+s}{2}\r)~,
\ee
where we use the notation
\begin{equation}
\Gamma(a\pm b) \equiv \Gamma(a+b)\Gamma(a-b)~.
\end{equation}
The propagators $G^{ll}$, $G^{rr}$ and $G^{lr}$ correspond to the anti-time-order propagator, the time-ordered propagator and the Wightman function respectively.\footnote{This can be seen as follows: the $i\eps$ prescription is determined by the short distance limit, where $s-1\sim(t_2-t_1)^2-x_{12}^2$. Then $G^{rr}$ becomes the flat-space time-ordered (Feynman) propagator, which depends on $-(t_1-t_2)^2+x_{12}^2+i\eps$ (this corresponds to the standard $p^2-i\eps$ prescription in the momentum space). A similar argument works also for $G^{ll}$  and $G^{lr}$.}
 The $i\eps$ prescription in \eqref{ininpert}-\eqref{dSprops} is obtained if we add imaginary parts to the time variables as follows:
 \be
 \label{ieps}
 \eta^l=\eta^l(1+i\eps),\quad  \eta^r=\eta^r(1-i\eps)\,.
 \ee
As long as we are describing the same state in dS, the answer for correlation functions is, of course, independent of the coordinate system used. In Appendix~\ref{App:global} we explain how our discussion can be repeated in global coordinates.

The hypergeometric function in \eqref{defW} has a singularity at $s=1$. This corresponds to the interval between the two points becoming null. For $s<1$ the points are space-like separated and all propagators agree and are real, while in the time-like region $s>1$ their imaginary part is different due to different $i\eps$ prescriptions. These properties follow from choosing the principal branch of the hypergeometric function, for which we have a branch-cut on the real axis for $s> 1$.

The asymptotic behavior of the propagators is obtained from the large-$s$ expansion of \eqref{defW}:
\be
W_\nu(s)\underset{|s|\to\infty}{\approx} \frac{1}{4 \pi^{\ddt+1}}\l(\frac{\G\l(\ddt-i\nu\r)\G(i \nu)}{\l(-2 s\r)^{\ddt-i \nu}}+\frac{\G\l(\ddt+i\nu\r)\G(-i \nu)}{\l(-2 s\r)^{\ddt+i \nu}}\r)\,.\,
\ee
Here we only showed the two leading power-law behaviors, each of which is accompanied by a series of subleading integer-shifted powers. For light fields the power $\frac{d}{2}-i\nu$ dominates, and this implies that in the limit of large time the light field will correspond to an operator of definite scaling dimension $\Delta_-=\ddt-i\nu$.
On the other hand, for heavy fields both powers are of the same absolute value. For our discussion it will be convenient to consider correlators of operators that have definite scaling dimension, and this can be achieved for the heavy fields by inserting appropriate linear combinations of the fields and their first derivatives with respect to time (i.e. conjugate momenta). 

It is convenient to define the ``bulk-to-boundary'' propagator
\begin{align}
\begin{split}
\label{KdS}
K^{l(r)}_\nu(s) & = \frac{N_\nu}{ \l(-2 s+(-)i\eps\r)^{\ddt-i \nu}}\,,\\
N_\nu & \equiv \frac{1}{4 \pi^{\ddt+1}}\G\l(\tfrac{d}{2}-i\nu\r)\G(i \nu)~.
\end{split}
\end{align}
This propagator will be used for the Wick contraction with field insertions on a fixed time slice at a large time, which in the Poincar\'e coordinates is parametrized by $\eta_c\approx 0$. So more explicitly and suppressing the $i\epsilon$ prescription we have
\begin{equation}\label{eq:Ketac}
K_\nu(s) =N_\nu \left(\frac{-\eta_c \eta_1}{ \eta_1^2 - x_{12}^2}\right)^{\ddt-i \nu}~.
\end{equation}
Here $(\eta_1,x_1)$ are the coordinates of the ``bulk'' point and $x_2$ is the spatial coordinate of the insertion on the time slice at $\eta_c$. This expression for the ``bulk-to-boundary'' propagator is valid both for light fields and for insertions of heavy fields with dimension $\Delta_-$, while in the case of heavy field insertion with dimension $\Delta_+ = \frac{d}{2}+i\nu$ we simply need to consider the complex conjugate. 

\begin{figure}[t] 
	\centering
		\includegraphics[width=.8\linewidth,angle=0]{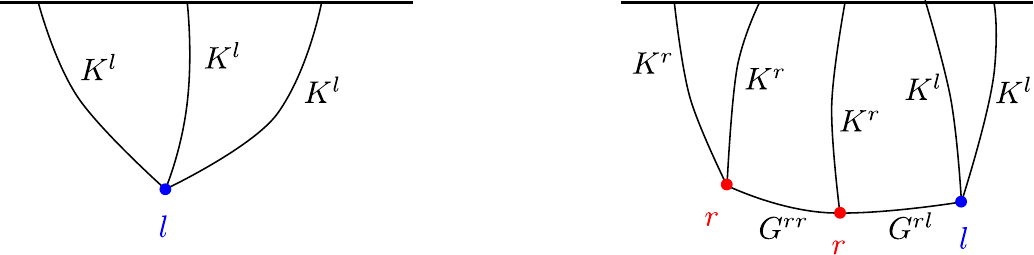}
	\caption{\small Examples of dS Feynman diagrams for the $\phi^3$ theory. Vertices are labeled by $r/l$ to denote if they are coming from the expansion of the time-ordered ``right'' time-evolution operator, or the anti time-ordered ``left'' one, in eq. \eqref{ininpert}. The propagators connecting the vertices to the insertions at the future boundary, or extending between two vertices, depend on the $r/l$ labels through the different $i\epsilon$ prescriptions.}
	\label{fig:phi3dS}
\end{figure}

The Feynman rules of the in-in perturbation theory in dS reduce to connecting a set of left and right vertices, as well as external operators, by means of the above defined propagators. Some examples of Feynman diagrams for the $\phi^3$ theory are showed in figure \ref{fig:phi3dS}.
Even though the perturbative formalism in dS space is straightforward in principle, practical calculations quickly become unwieldy. Traditionally, one goes to momentum space with respect to spacial coordinates, and keeps the time integrals explicit. Even for simple tree-level diagrams a direct evaluation of these integrals quickly becomes impossible. As a consequence, various alternative approaches have been developed, e.g.  based on analyticity and factorization properties in momentum space \cite{Arkani-Hamed:2015bza, Arkani-Hamed:2018kmz, Baumann:2019oyu, Baumann:2020dch}, and on the Mellin-Barnes representation \cite{Sleight:2019mgd, Sleight:2019hfp, Sleight:2020obc, Sleight:2021iix}. A further possibility is to employ the fact that the Euclidean version of dS space is a sphere and to perform the calculations on the sphere and then analytically continue the answers to dS \cite{Marolf:2010zp, Marolf:2010nz, Marolf:2011sh}. This approach allows to calculate certain diagrams, as well as to infer some important general properties of correlators which we are going to use later, see the appendix \ref{app:SR}. However, a direct evaluation of higher-point correlation functions on the sphere also appears impossible. The reason is that the sphere lacks an asymptotic boundary and one expects significant simplifications when computing asymptotic observables.

Instead, we are going to use a relation of de Sitter space to a different Euclidean manifold, the Euclidean Anti-de Sitter space (EAdS). This relationship was first explored in the context of the wave function calculations \cite{Maldacena:2002vr, Harlow:2011ke, Mata:2012bx, Anninos:2014lwa}, and more recently was used in the Mellin-Barnes  approach to dS correlators \cite{Sleight:2019mgd, Sleight:2019hfp, Sleight:2020obc}. In this series of papers the connection between AdS and dS tree-level exchange diagrams was established, which allowed the authors to obtain for the first time the explicit expressions for arbitrary masses and spins. Our claim is that any Feynman diagram which appears in the in-in formalism can be expressed as a linear combination of Witten diagrams in EAdS. Moreover, there is a local field theory on EAdS with twice the number of fields, as compared to the original theory, which generates all such diagrams. This implies that one can apply computational techniques developed for EAdS calculations, including the harmonic analysis on hyperbolic spaces. Before we discuss this in details, let us briefly review some basics of perturbation theory in EAdS for scalar field theories, in which we will mostly follow the notations of \cite{Carmi:2018qzm}. 

\subsection{Review of EAdS Feynman rules}
Let us consider the field theory in EAdS for a scalar particle of mass $m_{\rm AdS}$. The role of the propagator is played by the Green's function
\begin{align}
\begin{split}
\label{GAdS}
G_\nu^{\rm AdS}(s^{\rm AdS})& = \frac{\G\l(\ddt+i\nu\r)}{2\pi^\ddt \G\l(1+i\nu\r)\l(-2(s^{\rm AdS}+1)\r)^{\ddt+i\nu}}\\ &\hspace{4cm} \F\l(\ddt+i\nu,\half+i\nu,1+2i\nu,\frac{2}{s^{\rm AdS}+1}\r)\,.
\end{split}
\end{align}
Here the two-point invariant $s^{\rm AdS}$ is defined in \eqref{sAdSdef}
and $\nu$ is related to the mass of the particle by $\nu=\sqrt{-m_{\rm AdS}^2-\f{d^2}{4}}$.
At infinity this propagator behaves as $|s|^{-\ddt-i\nu}$, which corresponds to the ``ordinary'' boundary conditions. For masses above the BF bound, normally studied in AdS, $\nu$ is purely imaginary; however, below we are going to assume more general complex $\nu$'s with ${\text{Im}}\, \nu<0$. We will also need the propagator with ``alternate'' boundary conditions for a field with the same mass, given by  $G_{-\nu}^{\rm AdS}(s^{\rm AdS})$.

Operators inserted at the boundary of AdS are connected to (canonically normalized) bulk fields by the bulk-to-boundary propagators which are defined as 
\begin{align}
\begin{split}
\label{KAdS}
K_{\nu}^{\text{AdS}}(X_1,P_2) & = N^{\rm AdS}_\nu\left(\frac{z_1}{z_1^2 + x_{12}^2}\right)^{\frac{d}{2} + i \nu}\\
N^{\rm AdS}_\nu & \equiv \l(\frac{\Gamma(\frac{d}{2} + i\nu)}{2 \pi^{d/2} \Gamma(1+i\nu)}\r)^{\half}~.
\end{split}
\end{align}
Here $X_1$ is a bulk point with Poincar\'e coordinates $(z_1,x_1)$, $P_2$ is a boundary point with coordinate $x_2$, and the coefficient $N^{\rm AdS}_\nu$ is such that the boundary operators have unit-normalized two-point functions.
Note that, as compared to \eqref{KdS} we multiplied the propagator by a divergent factor $z_2^{-\frac{d}{2} -i \nu}$ in order to achieve a finite normalization of boundary operators. 

Another important object is the EAdS harmonic function 
\be
\label{OmegaAdS}
\Omega_\nu^{\rm AdS}(s^{\rm AdS})=\frac{W_\nu(s^{\rm AdS})}{\G(\pm i\nu)}\,,
\ee
where $W_\nu$ is defined in \eqref{defW}. Note that $s^{\rm AdS}<-1$, so no $i\eps$-prescription is necessary to define \eqref{GAdS} and \eqref{OmegaAdS}.
There are the following relations among the propagators and the harmonic function valid for ${\rm Im}\,\nu<0$:
\beg
\label{GOmegarel}
\Omega_\nu^{\rm AdS}=\frac{i\nu}{2\pi}\l(G^{\rm AdS}_\nu-G^{\rm AdS}_{-\nu}\r)\\
G^{\rm AdS}_\nu=\int_{-\infty}^{\infty} d\nu'\frac{1}{\nu'^2-\nu^2}\Omega_\nu^{\rm AdS}\\
G^{\rm AdS}_{-\nu}=\int_{-\infty}^{\infty} d\nu'\frac{1}{\nu'^2-\nu^2}\Omega_\nu^{\rm AdS}+\frac{2 \pi i}{\nu}\Omega^{\rm AdS}_\nu\,. 
\eeg

Perturbation theory in EAdS is generated by connecting interaction vertices and boundary sources by the propagators $G^{\rm AdS}_\nu$ and $K^{\rm AdS}_\nu$, or $G^{\rm AdS}_{-\nu}$ and $K^{\rm AdS}_{-\nu}$, depending on the boundary conditions for the fields. 

\subsection{Rotation to EAdS}\label{subsec:dStoAdS}

\begin{figure}[t] 
	\centering
		\includegraphics[width=0.33\linewidth,angle=0]{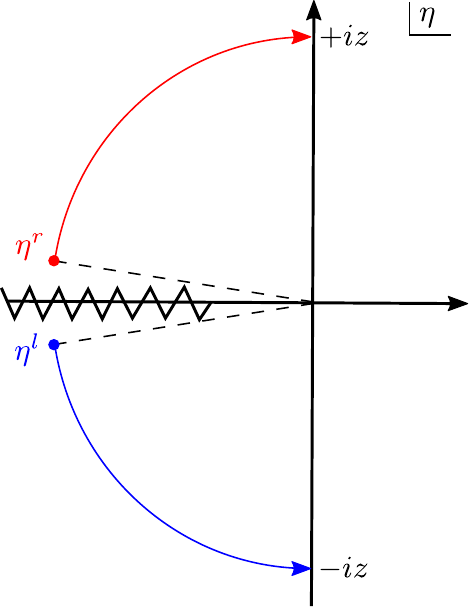}
	\caption{\small ``Wick rotation'' from dS to EAdS, in Poincar\'e coordinates. The branch-cut for real negative values of $\eta$ denotes the light-cone singularity and the associated discontinuity for time-like interval. The $r/l$ $i\epsilon$ prescriptions are illustrated with a displacement of the $\eta$ variable in the positive/negative imaginary direction. To connect to EAdS, we rotate the variable $\eta$ all the way to the imaginary axis avoiding the singularity on the real negative axis, i.e. clockwise for $\eta^r$ and anti-clockwise for $\eta^l$. This implies the identifications \eqref{eq:lrint}-\eqref{eq:llint} between the dS and EAdS two-point invariants.}
	\label{fig:EAdSrot}
\end{figure}

We will now show that integrals over vertex positions corresponding to any Feynman diagram in dS can be ``Wick rotated'' to become a linear combination of EAdS Feynman diagrams, for which the EAdS harmonic analysis can be used. As we discussed in subsection \ref{subsec:prop}, there are three different propagators appearing in in-in perturbation theory. Following \eqref{ieps}, the rotation we want to perform in order to avoid crossing any singularities of Feynman diagrams is
\be
\label{Wick}
\eta^l\to e^{\f{i\pi}{2}}\eta^l\,,\qquad \eta^r\to e^{-\f{i\pi}{2}}\eta^r\,,
\ee
and we then identify the absolute value of $\eta$ with EAdS radial coordinate $z$, as illustrated in figure \ref{fig:EAdSrot}.
Let us start with $G^{lr}_\nu$. This propagator is simply related to the EAdS harmonic function  \cite{Sleight:2019hfp}. Indeed, under \eqref{Wick} we get
\be\label{eq:lrint}
s(X_1^l,X_2^r)\to s^{\rm AdS}(X_1,X_2)\,, 
\ee
and consequently, using \eqref{OmegaAdS}, 
\be
\label{splitW}
G^{lr}_\nu(s)\to\G(\pm i\nu)\Omega_\nu^{\rm AdS}(s^{\rm AdS})=\frac{i\nu \G(\pm i\nu) }{2\pi}\l(G^{\rm AdS}_\nu (s^{\rm AdS})-G^{\rm AdS}_{-\nu}(s^{\rm AdS})\r)\,.
\ee
As a result, any left-right propagator after the Wick rotation gets replaced by the difference of two EAdS propagators with different boundary conditions, or simply by the harmonic function, up to an overall coefficient.

The situation is somewhat more subtle with the right-right and left-left propagators. Indeed, under \eqref{Wick}
\be\label{eq:llint}
s(X_1^{l(r)},X_2^{l(r)})\to - s^{\rm AdS}(X_1^{\rm AdS},X_2^{\rm AdS})\,, 
\ee
so we don't simply get the desired EAdS propagators. On the other hand, since the wave function of a QFT in dS can be obtained from a similar analytic continuation \cite{Maldacena:2002vr, Harlow:2011ke, Mata:2012bx, Anninos:2014lwa}, we should expect that left-left and right-right propagators --which basically encode the time evolution of the state in the interaction picture-- can also be obtained from EAdS. At this point it is useful to remember the following identities between the hypergeometric functions:
\be
\begin{aligned}
{}_2 F_{1}(a,b,c,z)=&(1-z)^{-a}{}_2 F_{1}\left(a,c-b,c,\frac{z}{z-1}\right)\comma\\
{}_2 F_{1}(a,b,c,z)=&\frac{\Gamma(c)\Gamma(c-a-b)}{\Gamma(c-a)\Gamma(c-b)}{}_2F_{1}(a,b,a+b+1-c,1-z)+\\
&\frac{\Gamma(c)\Gamma(a+b-c)}{\Gamma(a)\Gamma(b)}(1-z)^{c-a-b}{}_2F_{1}(c-a,c-b,1+c-a-b,1-z)\period
\end{aligned}
\ee
With their help we see that 
\beg
\label{splitGreen}
G_\nu^{ll}(s)\to\frac{i \nu}{2 \pi} {\G(\pm i\nu)}\l(G^{\rm AdS}_{\nu}(s^{\rm AdS})e^{ i \pi (\ddt+i\nu)}-G_{-\nu}^{\rm AdS}(s^{\rm AdS})e^{ i \pi (\ddt-i\nu)}\r)\,,\\
G_\nu^{rr}(s)\to\frac{i \nu}{2 \pi} {\G(\pm i\nu)}\l(G^{\rm AdS}_{\nu}(s^{\rm AdS})e^{ -i \pi (\ddt+i\nu)}-G_{-\nu}^{\rm AdS}(s^{\rm AdS})e^{- i \pi (\ddt-i\nu)}\r)\,,
\eeg
and thus these propagators also get replaced by linear combinations of AdS Green's functions. Equivalent relations for the Mellin-Barnes representations of the propagators were derived in \cite{Sleight:2020obc}.

We now apply the rotation to the external legs of the dS diagrams. Compare the definitions of the bulk-to-boundary propagators \eqref{KdS}-\eqref{eq:Ketac} and \eqref{KAdS} we get
\beg
\label{Krot}
K^{l}_\nu(s)\to(-\eta_c)^{\ddt-i\nu} e^{i\f{\pi}{2}\l(\ddt-i\nu\r)}\frac{N_\nu}{N^{\rm AdS}_{-\nu}}K^{\rm AdS}_{-\nu}(X,P)\\
K^{r}_\nu(s)\to(-\eta_c)^{\ddt-i\nu} e^{-i\f{\pi}{2}\l(\ddt-i\nu\r)}\frac{N_\nu}{N^{\rm AdS}_{-\nu}}K^{\rm AdS}_{-\nu}(X,P)\,.
\eeg
Note the explicit dependence on the time at which the external operators are inserted. We chose such normalization to emphasize that in dS we usually compute expectation values of bulk fields inserted at a large time $\eta_c$, rather than boundary operators.
 
It remains to continue analytically the integration measure. For each left or right vertex in dS we have, correspondingly,
\be
i\int \frac{d\eta_l}{(-\eta^l)^{d+1}}\to e^{-i\f{\pi}{2}(d-1)}\int \frac{d z}{z^{d+1}}\,,\quad -i\int \frac{d\eta_r}{(-\eta^r)^{d+1}}\to e^{i\f{\pi}{2}(d-1)}\int \frac{d z}{z^{d+1}}
\ee
At this point we have already shown that any in-in diagram gets replaced by a linear combination of EAdS Witten diagrams, which is enough for our computational purposes. In section \ref{sec:pert} we will see examples of application of this technique. For now we will proceed a bit further with the formal developments. Namely, we will show that the set of EAdS diagrams, obtained from a given theory in dS, corresponds to a (non-unitary) QFT defined directly on EAdS space. 

Before moving on, let us comment that the technique illustrated here basically relies on rotating to EAdS, and then using the decomposition in AdS harmonics to carry the calculation through. It would perhaps seem more natural to use a decomposition directly in dS harmonics. The reason we do not do that is simply that harmonic analysis on dS space contains additional subtleties as compared to EAdS. We give some more details about this in the appendix \ref{app:Gelfand}. 

\subsection{EAdS Lagrangian}\label{subsec:AdSL}

Our argument in this section will be purely perturbative, at the level of Feynman diagrams, but to any order in perturbation theory. For simplicity we will demonstrate this procedure in the example of a single scalar field with the Lagrangian
\be
{\cal{L}}=-\frac{1}{2} \d\phi\d\phi-\half m^2\phi^2-V(\phi)~.
\ee
The objective is to obtain an EAdS Lagrangian whose associated perturbative expansion precisely matches the one of this theory in dS, thanks to the relation between propagators that we derived in the previous subsection.

Let us start with the internal lines. It is standard in the in-in calculations to double the field content by introducing independent fields propagating on the left and on the right  sides of the contour. As we saw above, after our Wick rotation each propagator gets replaced by a pair of EAdS propagators corresponding to different boundary conditions. We thus first replace a single dS field by four AdS fields. Namely we take
\be
V^{\rm AdS}\l(\phi^l_+,\phi^l_-,\phi^r_+,\phi^r_-\r)=e^{-i\frac{\pi}{2}(d-1)}V(\phi^l_++\phi^l_-)+e^{i\frac{\pi}{2}(d-1)}V(\phi^r_++\phi^r_-)\,,
\ee 
and choose the ``$+$'' fields to have the following propagators:
\beg
\label{GM1}
\begin{pmatrix}
    G_+^{ll}(s^{\rm AdS}) & G_+^{lr}(s^{\rm AdS}) \\
    G_+^{rl}(s^{\rm AdS}) &  G_+^{rr}(s^{\rm AdS})
 \end{pmatrix}=\frac{i \nu}{2 \pi} {\G(\pm i\nu)}G_\nu^{\rm AdS}(s^{\rm AdS})
 \begin{pmatrix}
    e^{i\pi (\ddt+i\nu)} & 1 \\
    1 &  e^{-i\pi (\ddt+i\nu)} 
 \end{pmatrix}\,,
\eeg
while for the ``$-$'' fields we take 
\beg
\label{GM2}
\begin{pmatrix}
    G_-^{ll}(s^{\rm AdS}) & G_-^{lr}(s^{\rm AdS}) \\
    G_-^{rl}(s^{\rm AdS}) &  G_-^{rr}(s^{\rm AdS})
 \end{pmatrix}=-\frac{i \nu}{2 \pi} {\G(\pm i\nu)}G_{-\nu}^{\rm AdS}(s^{\rm AdS})
 \begin{pmatrix}
    e^{i\pi (\ddt-i\nu)} & 1 \\
    1 &  e^{-i\pi (\ddt-i\nu)} 
 \end{pmatrix}\,,
\eeg
and there is no mixing between the fields with different boundary conditions. This theory with four fields exactly reproduces the original dS theory. Next note that the matrices in the propagators above are degenerate. It means that only one linear combination actually propagates in each sector. 
Denoting these linear combinations as $\Phi_+$ and $\Phi_-$ we conclude that all Wick rotated Feynman diagrams are generated by the following EAdS Lagrangian 
\begin{align}
\begin{split}
\label{LAdS}
& {\cal{L}}^{\rm AdS}(\Phi_+,\Phi_-) =
 -i\,{\sinh\pi\nu} \l(\d\Phi_+\d\Phi_+- m^2\Phi_+^2\r) +  i\,{\sinh\pi\nu} \l(\d\Phi_-\d\Phi_-- m^2\Phi_-^2\r)\, \\
&\hspace{-0.65cm} -\,e^{-i\pi\frac{d-1}{2}}V\l(  e^{i\f{\pi}{2}  (\ddt+i\nu)}\Phi_++e^{i\f{\pi}{2}  (\ddt-i\nu)}\Phi_-\r)-e^{i\pi\frac{d-1}{2}}V\l(  e^{-i\f{\pi}{2}  (\ddt+i\nu)}\Phi_++e^{-i\f{\pi}{2}  (\ddt-i\nu)}\Phi_-\r)\,,
\end{split}
\end{align}
where the following boundary conditions for the fields are implemented to match the propagators in \eqref{GM1}, \eqref{GM2}:
\be
\Phi_{\pm}\approx z^{\ddt \pm i\nu}\,.
\ee
The first interaction term originates from the ``left'' vertices and second from the ``right'' vertices in dS.

Now we need to discuss the mapping of external fields between our dS and AdS theories. We look at the asymptotic behavior of the dS propagators, assuming that the external fields are light. At the leading order we can also consider heavy fields in external states. In this case the operator with definite scaling dimension requires a linear combination of $\phi$ and its time derivative. These fields are more subtle, however, due to the presence of contact terms in their correlation functions,  \cite{Matteo,Hogervorst:2021uvp}, as well as due to the fact that they move away from the principal series once interactions are included, see a related comment in section \ref{sec:Unitarity}. We see that to reproduce correlation functions in dS we need to insert the operators dual to the field $\Phi_-$ on the boundary of EAdS. Let us call this operator $\O_-$, which we assume to have a unit-normalized two-point function. We then need to find the normalization constant $\alpha$, such that 
\be
\l\langle \alpha \O_-(x_1)\alpha\O_-(x_2)\ldots\r\rangle_{\rm AdS}=\l\langle\phi(x_1,\eta_c)\phi(x_2,\eta_c)\ldots\r\rangle_{\rm dS},.
\ee
Note that the relative factors resulting from the rotation of $K^l$ and $K^r$ \eqref{Krot} are consistent with those in \eqref{LAdS}. Comparing the normalizations we see that 
\be
\alpha=(-\eta_c)^{\ddt-i\nu} \frac{(2 i{\sinh\pi\nu} )^{\half}N_\nu}{N^{\rm AdS}_{-\nu}}=(-\eta_c)^{\ddt-i\nu}\sqrt{N_{\nu}}\,.
\ee
It is straightforward to check that with this normalization disconnected contributions to the correlator (those where external fields are contracted with themselves) also agree in two theories.

Let us discuss the Lagrangian \eqref{LAdS} in some details. First of all note that the fields have opposite sign kinetic terms, so regardless of the value of $\nu$ one of them is always a ``ghost'', and the theory is not unitary in the EAdS sense.\footnote{The dS implications of unitarity are discussed in the later section \ref{sec:Unitarity}.} This is not a problem because we are not trying to interpret this Lagrangian as the Euclidean rotation of a theory on Lorentzian AdS. Next, the signs of the mass terms tell us that both fields have negative mass-squared from the EAdS point of view. Heavy dS fields, i.e. $m^2 > \frac{d^2}{4}$, correspond to EAdS fields below the BF bound. Light dS fields are above the bound, still for $0\leq m^2<\frac{d^2}{4}-1$ they lie outside the range where alternate boundary conditions are allowed in AdS \cite{Klebanov:1999tb}. These subtleties make the use of this formulation problematic beyond perturbation theory; however, it is does not lead to any difficulty at any order in perturbation theory.  

For light fields \eqref{LAdS} corresponds to a real, as opposed to complex, theory in the classification of \cite{Gorbenko:2018ncu}, since the Lagrangian is actually a real functional. This of course agrees with the fact that correlation functions in dS are real. The same is true for heavy fields, assuming complex conjugation exchanges $\Phi_+$ and $\Phi_-$.
\section{Analyticity}
\label{sec:Analyticity}
The EAdS Lagrangian derived in the previous section implies that quantum field theory in a rigid dS are in many ways similar to quantum field theory in (E)AdS, at least in perturbation theory. In particular, it suggests that the late-time correlation functions in dS admit an expansion akin to the one for the boundary correlation functions in EAdS; namely an expansion into a discrete sum in terms of conformal blocks. In this section, we focus on the late-time four-point function and examine various implications of this, at the same time clarifying important differences between (E)AdS and dS. The general discussions in this section will be borne out by explicit computations in section \ref{sec:pert}.
\subsection{Review of conformal partial waves and boundary correlators in EAdS}\label{subsec:CPWAdS}
Before discussing dS, let us first review basic properties of conformal field theory and boundary four-point functions in EAdS.
\paragraph{Conformal partial wave and conformal block.} In order to discuss the properties of the boundary four-point function, it is useful to introduce special functions called conformal partial waves (see \cite{Caron-Huot:2017vep,Simmons-Duffin:2017nub,Karateev:2018oml} for more details\footnote{For readers' convenience, let us provide a dictionary translating our notations to the notations in Appendix A of \cite{Simmons-Duffin:2017nub}: 
\be
\widehat{\mathcal{F}}=\Psi_{\text{\cite{Simmons-Duffin:2017nub}}}\comma\qquad \mathcal{C}=K_{\text{\cite{Simmons-Duffin:2017nub}}}\comma\qquad\widehat{\mathcal{K}}=G_{\text{\cite{Simmons-Duffin:2017nub}}}\period
\ee
}). The conformal partial waves are solutions to the {\it Casimir differential equation}, which are single-valued in $x_j$,
\be\label{eq:difeqcpw}
\mathcal{D}_{12}\cdot \widehat{\mathcal{F}}_{\Delta^{\prime},J}^{\{\Delta_j\}}(x_1,x_2,x_3,x_4)=c_{2}^{(\Delta^{\prime},J)}\widehat{\mathcal{F}}_{\Delta^{\prime},J}^{\{\Delta_j\}}(x_1,x_2,x_3,x_4)\comma
\ee
with
\be\label{eq:difc}
\mathcal{D}_{12}\equiv -\frac{1}{2}\left(\mathcal{L}_1^{AB}+\mathcal{L}_2^{AB}\right)\left(\mathcal{L}_{1,AB}+\mathcal{L}_{2,AB}\right)\period
\ee 
Here $\mathcal{L}_{j}^{AB}$ is a differential operator of $x_j$ which describes an action of a generator of~$SO(d+1,1)$. See \eqref{eq:differential} for definitions. The eigenvalue $c_{2}^{\Delta^{\prime},J}$ is given in terms of the conformal dimension $\Delta^{\prime}$ and spin $J$ as
\be\label{eq:eigenc}
c_{2}^{\Delta^{\prime},J}=\Delta^{\prime}(\Delta^{\prime}-d)+J(J+d-2)\period
\ee
Another important quantity is the {\it conformal block} $\widehat{\mathcal{K}}_{\Delta^{\prime},J}^{\{\Delta_j\}}$, with which the conformal partial wave $\widehat{\mathcal{F}}_{\Delta^{\prime},J}^{\{\Delta_j\}}$ can be expressed as 
\be\label{eq:CPW}
\widehat{\mathcal{F}}^{\{\Delta_j\}}_{\Delta^{\prime},J}(x_i)=\mathcal{C}_{d-\Delta^{\prime},J}^{\Delta_3,\Delta_4}\widehat{\mathcal{K}}_{\Delta^{\prime},J}^{\{\Delta_j\}}(x_i)+\mathcal{C}_{\Delta^{\prime},J}^{\Delta_1,\Delta_2}\widehat{\mathcal{K}}_{d-\Delta^{\prime},J}^{\{\Delta_j\}}(x_i)\comma
\ee
with
\be
\mathcal{C}_{\Delta,J}^{\Delta_1,\Delta_2}=\frac{\pi^{\frac{d}{2}}\Gamma(\Delta-\frac{d}{2})\Gamma (\Delta+J-1)\Gamma (\frac{d-\Delta+\Delta_1-\Delta_2+J}{2})\Gamma (\frac{d-\Delta+\Delta_2-\Delta_1+J}{2})}{(-2)^{J}\Gamma (\Delta-1)\Gamma (d-\Delta+J)\Gamma (\frac{\Delta+\Delta_1-\Delta_2+J}{2})\Gamma (\frac{\Delta+\Delta_2-\Delta_1+J}{2})}\period
\ee
In the literature, it is customary to define the conformal partial waves and the conformal blocks as functions of conformal cross ratios $(z,\bar{z})$ alone. This can be achieved by stripping off a universal prefactor from $\widehat{\mathcal{F}}$ and $\widehat{\mathcal{K}}$: 
\be\label{eq:stripping}
\begin{aligned}
\widehat{\mathcal{F}}^{\{\Delta_j\}}_{\Delta^{\prime},J}(x_i)&=\frac{1}{(x_{12}^2)^{\frac{\Delta_1+\Delta_2}{2}}(x_{34}^2)^{\frac{\Delta_3+\Delta_4}{2}}}\left(\frac{x_{14}^2}{x_{24}^2}\right)^{\frac{\Delta_2-\Delta_1}{2}}\left(\frac{x_{14}^2}{x_{13}^2}\right)^{\frac{\Delta_3-\Delta_4}{2}}\mathcal{F}^{\{\Delta_j\}}_{\Delta^{\prime},J}(z,\bar{z})\comma\\
\widehat{\mathcal{K}}_{\Delta^{\prime},J}^{\{\Delta_j\}}(x_i)&=\frac{1}{(x_{12}^2)^{\frac{\Delta_1+\Delta_2}{2}}(x_{34}^2)^{\frac{\Delta_3+\Delta_4}{2}}}\left(\frac{x_{14}^2}{x_{24}^2}\right)^{\frac{\Delta_2-\Delta_1}{2}}\left(\frac{x_{14}^2}{x_{13}^2}\right)^{\frac{\Delta_3-\Delta_4}{2}}\mathcal{K}_{\Delta^{\prime},J}^{\{\Delta_j\}}(z,\bar{z})\period
\end{aligned}
\ee
The resulting function $\mathcal{K}$ is what is known as the conformal block in the literature and we use the standard normalization throughout this paper (see e.g.~\cite{Simmons-Duffin:2017nub,Karateev:2018oml}).

\paragraph{Expansion of four-point functions.}
The boundary four-point functions in EAdS are known to admit the following representations, which we call the {\it spectral decomposition}\footnote{In this section, we mostly discuss the four-point functions of identical operators $\{\Delta_j\}=\{\Delta_{\mathcal{O}},\Delta_{\mathcal{O}},\Delta_{\mathcal{O}},\Delta_{\mathcal{O}}\}$ and therefore omit writing the superscript $\{\Delta_j\}$.}:
\be\label{eq:EAdSCP}
\begin{aligned}
\langle \mathcal{O}(x_1)\mathcal{O}(x_2)\mathcal{O}(x_3)\mathcal{O}(x_4)\rangle&=\frac{1}{|x_{12}|^{2\Delta_{\mathcal{O}}}|x_{34}|^{2\Delta_{\mathcal{O}}}}\sum_J\int^{\frac{d}{2}+i\infty}_{\frac{d}{2}}\frac{d\Delta^{\prime}}{2\pi i}\rho_{J}(\Delta^{\prime})\mathcal{F}_{\Delta^{\prime},J}(z,\bar{z})\\
&=\frac{1}{|x_{12}|^{2\Delta_{\mathcal{O}}}|x_{34}|^{2\Delta_{\mathcal{O}}}}\sum_J\int^{\frac{d}{2}+i\infty}_{\frac{d}{2}-i\infty}\frac{d\Delta^{\prime}}{4\pi i}\rho_{J}(\Delta^{\prime})\mathcal{F}_{\Delta^{\prime},J}(z,\bar{z})\comma
\end{aligned}
\ee
where $z$ and $\bar{z}$ are conformal cross ratios
\be\label{eq:CRdef}
z\bar{z}=\frac{|x_{12}|^2|x_{34}|^2}{|x_{13}|^2|x_{24}|^2}\comma\qquad (1-z)(1-\bar{z})=\frac{|x_{14}|^2|x_{23}|^2}{|x_{13}|^2|x_{24}|^2}\comma
\ee
and $\rho_J(\Delta^{\prime})$ is called the {\it spectral density}. In the second line of \eqref{eq:EAdSCP}, we used the ``shadow symmetry'' of the conformal partial wave and the spectral density, $\mathcal{F}_{\Delta^{\prime},J}=\mathcal{F}_{d-\Delta^{\prime},J}$ and $\rho_{J}(\Delta)=\rho_{J}(d-\Delta)$, to extend the range of integration.
The representations \eqref{eq:EAdSCP} (often called the {\it conformal partial wave decomposition}) can be derived directly from the harmonic analysis of the Euclidean conformal group\footnote{See also \cite{Hogervorst:2017sfd,Rutter:2020vpw} for a related, but different expansion, called the $\alpha$-space expansion.}. See for instance \cite{Caron-Huot:2017vep,Simmons-Duffin:2017nub,Karateev:2018oml} for more details. 

\begin{figure}[t] 
	\centering
		\hspace{-1cm}\includegraphics[width=0.6\linewidth,angle=0]{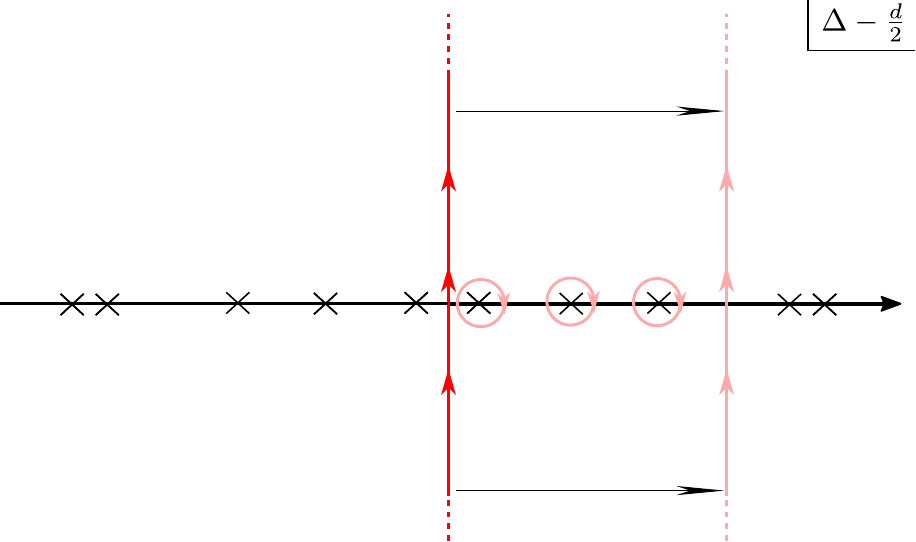}
	\caption{\small A contour deformation which produces the conformal block expansion \eqref{eq:confblock} from the spectral decomposition (or equivalently the conformal partial wave expansion) \eqref{eq:EAdSCP}. After rewriting the integrand, we can move the contour of integration to the right half plane. As a result, the integral picks up contributions from poles, which generate the conformal block expansion.}
	\label{fig:analyticityrhoAdS}
\end{figure}

Now, in conformal (field) theory on the boundary, it is customary to expand the four-point function into a discrete sum, called the conformal block expansion. Such an expansion can be derived from \eqref{eq:EAdSCP} by first decomposing $\mathcal{F}_{\Delta^{\prime},J}$ into a linear combination of conformal blocks $\mathcal{K}$ as (see \eqref{eq:CPW} for definitions)
$
\mathcal{F}_{\Delta^{\prime},J}=\mathcal{C}_{d-\Delta^{\prime},J}\mathcal{K}_{\Delta^{\prime},J}+\mathcal{C}_{\Delta^{\prime},J}\mathcal{K}_{\Delta^{\prime},J}
$,
where the first term decays exponentially on the right half plane $\Delta^{\prime}>\frac{d}{2}$ while the second term decays exponentially on the left half plane.
Inserting this expression into \eqref{eq:EAdSCP} and performing a change of variables ($\Delta^{\prime}\to d-\Delta^{\prime}$) in the second term, we obtain
\be\nonumber
\langle \mathcal{O}(x_1)\mathcal{O}(x_2)\mathcal{O}(x_3)\mathcal{O}(x_4)\rangle=\frac{1}{|x_{12}|^{2\Delta_{\mathcal{O}}}|x_{34}|^{2\Delta_{\mathcal{O}}}}\sum_J\int^{\frac{d}{2}+i\infty}_{\frac{d}{2}-i\infty}\frac{d\Delta^{\prime}}{2\pi i}\left(\rho_{J}(\Delta^{\prime})\mathcal{C}_{d-\Delta^{\prime},J}\right)\mathcal{K}_{\Delta^{\prime},J}\period
\ee
We can then pull the contour to the right (thanks to the asymptotic fall-off of $\mathcal{K}$ at ${\rm Re}\Delta\gg 1$) and replace the right hand side with a sum over contributions from poles (see figure \ref{fig:analyticityrhoAdS}). 
The result is given by
\be\label{eq:confblock}
\langle \mathcal{O}(x_1)\mathcal{O}(x_2)\mathcal{O}(x_3)\mathcal{O}(x_4)\rangle=\frac{1}{|x_{12}|^{2\Delta_{\mathcal{O}}}|x_{34}|^{2\Delta_{\mathcal{O}}}}\sum_{\Delta^{\prime}}C(\Delta^{\prime})\mathcal{K}_{\Delta^{\prime},J}(z,\bar{z})\comma
\ee
where $\sum_{\Delta^{\prime}}$ is a sum over poles of $\rho_{J}(\Delta^{\prime})$, and $C(\Delta^{\prime})$ is given by
\be
C(\Delta^{\prime})=-{\rm Res}_{\Delta=\Delta^{\prime}}\left[\rho_{J}(\Delta)\mathcal{C}_{d-\Delta,J}\right]\period
\ee
Note that in the derivation we assumed that $\rho_J(\Delta)$ is a meromorphic function of $\Delta$. This is guaranteed in perturbation theory in EAdS. However, beyond perturbation theory, we need to resort to a different logic which we explain below.
\paragraph{State-operator correspondence in EAdS and OPE.} The existence of the (discrete) expansion \eqref{eq:confblock} and its finite radius of convergence can be established without relying on perturbation theory. To see this, we need to use the {\it state-operator correspondence} for quantum field theory in EAdS, which is a natural generalization of the analogous statement for unitary CFT, and was first articulated in \cite{Paulos:2016fap}. Here we recall basic facts of this referring to the orginal paper for details.
\begin{figure}[t] 
	\centering
		\hspace{-1cm}\includegraphics[width=0.8\linewidth,angle=0]{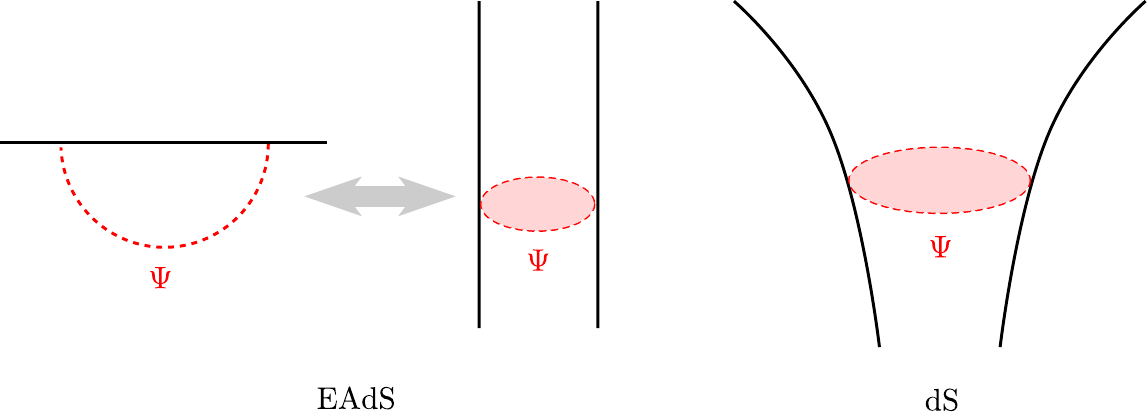}
	\caption{\small States in EAdS and dS. \textbf{Left:} States defined on a constant time slice in global AdS can be mapped to states defined on a hemisphere anchored at the boundary of AdS. The radius of the hemisphere shrinks to a point on the boundary under the action of dilatations. This provides an intuitive explanation of the state-operator correspondence. \textbf{Right:} States defined on a constant time slice in global dS. Unlike in AdS, there is no way to relate the time-slice to a point at future infinity using the action of the dS isometries. See more discussions in section \ref{subsec:dSanalytic}.}
	\label{fig:dSAdSstates}
\end{figure}

Consider insertions of local operators on the boundary of EAdS (see figure \ref{fig:dSAdSstates}). We then draw a hemisphere inside EAdS which is anchored on the boundary and surrounds some of those operators.  Performing a path integral in the interior of this hemisphere, we can replace the operators inside by a wave function defined on the hemisphere. Since this hemisphere can be mapped to a constant-time slice of global AdS by the AdS isometry, this wave function is in one-to-one correspondence with a state in global AdS. This argument can be applied both for single and multiple insertions of operators: For a single operator, this is nothing but the state-operator correspondence, which provides a map between an operator at the boundary to a state in global AdS. An intuitive (but  not very rigorous) way to understand this is as follows: by acting a dilatation transformation, which is one of the isometries of EAdS, we can change the radius of the hemisphere. In the limit where the hemisphere shrinks to a point, the wave function on the hemisphere has support only in an infinitesimal region around that point. This makes it intuitively clear that the wave function in the limit has as much information as an operator defined at that single point.

On the other hand, running this argument to multiple insertions, replacing them with a wave function on the hemisphere and converting it back to a sum of operators with definite scaling dimensions using the state-operator correspondence \cite{Rychkov:2016iqz,Simmons-Duffin:2016gjk} leads to the operator product expansion (OPE), which is given schematically by
\be
\mathcal{O}_1(x)\mathcal{O}_2(0)=\sum_{\mathcal{O}_3}c_{123}|x|^{\Delta_3-\Delta_1-\Delta_2}\mathcal{O}_3(0)+\cdots
\ee
Using this OPE inside the four-point function (and analyzing the constraints from conformal symmetry), we can establish the existence of the expansion \eqref{eq:confblock} at a fully non-perturbative level. This also allows us to interpret the sum as a sum over states in global AdS. Furthermore, by analyzing the series more carefully, we can also show that the expansion has a finite radius of convergence (see e.g.~\cite{Pappadopulo:2012jk,Hogervorst:2013sma}).

\subsection{Analytic structures of late-time correlators} \label{subsec:dSanalytic}
We now discuss the analytic structure of late-time correlation functions in dS.
\paragraph{Spectral density and spectral amplitude.} The starting point of our analysis is the spectral decomposition (or the conformal partial wave expansion) of the four-point function, which takes the same form as in EAdS:
\be\label{eq:dSCP}
\begin{aligned}
\langle \mathcal{O}(x_1)\mathcal{O}(x_2)\mathcal{O}(x_3)\mathcal{O}(x_4)\rangle&=\frac{1}{|x_{12}|^{2\Delta_{\mathcal{O}}}|x_{34}|^{2\Delta_{\mathcal{O}}}}\sum_J\int^{\frac{d}{2}+i\infty}_{\frac{d}{2}}\frac{d\Delta^{\prime}}{2\pi i}\rho_{J}(\Delta^{\prime})\mathcal{F}_{\Delta^{\prime},J}(z,\bar{z})\\
&=\frac{1}{|x_{12}|^{2\Delta_{\mathcal{O}}}|x_{34}|^{2\Delta_{\mathcal{O}}}}\sum_J\int^{\frac{d}{2}+i\infty}_{\frac{d}{2}-i\infty}\frac{d\Delta^{\prime}}{4\pi i}\rho_{J}(\Delta^{\prime})\mathcal{F}_{\Delta^{\prime},J}(z,\bar{z})\period
\end{aligned}
\ee
The existence of these representations can be proven both in perturbation theory and at a non-perturbative level: In perturbation theory, the EAdS Lagrangian derived in the previous section allows us to map all the diagrams to EAdS, and we therefore obtain the identical structure for the resulting four-point function. At a non-perturbative level, we can resort to a representation theory of the dS isometry as we discuss in detail in section \ref{sec:Unitarity}.

\begin{figure}[t] 
	\centering
	\begin{minipage}{0.45\hsize}
	\centering
		\includegraphics[scale=0.4]{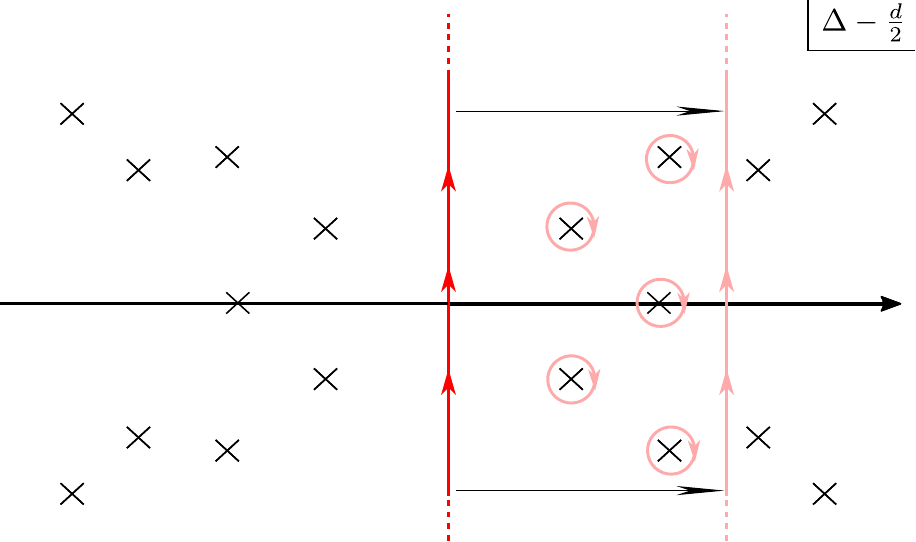}\\
		$\rho_J(\Delta)$
		\end{minipage}
		\begin{minipage}{0.45\hsize}
	\centering
		\includegraphics[scale=0.4]{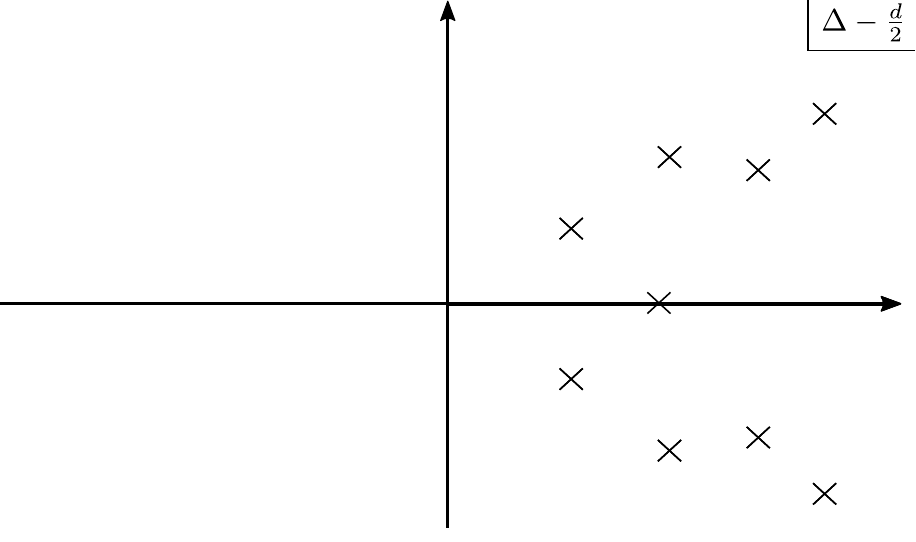}\\
		$f_J(\Delta)$
		\end{minipage}
	\caption{\small Analytic structures of the spectral density $\rho_J(\Delta)$ and the spectral amplitude $f_{J}(\Delta)$. {\bf  Left:} To all orders in perturbation theory, the spectral density of the late-time four-point function is also a meromorphic function thanks to the relation to EAdS discussed in section \ref{sec:EAdS}. We can therefore perform the same manipulation as in EAdS to produce a discrete sum over conformal blocks. Non-perturbatively, however, we currently do not have an argument for such an analytic property. {\bf Right:} Perturbative calculations often lead to a slightly different representation in which $\rho_J$ is replaced with a non-symmetric function $f_J$. In examples in perturbation theory, we can check that $f_J(\Delta)$ is analytic on the left half plane except for possible poles corresponding to the dS irreps.}
	\label{fig:analyticityrho}
\end{figure}

One interesting difference in dS is that, in perturbative calculations, we often land on a representation similar to the second line of \eqref{eq:dSCP}, but $\rho_J$ replaced with a function $f_{J}(\Delta^{\prime})$ which is not symmetric under $\Delta^{\prime} \to d-\Delta^{\prime}$. Instead, $f_J(\Delta^{\prime})$ has better analytic properties than $\rho_J$ (at least in perturbation theory): it is analytic on the left half plane ${\rm Re}\, \Delta<\frac{d}{2}$ except for possible poles corresponding to the unitary irreducible representations of dS isometries (such as the complementary and discrete series). This property is reminiscent of the partial wave amplitudes for flat space, namely the analyticity in the upper half plane, except for possible poles corresponding to bound states. Because of this reason, we call $f_{J}$ the {\it spectral amplitude}. It is related to $\rho_J (\Delta)$ by simple symmetrization:
\be
\rho_J(\Delta^{\prime})=\frac{f_{J}(\Delta^{\prime})+f_{J}(d-\Delta^{\prime})}{2}\period
\ee
The symmetrization produces singularities on both sides as can be seen in figure \ref{fig:analyticityrho}.

Unlike the spectral density $\rho_J$, at present we do not have a non-perturbative definition or a physical interpretation of $f_J$. One promising approach is to relate it to the physics of $S^{d+1}$: The reason for this optimism comes from the fact that an analog of the spectral amplitude can also be defined for the bulk two-point function, and there we can define it non-perturbatively by the analytic continuation of the coefficients of the harmonic expansion on $S^{d+1}$. See \eqref{eq:Amp2ptdef} in Appendix \ref{app:SR}. We feel it to be an important open problem to understand the physical meaning of $f_J$ and provide its non-perturbative definition.

\paragraph{Conformal block expansion.}
Now, {\it if the spectral density $\rho(\Delta)$ is a meromorphic function on the right half plane}, we can repeat the argument in EAdS, pull the contour to the right half-plane, and obtain  a conformal block expansion. In perturbation theory, this  is certainly true thanks to the mapping to EAdS we established. We therefore have a discrete sum that looks like a conformal block expansion (or equivalently an operator product expansion),
\be\label{eq:confblockdS}
\langle \mathcal{O}(x_1)\mathcal{O}(x_2)\mathcal{O}(x_3)\mathcal{O}(x_4)\rangle\overset{\text{perturbation}}{=}\frac{1}{|x_{12}|^{2\Delta_{\mathcal{O}}}|x_{34}|^{2\Delta_{\mathcal{O}}}}\sum_{\Delta^{\prime}}C(\Delta^{\prime})\mathcal{K}_{\Delta^{\prime},J}(z,\bar{z})\comma
\ee
The only difference from EAdS is that now the dimensions of the operators $\Delta^{\prime}$ are not real in general as we see in explicit examples in section \ref{sec:pert}. Apart from this difference, the analytic structure is identical to that for EAdS. We sketch the resulting analytic structure of $\rho(\Delta)$ in figure \ref{fig:analyticityrho}.

At this point, there are two important questions we can ask about \eqref{eq:confblockdS}:
\begin{itemize}
\item What is the physical interpretation of the sum $\sum_{\Delta^{\prime}}$?
\item Is the expansion valid also at a non-perturbative level?
\end{itemize}
Both of these questions are related to whether the state-operator correspondence exists in dS. Since this is a crucial difference between EAdS and dS, we will discuss it in more detail below.

\paragraph{(Im)possibility of state-operator correspondence in dS.} As explained in section \ref{subsec:CPWAdS}, the conformal block expansion in EAdS can be interpreted as a sum over states in the global AdS and this guaranteed the existence of the expansion at a non-perturbative level. So one might ask: {\it Is the conformal block expansion in dS related to a sum over states in dS?}

The answer is no. States in dS are defined on a constant-time slice in the global dS (or in the Poincare patch). Unlike the hemisphere discussed in section \ref{subsec:CPWAdS}, the constant-time slice in dS does not shrink to a point on the late-time surface under the action of dS isometries (see figure \ref{fig:dSAdSstates}).
This is an intuitive reason why the state-operator correspondence, at least its most straightforward generalization, fails in dS. We can also see this from representation theory: As mentioned above, the conformal block expansion in dS \eqref{eq:confblockdS} involves a sum over operators with arbitrary complex dimensions. These operators do not correspond to unitary irreducible representations of dS isometries and therefore cannot be interpreted in terms of states in the dS Hilbert space.

\begin{figure}[t]
\centering
\begin{minipage}{0.45\hsize}
\centering
\includegraphics[scale=0.35]{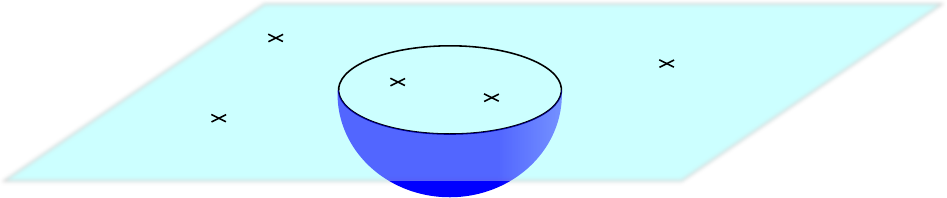}
\end{minipage}
\begin{minipage}{0.45\hsize}
\centering
\includegraphics[scale=0.35]{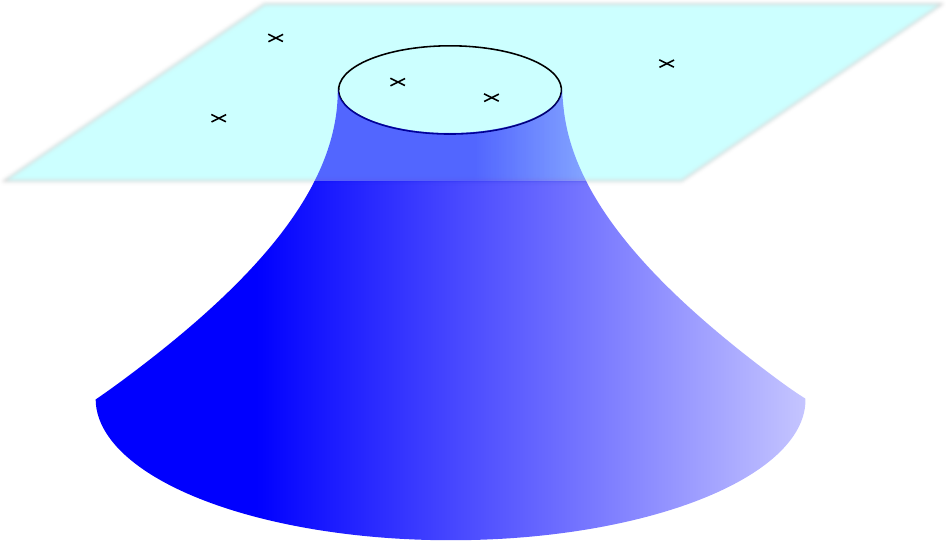}
\end{minipage}
\caption{\small Hemisphere in EAdS and its analog in dS. \textbf{Left:} By performing a path integral inside the hemisphere (the thick blue surface), we can replace the effect of insertions inside the hemisphere with a wave function defined on the surface. This establishes the state-operator correspondence. \textbf{Right:} On the other hand, an analog of the hemisphere in dS is given by a hyperboloid, which is non-compact and time-like. Therefore we cannot define a wave function on the surface in the standard sense.}\label{fig:HSdSAdS}
\end{figure}

That said, one could still ask if some unusual quantization of quantum field theory in dS can explain the sum $\sum_{\Delta^{\prime}}$ and implies a variant of a state-operator correspondence. For instance, one may try to consider an analog of the hemisphere in EAdS. This however does not work either because of the difference of the signature. In EAdS, the hemisphere centered at $(x^{\mu},z)=(0,0)$ in the Poincare coordinates is defined by
\be
|x|^{2}+z^2=R^2\comma
\ee
where $R$ is the radius of the hemisphere. It is invariant under the rotation around $x^{\mu}$ and the radius $R$ changes multiplicatively under the action of the dilatation. A surface that has similar transformation properties in dS is defined (in the Poincare coordinates) by
\be\label{eq:hyperboloid}
|x|^2-\eta^2=R^2\comma
\ee
which is a hyperboloid anchored at the late-time surface, see figure \ref{fig:HSdSAdS}. Although this transforms simply under the action of dS isometries, we cannot use it for defining states because the surface is time-like. 

Note that the arguments above do not exclude a possibility of choosing some space-like surface in dS (at the cost of sacrificing simple transformation properties under isometries) and defining an analog of state-operator correspondence. However at present, we do not know of any concrete realization of such an idea. Instead in the next subsection, we look more closely at the structure of the conformal block expansion in dS and suggest that the sum $\sum_{\Delta^{\prime}}$ may be interpreted as a sum over quasi-normal modes in a static patch of dS.

\subsection{OPE $=$ QNM?}\label{subsec:QNM}
We now discuss a possible interpretation of the OPE in terms of quasi-normal modes (QNM).

\begin{figure}[t] 
	\centering
		\hspace{-1cm}\includegraphics[width=0.25\linewidth,angle=-90]{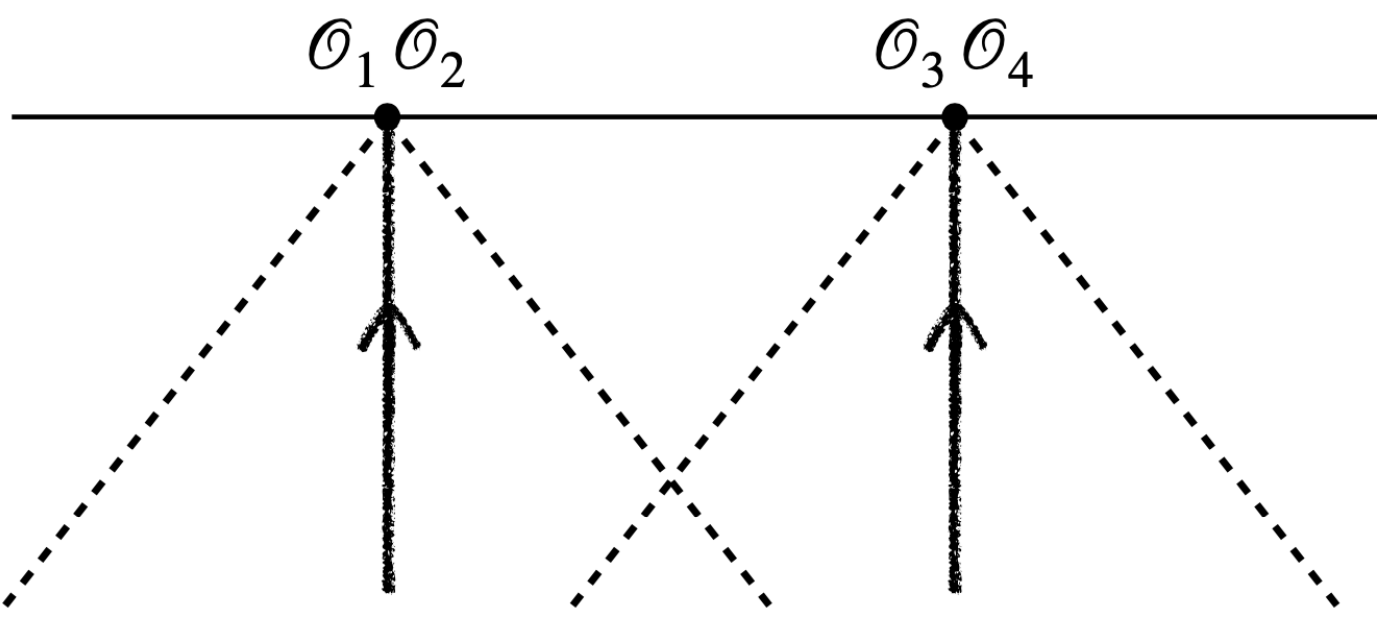}
	\caption{\small The evolution of operators in the bulk is indicated by the lines with arrows. Operators $O_1$ and $O_2$ are close by and evolve together, so do operators $O_3$ and $O_4$. Pairwise they evolve independently once their static patches stop to overlap and they get separated by cosmological horizons, indicated by the dashed lines.}
	\label{fig:qnm}
\end{figure}

Before discussing technical details, we give some intuitive (but not-meant-to-be-precise) arguments on why quasi-normal modes in a static patch of dS may be relevant for the state-operator correspondence and OPE. First, let us point out a simple geometric fact: the causal past of a point on the late-time surface includes
 a static patch of dS (see figure \ref{fig:patches}). This suggests that the physical information contained in an operator at that point is closely related to the physics in the static patch.  Second, the time evolution in the static patch governs the structure of the OPE and implies the existence of a convergence OPE expansion. To see this, let us consider the four-point functions at late time in which a pair of operators $\mathcal{O}_1$ and $\mathcal{O}_2$ are far separated from the other two ($\mathcal{O}_{3}$ and $\mathcal{O}_4$) and analyze them from the point of view of time evolution of particles dual to the operators, see figure~\ref{fig:qnm}. These particles interact most strongly when all of them are in the shared past lightcone of four operators (the overlapping region in figure \ref{fig:qnm}). As we time evolve such a state, the particles move away from that shared lightcone and get separated by the dS horizon. After that, the two sets of particles time-evolve independently by the static patch Hamiltonians and as a result the correlation between the two sets decays exponentially. Written more explicitly, this leads to an expression like
\be\label{eq:discretesumstatic}
\langle \mathcal{O}_1\mathcal{O}_2\mathcal{O}_3\mathcal{O}_4\rangle \sim \sum_{\Delta}e^{-2 \Delta t_{\ast}}\,,
\ee
where $t_{\ast}$ is the time (in the static patch coordinates) that passed since they were last in causal contact, and the factor of 2 comes from the independent evolutions inside the two  static patches. Since this expression orginates from the time-evolution in the static patch de Sitter, the discrete sum \eqref{eq:discretesumstatic} has a natural interpretation as the summation over the quasi-normal modes.  Now, because of the geometric structure of de Sitter, the time $t_{\ast}$ is related to the spatial distance between the two sets of operators $r$ as
\be
e^{-t_{\ast}}\sim \frac{1}{r}\,.
\ee
This suggests that the ``quasi-normal-mode expansion'' \eqref{eq:discretesumstatic} is related to the long distance expansion of the four-point function schematically given by
\be
\langle \mathcal{O}_1\mathcal{O}_2\mathcal{O}_3\mathcal{O}_4\rangle \sim \sum_{\Delta}r^{-2\Delta}\,.
\ee
In addition, the fact that the correlation rapidly decays once the particles are separated by the dS horizon suggests that the expansion has a good convergence property although this quantitative argument alone certainly is not enough for establishing such a claim or computing the radius of convergence.

We now attempt to make these simple intuitive arguments more quantitative by studying the structure of correlation functions. Our argument relies on the two facts: First, in examples in perturbation theory (and some examples beyond perturbation theory \cite{upcoming}), the spectral function $\rho_J$ for the late-time four-point function shares the same poles as the spectral function for the bulk two-point function. Second, the poles in the spectral function for the bulk two-point function control the behavior at large time-like separation and can be identified with the frequencies of quasi-normal modes in a static patch of dS.

The first point is more or less obvious in perturbation theory: take a tree-level exchange diagram as an example. The late-time four-point function can be computed by connecting pairs of boundary points to two bulk points by bulk-to-boundary propagators (see figure \ref{fig:dSexch}). Then the computation reduces to computing a integral of a product of bulk-to-boundary propagators and a bulk two-point function. We will perform such computations in sections \ref{sec:tree} and \ref{sec:compex}, and as one can see in \eqref{rho1loop}, the spectral function for the four-point function is given in terms of the bulk two-point function, and therefore shares the same pole structure. Let us also note that this is true in some examples beyond perturbation theory such as the large $N$ $O(N)$ model which we will discuss in an upcoming paper \cite{upcoming}.

The second point is rather technical and we refer to Appendix \ref{app:SR} for details. The main ideas are as follows: much like the boundary four-point functions, the bulk two-point functions in dS admits a spectral decomposition in terms of an integral along $\Delta=\frac{d}{2}+i\nu$ (plus discrete sums from other representations). See section \ref{sec:pos2pt} and appendix \ref{app:SR} for details. However, as we explain in Appendix \ref{app:SR}, such a representation is inappropriate for analyzing the asymptotic behavior of the time-like separated two-point function since the integral at large imaginary $\nu$ is only marginally convergent when the two-points are time-like separated. To overcome this problem, we need to follow what we did for deriving a conformal block expansion from the spectral decomposition; namely rewrite the integral, pull the contour to the right half plane and replace the integral with a sum over contributions from poles. The resulting expression has a better convergence even for a time-like separation and the positions of poles control the exponential decays in a static patch, which can be identified with quasi-normal modes.\footnote{We note some parallels between this discussion and the calculation of entropies and partition functions of various fields in dS done in \cite{Anninos:2020hfj}, where QNMs also manifest themselves in the analytic properties of the spectral density. It would be interesting to investigate the connection further.}

Combining these two observations, we conclude that there is a possibility of interpreting the OPE series in the late-time four-point functions in terms of quasi-normal modes in dS. However, at present, this is still a conjecture although it is well-motivated by what we learned in several examples. It is important to come up with non-perturbative justification of this statement. Furthermore, even if this conjecture is correct, it does not immediately provide a state-operator correspondence for dS. This is because the Hilbert space interpretation of the quasi-normal modes is tricky, the main difficulty being that
quasi-normal modes have, by definition, complex frequencies and therefore cannot be viewed as eigenstates of some physical observables. Nevertheless, there was an interesting proposal on the Hilbert space interpretation of quasi-normal modes for free scalar \cite{Jafferis:2013qia}, see also \cite{Anninos:2010gh}. It would be important to push the idea further and try to generalize it to interacting quantum field theory. 

Essentially the same statement was made in \cite{Arkani-Hamed:2015bza}, which observed that the power-laws appearing in the Taylor expansion of correlators in the squeezed limit in momentum space are also related to the QNMs of the static patch.

\subsection{Implications and physical interpretations}\label{sec:FAQ}
Let us now discuss implications of the analyticity of correlation functions and address some potentially confusing points.
\paragraph{Operators in late-time CFT and principal series.} In the past, it has been sometimes argued that the late-time CFT which holographically describes dS should have operators with dimension $\frac{d}{2}+i\nu$, namely operators corresponding to the principal series representation of the Euclidean conformal group. The implicit assumption behind this is that operators in the boundary CFT must belong to unitary representations of the bulk isometry. This is certainly true in AdS, but not in dS. In dS, what are classified by unitary representations are {\it states}, not the operators, and due to the absence of the state-operator correspondence, operators in the late-time CFT need not be in such unitary representations. We will see in several examples in section \ref{sec:Unitarity} that the conformal block expansion of late-time four-point functions involves operators with complex dimensions that do not belong to unitary representations of the Euclidean conformal group\footnote{In particular, in general we expect operators with arbitray large real part of the scaling dimension.}.

\begin{figure}[t] 
	\centering
		\includegraphics[width=0.4\linewidth,angle=-90]{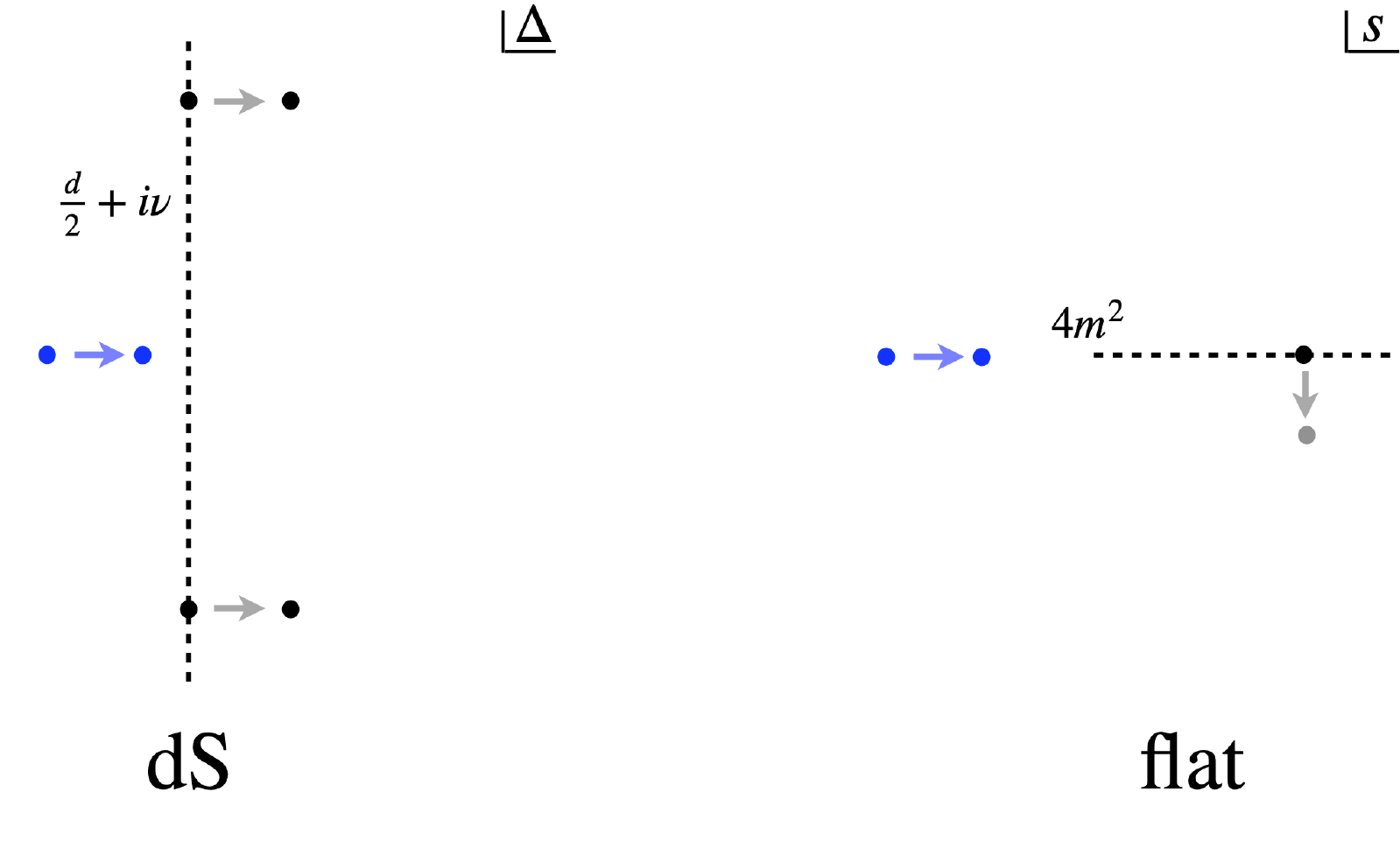}
	\caption{\small {\bf Left:} Poles in the spectral amplitude $f_J$ for the late-time four-point function. An exchange of a heavy particle produces a pair of poles (the black dots) on the principal series ${\rm Re}\Delta = \frac{d}{2}$ (the dashed line) at tree level. At finite coupling, they move off the principal series to the right and become resonance. On the other hand, a light particle gives a pole (the blue dot) in the region, ${\rm Re}\Delta<\frac{d}{2}$ and ${\rm Im}\Delta=0$, corresponding to the complementary series. Such a pole will continue to remain in the complementary series even after turning on the interaction weakly.  {\bf Right:} Analogous features in the partial wave amplitude in flat space. A pole on the two-particle branch cut (the black dot) moves to the second sheet after turning on the interaction while a pole below the threshold (the blue dot) is stable even after turning on the interaction weakly.}
	\label{fig:poleshift}
\end{figure}

In fact, we expect something stronger is true: 
\begin{itemize}
\item[]{\it For generic interacting quantum field theory in dS, we expect that there are no operators in the late-time CFT corresponding to the principal series representation}.
\end{itemize}
To see this, suppose we have a massive scalar particle in dS corresponding to the principal series. At tree-level, such a scalar shows up in an exchange diagram and therefore the spectral decomposition of the four-point function contains a pole precisely on the principal series. However, once we turn on the interaction, the mass of the particle gets renormalized and the pole moves off the principal series as can be seen in an example discussed in \ref{sec:compex} (see also figure \ref{fig:poleshift}). 
This is simply a dS analog of a process in flat space in which a stable particle becomes a resonance because of the interaction. In flat space, this almost always happens if the pole of a particle is on top of a continuous spectral density and we expect the same should be true in dS.

On the other hand, the situation is different for light particles belonging to the complementary series (see figure \ref{fig:poleshift}). Because of dS unitarity, a pole corresponding to such a particle cannot move away from the line of the complementary series until it hits $\Delta=\frac{d}{2}$, where it merges the principal series. In this sense, they are analogs of bound states below the two-particle threshold in flat space.

\paragraph{Possible analogy with Liouville.}Let us also point out that there is some analogy with Liouville field theory in two dimensions.  In Liouville field theory, the Hilbert space is spanned by normalizable modes that have Liouville momentum $\alpha=\frac{Q}{2}+i\nu$ with $Q$ being the Liouville background charge. However, when analyzing correlation functions, it is customary to use vertex operators $e^{\alpha\varphi}$ with $\alpha$ being a real number. Much like what we saw in dS, there is no state-operator correpondence for such vertex operators since they correspond to non-normalizable modes in the Hilbert space. See related discussions in the conclusion section for potential holographic microscopic realizations of dS.

\paragraph{Two-particle states and resonances.} For a free scalar in EAdS, the CFT on the boundary will contain operators dual to multi-particle states in addition to a single-particle state. As long as the mass of the particle is non-tachyonic\footnote{More precisely speaking, it must be above the Breitenlohner-Freedman bound.}, all these operators belong to unitary representations of the Lorentzian conformal group. Within perturbation theory, these operators survive even after we turn on interactions. For instance, the existence of the  double-trace operators can be confirmed by performing the conformal block expansion of simple EAdS Feynman four-point diagrams such as a contact diagram or an exchange diagram. 
\begin{figure}[t] 
	\centering
		\hspace{1cm}\includegraphics[width=0.5\linewidth,angle=0]{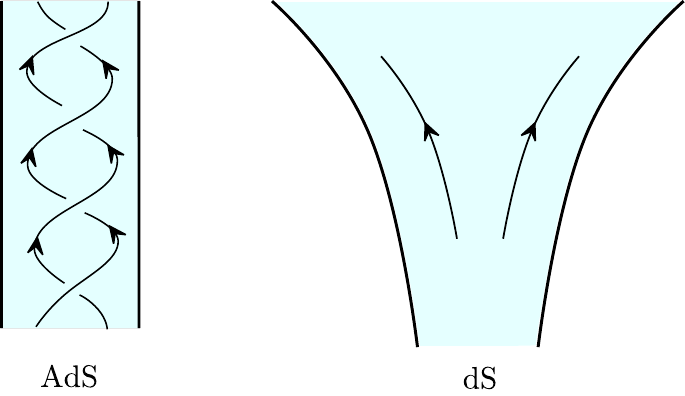}
		\caption{\small Two-particle states in AdS and dS. In AdS, the double-trace operators correspond to states in which two particles are orbiting each other. This is possible thanks to the gravitational potential in AdS. In dS, such states cannot exist as stable states and become resonances, owing to the expansion of dS.}
	\label{fig:dSAdS2pt}
\end{figure}

The situation is different in dS in several ways. Consider a free heavy scalar in dS whose dual operator belongs to the principal series $\frac{d}{2}+i\nu$. The first difference is that, because of the doubling of fields discussed in section \ref{sec:EAdS}, we will have three series of two-particle states, whose dimensions are
\be
2\left(\frac{d}{2}+i\nu\right)+n\comma\quad 2\left(\frac{d}{2}-i\nu\right)+n\comma\quad \left(\frac{d}{2}+i\nu+\frac{d}{2}-i\nu\right)+n=d+n \comma 
\ee
where $n$ is a positive integer. These poles can also be seen in Feynman diagrams in interacting theories as we see in section \ref{sec:pert}. Second, unlike EAdS, none of these two-particle states corresponds to a unitary representation. Instead, they are analogs of resonances in flat space. This may be slightly counter-intuitive since the resonances in flat space arise only after turning on the interaction while here they exist already in free theory. Intuitively this can be understood as a consequence of the expansion of dS: In EAdS, these states correspond to two particles orbiting each other. Such states cannot exist as stable states in dS since the dS space is expanding and particles tend to move away from each other (see figure \ref{fig:dSAdS2pt}). This is a basic physical reason why such states can only make sense as resonances.
\begin{figure}[t] 
	\centering
		\hspace{1cm}\includegraphics[width=0.7\linewidth,angle=0]{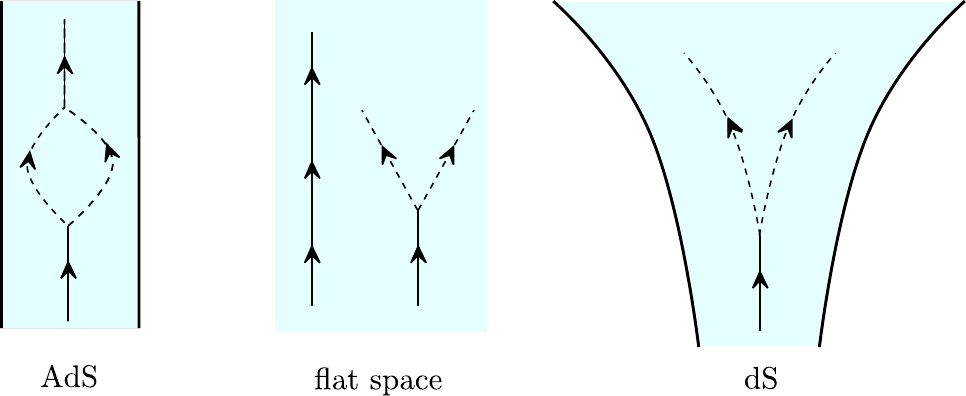}
	\caption{\small Decay of a particle in AdS, flat-space and dS. In AdS, the gravitational potential bounces back everything and particles can never be unstable. In flat space, most of heavy particles can decay while some particles (bound states below the threshold) can remain stable. In dS, particles are more likely to decay because of the expansion of the spacetime.}
	\label{fig:dSAdSflatdecay}
\end{figure}

\paragraph{Almost everything is a resonance in de Sitter.}More generally, there is an interesting qualitative contrast among AdS, flat space and dS (see also figure \ref{fig:dSAdSflatdecay}). In AdS, all the states are stable because the gravitational potential near the boundary bounces back everything and a particle can never decay\footnote{See \cite{Carmi:2018qzm} for more details on this point, in particular about how the AdS space reproduces the resonance physics in flat space when the flat-space limit is taken. }. On the other hand, in flat space, particles can decay and only a few states are stable while others are resonances. By contrast, in dS, almost all the states are unstable resonances because of the dS expansion as we saw above.

\section{Unitarity}
\label{sec:Unitarity}
In quantum field theory in flat Minkowski space and in AdS, unitarity can be translated to certain positivity constraints on the basic observables, namely the S-matrix and the boundary correlation functions. Those positivity constraints are essential to formulating the numerical bootstrap for these obervables and also to obtaining bounds on the low-energy Wilson coefficients (see \cite{Bellazzini:2020cot,Tolley:2020gtv,Arkani-Hamed:2020blm,Caron-Huot:2020cmc,Sinha:2020win,Chiang:2021ziz,Caron-Huot:2021rmr,Caron-Huot:2021enk,Guerrieri:2021ivu,Kundu:2021qpi} for very recent developments in this direction).

In this section, we study constraints from unitarity on the cosmological correlators. Unlike the discussions in the preceding sections, the derivation of these constraints does not rely on perturbation theory and therefore is valid at a nonperturbative level. Since the argument involves several steps, let us first give a preview of the outcome. Both in EAdS and dS, unitarity implies positivity, however of different quantities: As reviewed in section \ref{subsec:CPWAdS}, the boundary correlation functions in EAdS can be expanded into a discrete series called the conformal block expansion\footnote{Here we assumed for simplicity that all the operators are identical, but a similar expansion exists also for non-identical operators.},
 \be\label{eq:AdSexpansion}
 \langle \mathcal{O}(x_1)\mathcal{O}(x_2)\mathcal{O}(x_3)\mathcal{O}(x_4)\rangle=\frac{1}{|x_{12}|^{2\Delta_{\mathcal{O}}}|x_{34}|^{2\Delta_{\mathcal{O}}}}\sum_{\Delta^{\prime}}C_{\Delta^{\prime}}\mathcal{K}_{\Delta^{\prime},J}(z,\bar{z})\,.
 \ee
 In this expansion, imposing unitarity amounts to requiring the coefficients of the expansion $C_{\Delta^{\prime}}$ to be positive. The discrete sum in \eqref{eq:AdSexpansion} comes from summing over all possible unitary representations of the AdS isometry group, which is the {\it Lorentzian} conformal group. By contrast, late-time correlation functions in dS can be expanded into a sum and an integral over all possible unitary representations of the {\it Euclidean} conformal group, which is the isometry of dS. The resulting expression would look like\footnote{Our derivation of positivity constraints can be generalized also to other representations, but for most of the analyses performed in this paper, it is enough to focus on the principal series representation, which we will do also in what follows.} 
 \be\label{eq:EuInv}
 \begin{aligned}
 &\langle \mathcal{O}(x_1)\mathcal{O}(x_2)\mathcal{O}(x_3)\mathcal{O}(x_4)\rangle=\\
 &\frac{1}{|x_{12}|^{2\Delta_{\mathcal{O}}}|x_{34}|^{2\Delta_{\mathcal{O}}}}\left[\sum_{J}\int_{\frac{d}{2}}^{\frac{d}{2}+i\infty} \frac{d\Delta^{\prime}}{2\pi i}\rho_J(\Delta^{\prime})\mathcal{F}_{\Delta^{\prime}}(z,\bar{z})+(\text{others})\right]\,,
 \end{aligned}
 \ee
 Here the terms written down correspond to the principal series representation and can be identified with the spectral decomposition discussed in section \ref{subsec:dSanalytic}, while the terms denoted by $(\text{others})$ correspond to other representations such as the complementary series or the discrete series. Then the unitarity constraint becomes the positivity of the coefficient,
 \be
 \rho_J(\Delta^{\prime})\geq 0\period
 \ee
 
 Note that the boundary four-point functions in AdS (or more generally the four-point functions in unitary CFT) also admits an expansion like \eqref{eq:EuInv} as reviewed in section \ref{subsec:CPWAdS}. However in that case, the coefficients $\rho_J(\Delta)$  need not be positive.  
 
 \subsection{Recap: two-point function in flat space} As a warm-up, let us first recap the derivation of unitarity constraints on the two-point function in flat space
 \be\label{eq:flat2pt1}
 \mathcal{G}(x_1,x_2)\equiv \langle \Omega | \phi (x_1)\phi (x_2)|\Omega\rangle\period
 \ee
 Here $|\Omega\rangle$ is the Poincare-invariant vacuum state. As the first step, we introduce a projection operator to a sector with momentum $p$ 
 \be\label{eq:expPp}
 \mathcal{P}_{p}\equiv \sum_{p_{\psi}=p}|\psi\rangle\langle \psi|\comma 
 \ee
 where $\sum_{p_{\psi}=p}$ means a sum over all possible states with momentum $p$, $p^0>0$, $p_\mu p^\mu<0$. We then write a resolution of the identity
 \be
 {\bf 1}=\int d^{d}p\,\, \mathcal{P}_p\comma
 \ee
 and insert it to \eqref{eq:flat2pt1}.
 As a result we obtain
 \be
 \begin{aligned}
 \mathcal{G}(x_1,x_2)&=\int d^{d}p\,\langle \Omega |\phi(x_1)\mathcal{P}_{p}\phi(x_2)|\Omega\rangle\period
 \end{aligned}
 \ee
 
 Next we impose the symmetry constraint; namely we write the operator $\phi(x)$ as $e^{i\hat{P}\cdot  x}\phi(0)e^{-i\hat{P}\cdot x}$ with $\hat{P}_{\mu}$ being the translation operator. Using the translational invariance of the vacuum state and the property of the projection operator $\hat{P}_{\mu}\mathcal{P}_{p}=\mathcal{P}_{p}\hat{P}_{\mu}=p_{\mu}\mathcal{P}_{p}$, we can determine the coordinate dependence of $\langle \Omega |\phi(x_1)\mathcal{P}_{p}\phi(x_2)|\Omega\rangle$ to be
 \be
 \langle \Omega |\phi(x_1)\mathcal{P}_{p}\phi(x_2)|\Omega\rangle=C(p)e^{ip \cdot (x_1-x_2)}\period
 \ee
 Here $C(p)$ is a constant of proportionality given by 
 \be
 C(p)\equiv \langle \Omega |\phi(0)\mathcal{P}_{p}\phi(0)|\Omega\rangle\period
 \ee

Finally we relate $C(p)$ to a sum of squares   using \eqref{eq:expPp}:
 \be
 C(p)=\sum_{p_{\psi}=p}\left|\langle \Omega|\phi(0)|\psi\rangle\right|^2\period
 \ee
 This makes manifest that the coefficients of the expansion
 \be
 G(x_1,x_2)=\int d^{d}p\,C(p)\,e^{i p(x_1-x_2)}\comma
 \ee
 must be positive $(C(p)\geq 0)$ in unitary theories. In this discussion we ignored the $i\eps$ prescription, however, note that the position-space two-point function we study is the Wightman one. The time-ordered correlator need not be positive in momentum space, see \cite{Gillioz:2018mto} for explicit examples of this in CFTs.
 
 As this exercise shows, the derivation of the positivity consists of three main steps:
 \begin{enumerate}
 \item Decompose the correlation function into contributions from each irreducible representation by inserting a resolution of identity $({\bf 1}=\int d^{d}p\,\mathcal{P}_{p})$.
 \item  Impose the symmetry constraint and express the contribution from each representation as a dynamical prefactor $(C(p))$, which is theory-dependent, times a universal factor determined purely by the symmetry $(e^{ip\cdot (x_1-x_2)})$.
 \item Relate the dynamical prefactor to a sum of squares of overlaps and establish positivity.
 \end{enumerate}
 In the rest of this section, we demonstrate that the same logic can be applied to correlation functions in dS.
 
 \subsection{Positivity of bulk two-point function in de Sitter} \label{sec:pos2pt}
 
 We now explain how the argument above can be generalized to the ``bulk'' two-point function in dS; namely the two-point function at finite time. 
 
\begin{figure}[t] 
	\centering
		\includegraphics[width=0.35\linewidth,angle=-90]{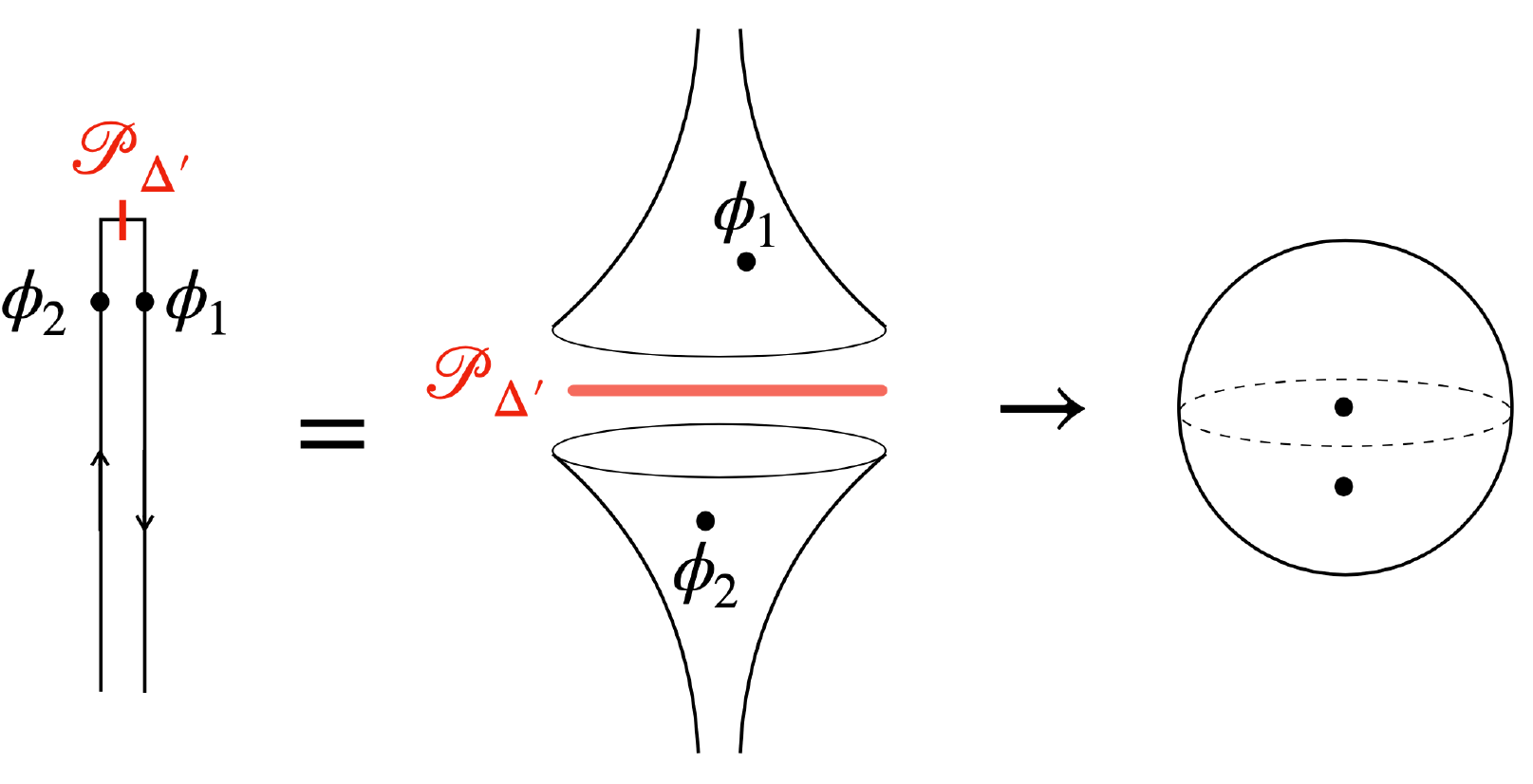}
	\caption{\small A pictorial explanation of a resolution of identity and the proof of positivity. {\bf Left:} The insertion of the projector corresponds to cutting the in-in path integral and inserting a set of states corresponding to the representation $\Delta^{\prime}$. {\bf Middle:} Geometrically, this is equivalent to considering two sets of dS and inserting a resolution of identity. {\bf Right:} To prove the positivity, we analytically continue the two points to sphere and place them symmetrically across the equator (the dashed circle).}
	\label{fig:2ptcutting}
\end{figure}

 Let $X_1$ and $X_2$ be space-like separated points in dS and consider the two-point function
 \be
 \mathcal{G}(X_1,X_2)\equiv \langle \Omega | \phi (X_1)\phi (X_2)|\Omega\rangle\period
 \ee
Here we took the two operators to be identical scalar operators\footnote{Note that $\phi$'s do not need to be elementary fields for the following discussion.} for simplicity and $|\Omega\rangle$ is the Bunch-Davies vacuum, which is invariant under the dS isometry. As with the previous subsection, the first step is to write a resolution of the identity. In the present case, the resolution of the identity is given by a sum and an integral of all possible unitary irreducible representations of the dS isometry, which we reviewed in section \ref{subsec:basics}. Since $\phi$ is a scalar operator, here we only need the principal series and the complementary series
\be
{\bf 1}=\underbrace{\int_{\frac{d}{2}}^{\frac{d}{2}+i\infty} \frac{d\Delta^{\prime}}{2\pi i} \mathcal{P}_{\Delta^{\prime}}}_{\text{principal}}\,\,+(\text{complementary})\period
\ee
Inserting this to the two-point function, we obtain
\be\label{eq:2ptdSdecomp}
\mathcal{G}(X_1,X_2)=\int_{\frac{d}{2}}^{\frac{d}{2}+i\infty} \frac{d\Delta^{\prime}}{2\pi i} \langle \Omega | \phi (X_1)\mathcal{P}_{\Delta^{\prime}}\phi (X_2)|\Omega\rangle +(\text{complementary})\period
\ee
Let us make a remark on this expression. As reviewed in section \ref{subsec:prop}, the correlation functions in dS are defined by a path integral along the in-in (or Schwinger-Keldysh) contour. From this path-integral point of view, the insertion of the projector in \eqref{eq:2ptdSdecomp} corresponds to splitting the left and the right contours of the in-in path integral and inserting a complete set of states at the late-time point where the two contours are glued. One can also visualize this as two dS spacetimes, each of which contains a single operator, glued together with the insertion of the projector in the middle (see figure \ref{fig:2ptcutting}).  

We next impose the symmetry constraints. To do so, we act the quadratic Casimir of the dS isometry to $\phi(X_1)$. This amounts to considering a double commutator
\be\label{eq:doublecomm}
 \langle \Omega |\left(\mathcal{C}_2\cdot \phi (X_1)\right)\mathcal{P}_{\Delta^{\prime}}\phi (X_2)|\Omega\rangle\comma
\ee
where
\be
\mathcal{C}_2\cdot \phi (X)\equiv- \frac{1}{2}\sum_{A,B}[\hat{M}^{AB},[\hat{M}_{AB},\phi (X)]]\comma
\ee
and $\hat{M}_{AB}$'s are generators of the isometry. Using the invariance of the Bunch-Davies vacuum $\langle \Omega|\hat{M}_{AB}=0$ and the fact that $\mathcal{P}_{\Delta^{\prime}}$ projects to an eigenspace of the Casimir, the equation \eqref{eq:doublecomm} can be re-expressed as
\be
-\langle \Omega|\phi (X_1)\left(\frac{1}{2}\sum_{A,B}\hat{M}^{AB}\hat{M}_{AB}\right)\mathcal{P}_{\Delta^{\prime}}\phi (X_2)|\Omega\rangle=\Delta^{\prime}(\Delta^{\prime}-d)\underbrace{\langle \Omega|\phi (X_1)\mathcal{P}_{\Delta^{\prime}}\phi (X_2)|\Omega\rangle}_{\equiv g_{\Delta^{\prime}}(X_1,X_2)}\period
\ee
On the other hand, the action of the dS isometry $\hat{M}_{AB}$ on $\phi (X)$ can be replaced by differential operators acting on $X$ (see \eqref{eq:differential}). As a result, the double-commutator in \eqref{eq:doublecomm} coincides with the action of the dS Laplacian:
\be
\langle \Omega |\left(\mathcal{C}_2\cdot \phi (X_1)\right)\mathcal{P}_{\Delta^{\prime}}\phi (X_2)|\Omega\rangle=-\Box_{X_1}\left(\langle \Omega |\phi (X_1)\mathcal{P}_{\Delta^{\prime}}\phi (X_2)|\Omega\rangle\right)\period
\ee
Equating the two expressions, we find that $f_{\Delta^{\prime}}(X_1,X_2)$ satisfies the differential equation
\be
-\Box_{X_1}g_{\Delta^{\prime}}(X_1,X_2)=\Delta^{\prime}(\Delta^{\prime}-d)g_{\Delta^{\prime}}(X_1,X_2)\comma
\ee
which coincides with the differential equation for the Green's function\footnote{In principle, there is another independent solution to this equation since it is a second-order differential equation. However, the other solution has a singularity at $\zeta=4$ and therefore is not appropriate for expanding the bulk two-point function which is non-singular at that point.} in dS.  
We therefore obtain
\be\label{eq:fandG}
g_{\Delta^{\prime}}(X_1,X_2)=\rho(\Delta^{\prime})G_{\Delta^{\prime}}(X_1,X_2)\comma
\ee
where $\rho(\Delta)$ is a constant of proportionality and $G_{\Delta}$ is the Green's function given by
\be
G_{\Delta}(X_1,X_2)= \frac{\Gamma (\Delta)\Gamma(d-\Delta)}{(4\pi)^{\frac{d+1}{2}}\Gamma\left(\frac{d+1}{2}\right)}{}_2F_{1}\left(\Delta,d-\Delta,\frac{d+1}{2},\frac{1+s (X_1,X_2)}{2}\right) \period
\ee 
Combining \eqref{eq:2ptdSdecomp} and \eqref{eq:fandG}, we obtain a decomposition of the two-point function,
\be\label{eq:deSitterdecomp2pt}
\mathcal{G}(X_1,X_2)=\int_{\frac{d}{2}}^{\frac{d}{2}+i\infty}\frac{d\Delta^{\prime}}{2\pi i}\rho(\Delta^{\prime})G_{\Delta^{\prime}}(X_1,X_2)+\text{(complementary)}\period
\ee 
Let us make a few comments on this decomposition. The expression \eqref{eq:deSitterdecomp2pt} takes a form similar to the expansion of the two-point function in AdS in terms of harmonic functions. However, performing harmonic analysis directly in dS is a nontrivial task since the dS space is a Lorentzian manifold. Instead, the easiest way to relate \eqref{eq:deSitterdecomp2pt} to harmonic analysis is to first use harmonic analysis on the sphere and perform an analytic continuation. This is a standard procedure discussed in the literature \cite{Marolf:2010zp,Marolf:2010nz}, which we review in Appendix \ref{app:SR}. It is an interesting and important problem to derive \eqref{eq:deSitterdecomp2pt} directly from harmonic analysis in dS, the first steps of which are explained in Appendix \ref{app:Gelfand}.

The last step is to relate $\rho(\Delta)$ to a sum of squares of overlaps. For this purpose, we invert the relation \eqref{eq:fandG} and write
\be\label{eq:Casratio}
\rho(\Delta^{\prime})=\frac{g_{\Delta^{\prime}}(X_1,X_2)}{G_{\Delta^{\prime}}(X_1,X_2)}\period
\ee
The equality holds at any $X_{1,2}$ and the dependence on $X_{1,2}$ drops out upon taking the ratio on the right hand side. In particular, it is convenient to perform the analytic continuation to a sphere and place the two points symmetrically across the equator (see figure \ref{fig:2ptcutting}). As is clear from the figure, this can be interpreted as a sum of  squares of overlaps and is therefore positive:
\be\label{eq:denompos}
\left.g_{\Delta^{\prime}}(X_1,X_2)\right|_{\text{sphere, symmetric}}=\sum_{\psi\in R_{\Delta^{\prime}}}\left|\langle \Omega|\phi (X)|\psi\rangle\right|^2\geq 0 \period
\ee
Here we used a representation of the projector 
\be
\mathcal{P}_{\Delta^{\prime}}=\sum_{\psi\in R_{\Delta^{\prime}}}|\psi\rangle\langle \psi|\comma
\ee
where $\psi\in R_{\Delta^{\prime}}$ means a sum over states belonging to the principal series representation corresponding to $\Delta^{\prime}$. As $X$ approaches the equator of the sphere, the sum on the right hand side of \eqref{eq:denompos} is expected to diverge since it corresponds to a short distance limit of the two points $X_{1,2}$.
On the other hand, the numerator of \eqref{eq:Casratio} can be expanded in the small distance limit as
\be
\lim_{X_1\to X_2}G_{\Delta^{\prime}}(X_1,X_2)\sim \frac{\zeta^{\frac{1-d}{2}}}{2\pi^{\frac{d+1}{2}}(d-1)}+\cdots\comma
\ee
where $\zeta$ is the chordal distance given by \eqref{eq:chordal}.
As can be seen from the expansion, it is singular but {\it positive} in the limit $\zeta\to 0$  since we are approaching the singularity from a space-like direction $(\zeta\to  0^{+})$. We can then rewrite \eqref{eq:Casratio} as
\be
\rho(\Delta^{\prime})=\lim_{X\to\,  \text{equator}}\left[\left(\frac{\zeta^{\frac{1-d}{2}}}{2\pi^{\frac{d+1}{2}}(d-1)}\right)^{-1}\sum_{\psi\in R_{\Delta^{\prime}}}\left|\langle \Omega|\phi (X)|\psi\rangle\right|^2\right]\period
\ee 
Here the limit is defined on a sphere, where the two points approach symmetrically to the equator and the distance $\zeta$ vanishes. As argued above, this expression is a limit of positive quantity and therefore remains to be positive even after taking the limit. This establishes the claim we wished to make,
\be
\rho(\Delta^{\prime})\geq 0\period
\ee
\subsection{Positivity of late-time four-point functions in de Sitter} Let us now discuss the late-time four-point function in dS,
\be
\mathcal{G}(x_1,x_2,x_3,x_4)=\langle \Omega | \mathcal{O}_1 (x_1)\mathcal{O}_2 (x_2)\mathcal{O}_3 (x_3)\mathcal{O}_4 (x_4)|\Omega\rangle\comma
\ee
where all the operators are inserted at late time $(\eta_c\to 0^{-})$.
In what follows, we take the operators to be pair-wise complex conjugate, $\mathcal{O}_1=\mathcal{O}_3^{\dagger}$ and $\mathcal{O}_2=\mathcal{O}_4^{\dagger}$, for simplicity. Furthermore, we assume that they transform as scalar conformal primaries in the complementary series under the dS isometry and denote the dimensions of $\mathcal{O}_j$ by $\Delta_j$. At the leading order in perturbation theory we can generalize this discussion to heavy fields that have complex dimensions in the principal series assuming  ($\Delta_1=(\Delta_3)^{\ast}$ and $\Delta_2=(\Delta_4)^{\ast}$). There are, however, several subtleties associated to such heavy field insertions. First of all, a two-point function of heavy operators $\mathcal{O}^{\dagger}(x)$ and $\mathcal{O}^{\dagger}(y)$ contains a local term proportional to $\delta^d(x-y)$. Moreover, this term depends on the ordering of operators, see \cite{Matteo,Hogervorst:2021uvp}, and it is important to include it to preserve the positivity property we discuss below. For free-theory dimensions this term is consistent with conformal invariance of correlators, since $\Delta+\Delta^{\ast}=d$. However, when interactions are included, all principal series operators get anomalous dimensions, so that the real part is always greater than $d/2$. As a result, they even stop to belong to a unitary representation of the dS isometry group. Nevertheless, local terms are expected to survive in the correlation functions. For these reasons at the moment it is not clear how to think about operators corresponding to heavy fields in the external states in an interacting theory, while maintaining conformal invariance and unitarity. Presumably they should be treated in the same way as unstable particles, or resonances, are treated in flat space QFTs and in quantum mechanics. The argument of this section can also be generalized to the case with four distinct operators, following what has been done in the conformal bootstrap (see e.g.~\cite{Kos:2014bka}). We will briefly comment on this at the end of this subsection.

\begin{figure}[t] 
	\centering
		\includegraphics[width=0.35\linewidth,angle=-90]{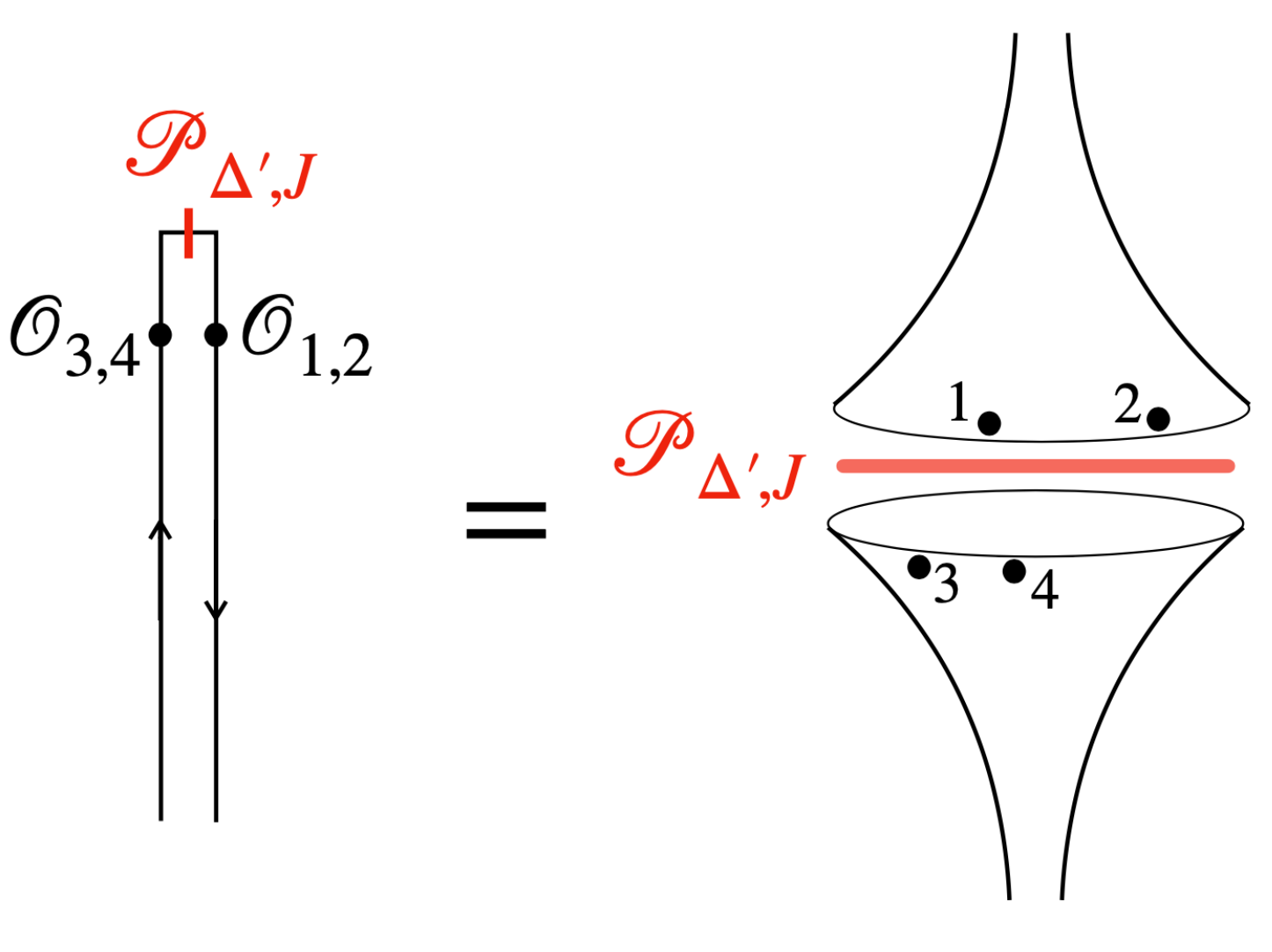}
	\caption{\small A resolution of identity for four-point functions. To prove the positivity, it is convenient to place $\mathcal{O}_{1,2}$ and $\mathcal{O}_{3,4}$ on different parts of the in-in contour so that the four-point function can be interpreted as an overlap of two states, shown on the right hand side of the figure. }
	\label{fig:4ptcutting}
\end{figure}

\paragraph{A resolution of the identity.}As in the previous discussion, we first insert a resolution of identity 
\be
{\bf 1}=\sum_{J=0}^{\infty}\int^{\frac{d}{2}+i\infty}_{\frac{d}{2}}\frac{d\Delta^{\prime}}{2\pi i }\mathcal{P}_{\Delta^{\prime},J}+\text{(rest)}\period
\ee
Unlike the bulk two-point function discussed in the previous subsection, the decomposition now involves a sum over angular momenta $J$. Here we wrote down the terms corresponding to the principal series representation and $(\text{rest})$ denotes all the other unitary representations of the Euclidean conformal group, such as the complementary series and the discrete series. As a result, we obtain a decomposition
\be\label{eq:4ptdecompdS}
\mathcal{G}(x_1,x_2,x_3,x_4)=\sum_{J=0}^{\infty}\int_{\frac{d}{2}}^{\frac{d}{2}+i\infty}\frac{d\Delta^{\prime}}{2\pi i}\langle \Omega | \mathcal{O}_1 (x_1)\mathcal{O}_2 (x_2)\mathcal{P}_{\Delta^{\prime},J}\mathcal{O}_3 (x_3)\mathcal{O}_4 (x_4)|\Omega\rangle+\text{(rest)}\period
\ee
As with the bulk two-point function, we can interpret this in terms of a path integral in the in-in formalism. For this purpose, it is convenient to think of $\mathcal{O}_3$ and $\mathcal{O}_4$ as insertions on the left part of the in-in contour while $\mathcal{O}_1$ and $\mathcal{O}_2$ as insertions on the right part of the contour. Then, $\mathcal{P}_{\Delta^{\prime},J}$ corresponds to a projection at the top of the contours, see figure \ref{fig:4ptcutting}.

\paragraph{Symmetry constraints.}The next step is to impose the constraints from symmetry. We again consider the action of the quadratic Casimir of the dS isometry, acting on the operators $\mathcal{O}_1$ and $\mathcal{O}_2$,
\be
\langle \Omega | \mathcal{C}_2\cdot \left(\mathcal{O}_1 (x_1)\mathcal{O}_2 (x_2)\right)\mathcal{P}_{\Delta^{\prime},J}\mathcal{O}_3 (x_3)\mathcal{O}_4 (x_4)|\Omega\rangle\period
\ee
where the action of the Casimir is given by
\be
\mathcal{C}_2\cdot \left(\mathcal{O}_1 (x_1)\mathcal{O}_2 (x_2)\right)\equiv -\frac{1}{2}\sum_{A,B}[\hat{M}^{AB},[\hat{M}_{AB},\mathcal{O}_1(x_1)\mathcal{O}_2(x_2)]]\period
\ee
Using the invariance of $\langle \Omega |$ under $\hat{M}_{AB}$, this can be rewritten as the action of the Casimir on $\mathcal{P}_{\Delta^{\prime},J}$. We thus get
\be\label{eq:casimireigen4pt}
\begin{aligned}
&\langle \Omega | \mathcal{C}_2\cdot \left(\mathcal{O}_1 (x_1)\mathcal{O}_2 (x_2)\right)\mathcal{P}_{\Delta^{\prime},J}\mathcal{O}_3 (x_3)\mathcal{O}_4 (x_4)|\Omega\rangle=\\
&\qquad c^{(\Delta^{\prime},J)}_2\underbrace{\langle \Omega | \mathcal{O}_1 (x_1)\mathcal{O}_2 (x_2)\mathcal{P}_{\Delta^{\prime},J}\mathcal{O}_3 (x_3)\mathcal{O}_4 (x_4)|\Omega\rangle}_{\equiv g_{\Delta^{\prime},J}(x_1,x_2,x_3,x_4)}\comma
\end{aligned}
\ee
where $c_2^{\Delta^{\prime},J}$ is the Casimir eigenvalue \eqref{eq:eigenc} given by
$
c_2^{(\Delta^{\prime},J)}=\Delta^{\prime} (\Delta^{\prime}-d)+J(J+d-2)\period
$
Alternatively, we can rewrite the action of the Casimir as a differential operator acting on $x_{1,2}$. To see this, we express $\mathcal{O}(x)$ as
\be
\mathcal{O} (x)=e^{i\hat{P}\cdot x}\mathcal{O}(0)e^{-i\hat{P}\cdot x}\comma
\ee
with $\hat{P}_{\mu}$ being the translation operator. We can then rewrite the action of $\hat{M}_{AB}$ by computing $\hat{M}_{AB}$ with $e^{i\hat{P}\cdot x}$ and using the fact that $\mathcal{O}$ is a conformal primary (namely it is annihilated by the special conformal transformation). This is a standard manipulation in deriving the Casimir differential equation for the conformal field theory in higher dimensions and we refer to standard reviews (for instance \cite{Simmons-Duffin:2016gjk}) for details. As a result, we obtain
\be
\begin{aligned}
&\langle \Omega | \mathcal{C}_2\cdot \left(\mathcal{O}_1 (x_1)\mathcal{O}_2 (x_2)\right)\mathcal{P}_{\Delta^{\prime},J}\mathcal{O}_3 (x_3)\mathcal{O}_4 (x_4)|\Omega\rangle= \mathcal{D}_{12}\cdot g_{\Delta^{\prime},J}(x_1,x_2,x_3,x_4)\comma
\end{aligned}
\ee
where $\mathcal{D}_{12}$ is a differential representation of the conformal Casimir given by \eqref{eq:difc}. Equating this with \eqref{eq:casimireigen4pt}, we find that $g_{\Delta^{\prime}, J}$ satisfies
\be\label{eq:4ptdifferential}
\mathcal{D}_{12}\cdot g_{\Delta^{\prime},J}(x_1,x_2,x_3,x_4)=c_2^{(\Delta^{\prime},J)}g_{\Delta^{\prime},J}(x_1,x_2,x_3,x_4)\period
\ee
This coincides with the differential equation satisfied by the conformal partial wave \eqref{eq:difeqcpw}.

Being a second-order differential equation for two independent variables, the equation \eqref{eq:4ptdifferential} admits several solutions. Among those solutions, we need to pick the one that has correct properties\footnote{Precisely speaking, we also need to impose the differential equation for the quartic Casimir. For details, see \cite{Simmons-Duffin:2017nub}.}. In particular, it must be single-valued in $x_j$ on the late-time surface and have the correct asymptotic behavior in the coincident point limit, $x_1\to x_2$ and $x_3\to x_4$. That is precisely the conformal partial wave $\widehat{\mathcal{F}}^{\{\Delta_j\}}_{\Delta^{\prime},J}$, which we reviewed in section \ref{subsec:CPWAdS}. We therefore conclude that the function $g_{\Delta^{\prime},J}(x_1,x_2,x_3,x_4)$ must be proportional to $\widehat{\mathcal{F}}$,
\be\label{eq:spectraldef4pt}
g_{\Delta^{\prime},J}(x_1,x_2,x_3,x_4)=\rho_{J}(\Delta^{\prime})\widehat{\mathcal{F}}^{\{\Delta_j\}}_{\Delta^{\prime},J}(x_1,x_2,x_3,x_4)\period
\ee
Comparing \eqref{eq:4ptdecompdS} and \eqref{eq:spectraldef4pt} with the spectral decomposition in section \ref{subsec:dSanalytic}, one can see that
the constant of proportionality $\rho_{J}(\Delta^{\prime})$ is identified with the {\it spectral density} discussed there. 

\paragraph{Positivity of the spectral density.}The last step of our derivation is to relate $\rho_{J}(\Delta^{\prime})$ to a sum of norms, which is positive.
For this purpose, we first decompose the left hand side of \eqref{eq:spectraldef4pt} further into the eigenstates of translation;
\be
g_{\Delta^{\prime},J}(x_1,x_2,x_3,x_4)=\int \frac{d^{d}p}{(2\pi)^{d}}\langle \Omega | \mathcal{O}_1 (x_1)\mathcal{O}_2 (x_2)\mathcal{P}_{\Delta^{\prime},J}^{(p)}\mathcal{O}_3 (x_3)\mathcal{O}_4 (x_4)|\Omega\rangle\comma
\ee
where $\mathcal{P}^{(p)}_{\Delta^{\prime},J}$ is a projection to the translation eigenstates with momentum $p^{\mu}$ inside the space invariant under $\mathcal{P}_{\Delta^{\prime},J}$. We next consider a ``symmetric configuration'' in which $x_3=x_1+y$ and $x_4=x_2+y$. Using $\mathcal{O}_3(x_3)\mathcal{O}_4(x_4)=e^{i\hat{P}\cdot y}\mathcal{O}_3(x_1)\mathcal{O}_4(x_2)e^{-i\hat{P}\cdot y}$, we then find
\be
\left.g_{\Delta^{\prime},J}\right|_{\rm sym}=\int \frac{d^{d}p}{(2\pi)^{d}}e^{ip\cdot y}\langle \Omega | \mathcal{O}_1 (x_1)\mathcal{O}_2 (x_2)\mathcal{P}_{\Delta^{\prime},J}^{(p)}\mathcal{O}_3 (x_1)\mathcal{O}_4 (x_2)|\Omega\rangle\period
\ee
By performing the inverse Fourier transformation, we get
\be
\int d^{d}y \, e^{-i p y}\left.g_{\Delta^{\prime},J}\right|_{\rm sym}=\langle \Omega | \mathcal{O}_1 (x_1)\mathcal{O}_2 (x_2)\mathcal{P}_{\Delta^{\prime},J}^{(p)}\mathcal{O}_3 (x_1)\mathcal{O}_4 (x_2)|\Omega\rangle\period
\ee
When the operators are pair-wise conjugate ($\mathcal{O}_1=\mathcal{O}_3^{\dagger}$ and $\mathcal{O}_2=\mathcal{O}_4^{\dagger}$), the right hand side can be expressed as a sum of squares of overlaps and is therefore positive:
\be\label{eq:gfourier}
\int d^{d}y \, e^{-i p y}\left.g_{\Delta^{\prime},J}\right|_{\rm sym}=\sum_{\substack{\psi\in R_{\Delta^{\prime},J}\\p_{\psi}=p}}\left|\langle \Omega | \mathcal{O}_1 (x_1)\mathcal{O}_2 (x_2)|\psi\rangle\right|^2\geq 0\period
\ee

Next we perform the same inverse Fourier transformation to $\widehat{\mathcal{F}}_{\Delta^{\prime},J}^{\{ \Delta_j\}}$. To achieve this, we use the integral representation of the conformal partial wave (see (3.3) in \cite{Simmons-Duffin:2017nub}),
\be\label{eq:doubleintpartial}
\begin{aligned}
\widehat{\mathcal{F}}_{\Delta^{\prime},J}^{\{ \Delta_j\}}=&\int \frac{d^{d}p}{(2\pi)^{d}}\int \frac{d^{d}x_5\, e^{-i p \cdot x_5}}{|x_{12}|^{\Delta_1+\Delta_2-\Delta^{\prime}}|x_{15}|^{\Delta_1+\Delta^{\prime}-\Delta_2}|x_{25}|^{\Delta_2+\Delta^{\prime}-\Delta_1}}\\
&\times \frac{d^{d}x_6\, e^{ip \cdot x_6}}{|x_{34}|^{\Delta_3+\Delta_4-(d-\Delta^{\prime})}|x_{36}|^{\Delta_3+(d-\Delta^{\prime})-\Delta_4}|x_{46}|^{\Delta_4+(d-\Delta^{\prime})-\Delta_3}}\times \hat{C}_{J}(\tilde{\eta})\comma
\end{aligned}
\ee
where $\tilde{\eta}$ is given by  
\be
\tilde{\eta}\equiv \frac{|x_{15}||x_{25}|}{|x_{12}|}\frac{|x_{36}||x_{46}|}{|x_{34}|}\left(\frac{\vec{x}_{15}}{x_{15}^2}-\frac{\vec{x}_{25}}{x_{25}^2}\right)\cdot \left(\frac{\vec{x}_{36}}{x_{36}^2}-\frac{\vec{x}_{46}}{x_{46}^2}\right)\comma
\ee
and $\hat{C}_J(\eta)$ is a Gegenbauer polynomial (see \cite{Simmons-Duffin:2017nub} for the normalization employed here). For the following discussions, we only need its positivity at $\eta=1$:
\be\label{eq:CJpositive}
\hat{C}_{J}(1)\geq 0\period
\ee
The representation \eqref{eq:doubleintpartial} is different from the one that appears in \cite{Simmons-Duffin:2017nub}, but one can easily check the equivalence by first performing the integration of $p$.

Now, in the symmetric configuration ($x_3=x_1+y$, $x_4=x_2+y$, $\Delta_1=\Delta_3^{\ast}$ and $\Delta_2=\Delta_4^{\ast}$), this can be rewritten by shifting the integration variable $x_6$ to $x_6+y$. As a result, we obtain
\be
\begin{aligned}
\left.\widehat{\mathcal{F}}_{\Delta^{\prime},J}^{\{ \Delta_j\}}\right|_{\rm sym}=&\int \frac{d^{d}p\, e^{i p \cdot y}}{(2\pi)^{d}}\int \frac{d^{d}x_5\, e^{-i p \cdot x_5}}{|x_{12}|^{\Delta_1+\Delta_2-\Delta^{\prime}}|x_{15}|^{\Delta_1+\Delta^{\prime}-\Delta_2}|x_{25}|^{\Delta_2+\Delta^{\prime}-\Delta_1}}\\
&\times \frac{d^{d}x_6\, e^{ip \cdot x_6}}{|x_{14}|^{\Delta_1^{\ast}+\Delta_2^{\ast}-(d-\Delta^{\prime})}|x_{16}|^{\Delta_1^{\ast}+(d-\Delta^{\prime})-\Delta_2^{\ast}}|x_{26}|^{\Delta_2^{\ast}+(d-\Delta^{\prime})-\Delta_1^{\ast}}}\times \hat{C}_{J}(\tilde{\eta}_{\rm sym})\comma
\end{aligned}
\ee
with
\be
\tilde{\eta}_{\rm sym}\equiv \frac{|x_{15}||x_{25}|}{|x_{12}|}\frac{|x_{16}||x_{26}|}{|x_{12}|}\left(\frac{\vec{x}_{15}}{x_{15}^2}-\frac{\vec{x}_{25}}{x_{25}^2}\right)\cdot \left(\frac{\vec{x}_{16}}{x_{16}^2}-\frac{\vec{x}_{26}}{x_{26}^2}\right)\period
\ee
Performing the inverse Fourier transform, we then get
\be\nonumber
\begin{aligned}
\int d^{d}y\, e^{-i p y}\left.\widehat{\mathcal{F}}_{\Delta^{\prime},J}^{\{ \Delta_j\}}\right|_{\rm sym}=&\int \frac{d^{d}x_5\, e^{-i p \cdot x_5}}{|x_{12}|^{\Delta_1+\Delta_2-\Delta^{\prime}}|x_{15}|^{\Delta_1+\Delta^{\prime}-\Delta_2}|x_{25}|^{\Delta_2+\Delta^{\prime}-\Delta_1}}\\
&\times \frac{d^{d}x_6\, e^{ip \cdot x_6}}{|x_{14}|^{\Delta_1^{\ast}+\Delta_2^{\ast}-(d-\Delta^{\prime})}|x_{16}|^{\Delta_1^{\ast}+(d-\Delta^{\prime})-\Delta_2^{\ast}}|x_{26}|^{\Delta_2^{\ast}+(d-\Delta^{\prime})-\Delta_1^{\ast}}}\times \hat{C}_{J}(\tilde{\eta}_{\rm sym})\period
\end{aligned}
\ee
It is a nontrivial task to evaluate this integral because of $\hat{C}_{J}(\tilde{\eta}_{\rm sym})$, which couples $x_5$ and $x_6$. However in the limit $|p|\gg 1$, the rapidly oscillating factor $e^{i p\cdot (x_5-x_6)}$ enforces $x_5\sim x_6$ and simplifies the computation. In particular the leading behavior can be computed by setting $x_5=x_6$ in $\hat{C}_J(\tilde{\eta}_{\rm sym})$, which amounts to setting $\tilde{\eta}_{\rm sym}=1$. As a result, we find in the limit
\be\label{eq:Ffourier}
\begin{aligned}
&\int d^{d}y\, e^{-i p y}\left.\widehat{\mathcal{F}}_{\Delta^{\prime},J}^{\{ \Delta_j\}}\right|_{\rm sym}\overset{|p|\gg 1}{\sim}\\
&\qquad \qquad \hat{C}_{J}(1)\times\left|\int \frac{d^{d}x_5\, e^{-i p \cdot x_5}}{|x_{12}|^{\Delta_1+\Delta_2-\Delta^{\prime}}|x_{15}|^{\Delta_1+\Delta^{\prime}-\Delta_2}|x_{25}|^{\Delta_2+\Delta^{\prime}-\Delta_1}}\right|^2\geq 0\period
\end{aligned}
\ee
Here we used the fact that $\Delta^{\prime}$ belongs to the principal series, namely $\Delta^{\prime}=(d-\Delta^{\prime})^{\ast}=\frac{d}{2}+i\nu$. This manifests the positivity of the first line in the limit $|p|\gg 1$. Combining this with the positivity for $g_{\Delta^{\prime},J}$ \eqref{eq:gfourier}, we conclude that the constant of proportionality $\rho_J(\Delta^{\prime})$ must also be positive:
\be
\rho_J(\Delta^{\prime})\geq 0\period
\ee
This proves the positivity of the spectral density.

\paragraph{Convergence of spectral integrals.} As was pointed out in \cite{Hogervorst:2021uvp} generically the integral over $\Delta^\prime$ in \eqref{eq:4ptdecompdS} is divergent. This is not an issue for the proof of positivity since one can define $\rho_J(\Delta')$ directly using the projector $\mathcal{P}_{\Delta^{\prime},J}$. An alternative and more practical way to compute $\rho_{J}(\Delta^{\prime})$ is to extract it from the four-point function through the Euclidean inversion formula \cite{Caron-Huot:2017vep}.  The convergence of this inversion formula was discussed in \cite{Simmons-Duffin:2017nub}. From there we see that the inversion integral converges for light external particles.\footnote{Note that potential divergences coming from the exchange of light operators in the $s$-channel need to be subtracted, they correspond to the complementary series contributions in \eqref{eq:4ptdecompdS}.} For heavy external particles the inversion formula can have divergences that require regularization and therefore it is not straightforward to compute $\rho_{J}(\Delta^{\prime})$ from the four-point function. We leave this problem for future investigations.


\paragraph{Generalizations.} Let us finally comment on the generalization of our positivity condition. In our derivation, we assumed that all the operators are pair-wise conjugate $\mathcal{O}_1=\mathcal{O}_3^{\dagger}$ and $\mathcal{O}_2=\mathcal{O}_4^{\dagger}$. However, the derivation can be extended to to the case of four different operators. For this, we need to consider a matrix of four-point functions
\be
\left(\begin{array}{cc}\langle \Omega|\mathcal{O}_1^{\dagger}\mathcal{O}_2^{\dagger}\mathcal{O}_1\mathcal{O}_2|\Omega\rangle&\langle \Omega|\mathcal{O}_1^{\dagger}\mathcal{O}_2^{\dagger}\mathcal{O}_3\mathcal{O}_4|\Omega\rangle\\\langle \Omega|\mathcal{O}_3^{\dagger}\mathcal{O}_4^{\dagger}\mathcal{O}_1\mathcal{O}_2|\Omega\rangle&\langle \Omega|\mathcal{O}_3^{\dagger}\mathcal{O}_4^{\dagger}\mathcal{O}_3\mathcal{O}_4|\Omega\rangle\end{array}\right)\comma
\ee
and unitarity translates to the positivity of a matrix of spectral functions. It would be interesting to work this out and check it in nontrivial examples. We can also write similar equations for higher-point functions. An important open question is whether such relations exhaust the consequence of unitarity.

\section{Perturbative calculations}\label{sec:pert}
Above we formulated several general requirements that QFT correlators on the boundary of dS space have to satisfy. We also developed a perturbative technique which allows to reduce a dS calculation to a set of EAdS Feynman diagrams. In this section we will perform some explicit calculations in order to demonstrate this technique and to check explicitly that the required properties are indeed satisfied. We will focus on a theory with two scalar fields, $\phi$ and $\sigma$ and a cubic interaction:
\be
{\cal{L}}=\half\d\phi\d\phi-\half m_\phi^2\phi^2+\half\d\sigma\d\sigma-\half m_\sigma^2\sigma^2+\half g\phi^2\sigma\,.
\ee
In this section we will assume that the field $\phi$ is light, while the field $\sigma$ could be either light or heavy.

Our strategy will be to first use Feynman rules of the in-in perturbation theory, reviewed in subsection \ref{subsec:prop}, and then reduce the Feynman diagrams to the EAdS ones as described in subsection \ref{subsec:dStoAdS}. We could have arrived at the same results by using directly the EAdS Lagrangian, as described in \ref{subsec:AdSL}; however, this would make the combinatorics for this particular problem slightly more complicated.

\subsection{Review of exchange diagram in EAdS}\label{subsec:exchAdS}
\begin{figure}[t] 
	\centering
		\hspace{-1cm}\includegraphics[width=0.2\linewidth,angle=0]{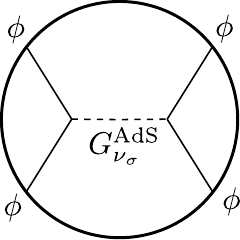}
	\caption{\small Exchange diagram in EAdS. Using the spectral representation of the bulk-to-bulk propagator $G^{\rm AdS}_{\nu_\sigma}$, and then the split representation of the harmonic function, one readily obtains the expansion of the four-point function of $\phi$ in the basis of conformal partial waves, eq. \eqref{AdStree}.}
	\label{fig:EAdSexch}
\end{figure}

Let us quickly review the calculation of the exchange diagram in figure \ref{fig:EAdSexch}\cite{DHoker:1998ecp, DHoker:1999mqo, Penedones:2010ue, El-Showk:2011yvt}:
\begin{align}
\begin{split}\label{eq:exchAdS}
& \l\langle\phi(P_1)\phi(P_2)\phi(P_3)\phi(P_4)\r\rangle\supset\int dX_1 dX_2 \\  & \hspace{2cm} K_{\nu_\phi}^{\text{AdS}}(X_1,P_1) K_{\nu_\phi}^{\text{AdS}}(X_1,P_2) K_{\nu_\phi}^{\text{AdS}}(X_2,P_3)K_{\nu_\phi}^{\text{AdS}}(X_2,P_4) G^{\rm AdS}_{\nu_\sigma} (X_1,X_2)\,,
\end{split}
\end{align}
where ordinary boundary conditions were assumed. A direct evaluation of the integrals over the bulk points, or just the integrals over the radial coordinates $z$, were we to use momentum representation for the boundary directions, are rather complicated. A simplifying method is to use the split representation of the harmonic function\cite{Moschella:2007zza, Cornalba:2008qf, Penedones:2010ue}:
\be
\Omega^{\rm AdS}_\nu(s^{\rm AdS})=\frac{\nu^2 N^{\rm AdS}_\nu N^{\rm AdS}_{-\nu} }{\pi}\int dP K^{\rm AdS}_\nu(X_1,P)K^{\rm AdS}_{-\nu}(X_2,P)\,.
\ee
Combining it with \eqref{GOmegarel} one gets a split representation also for the propagators. The latter can be thought of as the decomposition of the propagator in EAdS harmonics labeled by the boundary point and the spectral parameter $\nu'$. 

Now the diagram can be evaluated explicitly by first performing $d X_1$ and $d X_2$ and then the $dP$ integral, leading to \cite{Penedones:2010ue, Costa:2014kfa, Carmi:2018qzm} 
\begin{align}
\begin{split}
\label{AdStree}
\l\langle\phi(P_1)\phi(P_2)\phi(P_3)\phi(P_4)\r\rangle & \supset \frac{1}{x_{12}^{d+2i\nu_\phi} x_{34}^{d+2i\nu_\phi}}
\int_{-\infty}^{+\infty} \frac{d\nu'}{2\pi} \\
&\frac{1}{\nu'^2-\nu_\sigma^2}\frac{\G^2\l(\frac{d+4i\nu_\phi\pm2i\nu'}{4}\r)\G^2\l(\frac{d\pm2i\nu'}{4}\r)}{32 \pi^d\G^2\l(\ddt+i\nu_\phi\r)\G^2\l(1+i\nu_\phi\r)\G\l(\pm i\nu'\r)}{\cal F^{\{\nu_\phi\}}_{\nu'}}(z,\bar z).
\end{split}
\end{align}
Here ${\cal F^{\{\nu_\phi\}}_{\nu'}}(z,\bar z)$ is the (stripped) conformal partial wave, see eq.s \eqref{eq:CPW}-\eqref{eq:stripping}. Compared to eq.s \eqref{eq:CPW}-\eqref{eq:stripping} here we are changing the notation in the superscripts from the variable $\Delta_\phi = \frac{d}{2}+i\nu_\phi $ to the variable $\nu_\phi$.

\subsection{Tree-level four-point function}\label{sec:tree}
\begin{figure}[t] 
	\centering
		\hspace{-2cm}\includegraphics[width=1\linewidth,angle=0]{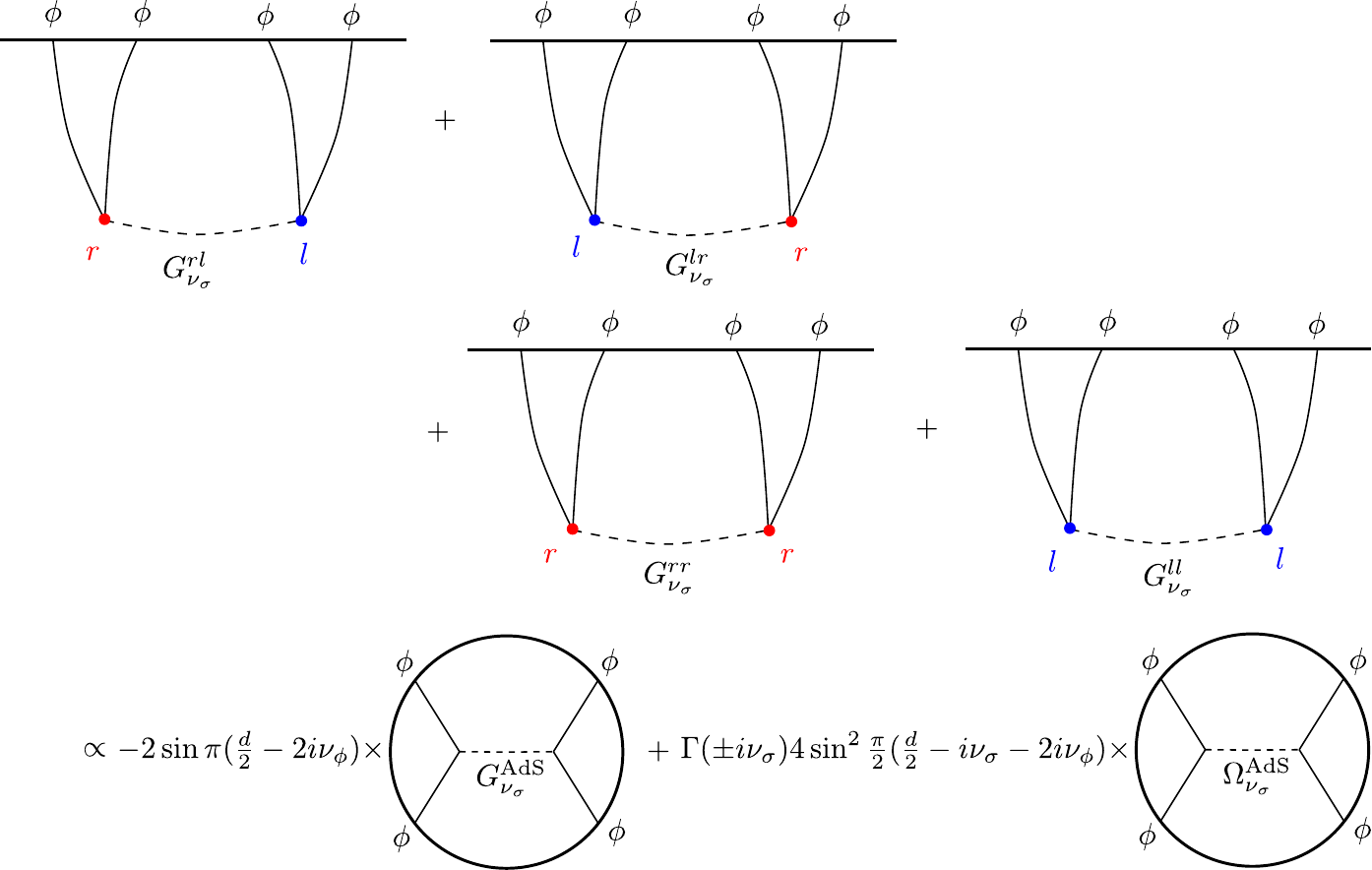}
	\caption{\small Exchange diagram in dS. There are four contributions, because each of the two vertices can be on the left or the right portion of the in-in contour. Relating each dS propagator to a combination of EAdS propagators and harmonic functions, the full exchange diagram in dS becomes proportional to the linear combination of EAdS diagrams in the last line, with some oscillatory dimension-dependent factors. One can then readily obtain the conformal partial wave decomposition in eq. \eqref{treefull}.}
	\label{fig:dSexch}
\end{figure}

We begin with the calculation of the tree-level s-channel exchange diagram which contributes to late-time the four-point function $\l\langle\phi(x_1,\eta_c)\phi(x_2,\eta_c)\phi(x_3,\eta_c)\phi(x_4,\eta_c)\r\rangle$, illustrated in figure \ref{fig:dSexch}. For now let us assume that $\sigma$ is a light field as well, $m_\sigma<\ddt$, or $i\nu_\sigma>0$. There are in total four distinct diagrams since each of the two interaction vertices can come from either the left or the right part of the contour. Explicitly we have
\beg
D^{ll}=-g^2 \int DX DY K_{\nu_\phi}^l(s_1)K_{\nu_\phi}^l(s_2)K_{\nu_\phi}^l(s_3)K_{\nu_\phi}^l(s_4)G^{ll}_{\nu_\sigma}(s_{XY})\,,\\
D^{rr}=-g^2 \int DX DY K_{\nu_\phi}^r(s_1)K_{\nu_\phi}^r(s_2)K_{\nu_\phi}^r(s_3)K_{\nu_\phi}^r(s_4)G^{rr}_{\nu_\sigma}(s_{XY})\,,\\
D^{lr}=g^2 \int DX DY K_{\nu_\phi}^l(s_1)K_{\nu_\phi}^l(s_2)K_{\nu_\phi}^r(s_3)K_{\nu_\phi}^r(s_4)G^{lr}_{\nu_\sigma}(s_{XY})\,,\\
D^{ll}=g^2 \int DX DY K_{\nu_\phi}^r(s_1)K_{\nu_\phi}^r(s_2)K_{\nu_\phi}^l(s_3)K_{\nu_\phi}^l(s_4)G^{lr}_{\nu_\sigma}(s_{XY})\,,
\eeg
where
\be
s_i=s(X,(x_i,\eta_c))~,~~i=1,\dots,4~,~~s_{XY}=s(X,Y)~.
\ee

We now perform the rotation of the integration contour of $X$ and $Y$, as in \eqref{Wick}, and use the relations \eqref{splitW}, \eqref{splitGreen}, and \eqref{Krot}. It is convenient to also use the relations between AdS propagators and harmonic functions, given in \eqref{GOmegarel} to express the answer in terms of the AdS Greens function with standard boundary conditions, $G^{\rm AdS}_\nu$, and the harmonic function $\Omega^{\rm AdS}_\nu$. Collecting the terms from all four diagrams together we get
\begin{align}
\begin{split}
\label{dSDiagrams}
& D^{ll}+D^{rr}+D^{lr}+D^{rl}\\
& =g^2\l(-\eta_c\r)^{4\l(\ddt-i\nu_\phi\r)}\l(\frac{N_{\nu_\phi}}{N^{\rm AdS}_{-\nu_\phi}}\r)^4\Bigg[\G\l(\pm i\nu_\sigma\r) 4 \sin^2\tfrac{\pi}{2}\l(\tfrac{d}{2}-i\nu_\sigma-2 i \nu_\phi\r){\cal I}_\Omega \\ & \hspace{9cm} -2 \sin \pi \l(\tfrac{d}{2}-2i \nu_\phi\r){\cal I}_G\Bigg]\,,
\end{split}
\end{align}
where we defined  
\beg
\hspace{-0.5cm}{\cal I}_\Omega=\int dX_1 dX_2 K_{-\nu_\phi}^{\text{AdS}}(X_1,P_1) K_{-\nu_\phi}^{\text{AdS}}(X_1,P_2) K_{-\nu_\phi}^{\text{AdS}}(X_2,P_3)K_{-\nu_\phi}^{\text{AdS}}(X_2,P_4) \Omega^{\rm AdS}_{\nu_\sigma} (X_1,X_2)\,,\\
\hspace{-0.5cm}{\cal I}_G=\int dX_1 dX_2 K_{-\nu_\phi}^{\text{AdS}}(X_1,P_1) K_{-\nu_\phi}^{\text{AdS}}(X_1,P_2) K_{-\nu_\phi}^{\text{AdS}}(X_2,P_3)K_{-\nu_\phi}^{\text{AdS}}(X_2,P_4) G^{\rm AdS}_{\nu_\sigma} (X_1,X_2)\,,
\eeg
and $P_i$ is an AdS boundary point with coordinate $x_i$. Note the relative minus sign in the subscript of the bulk-to-boundary propagators compared to \eqref{eq:exchAdS}, which follows from the relation \eqref{Krot} between the two bulk-to-boundary propagators.
${\cal I}_G$ is just the AdS exchange diagram evaluated in \eqref{AdStree}:
\beg
\hspace{-0.5cm}{\cal I}_G=
\frac{1}{x_{12}^{d-2i\nu_\phi} x_{34}^{d-2i\nu_\phi}}
\int \frac{d\nu'}{2\pi}\frac{1}{\nu'^2-\nu_\sigma^2}\frac{\G^2\l(\frac{d-4i\nu_\phi\pm2i\nu'}{4}\r)\G^2\l(\frac{d\pm2i\nu'}{4}\r)}{32 \pi^d\G^2\l(\ddt-i\nu_\phi\r)\G^2\l(1-i\nu_\phi\r)\G\l(\pm i\nu'\r)}{\cal F^{\{-\nu_\phi\}}_{\nu'}}(z,\bar z)~.
\eeg
The evaluation of ${\cal I}_\Omega$ is, in fact, just a step in evaluation of ${\cal I}_G$ and the result is \cite{Carmi:2018qzm}
\beg
{\cal I}_\Omega=
\frac{1}{x_{12}^{d-2i\nu_\phi} x_{34}^{d-2i\nu_\phi}}
\frac{\G^2\l(\frac{d-4i\nu_\phi\pm2i\nu_\sigma}{4}\r)\G^2\l(\frac{d\pm2i\nu_\sigma}{4}\r)}{64 \pi^{d+1}\G^2\l(\ddt-i\nu_\phi\r)\G^2\l(1-i\nu_\phi\r)\G\l(\pm i\nu_\sigma\r)}{\cal F^{\{-\nu_\phi\}}_{\nu_\sigma}}(z,\bar z)~.
\eeg
Substituting these expressions in \eqref{dSDiagrams} we finally get the complete tree-level $s$-channel contribution:
\begin{align}\label{treefull}
\begin{split}
& D^{s,{\rm tree}}_{\nu_\sigma}=D^{ll}+D^{rr}+D^{lr}+D^{rl} \\ & =g^2\l( \frac{\eta_c}{x_{12}}\r)^{d-2i\nu_\phi} \l(\frac{\eta_c}{x_{34}}\r)^{d-2i\nu_\phi}\frac{\G^4(i\nu_\phi)}{2^{11}\pi^{2d+5}}\\
& \times\Bigg( 2 \sin^2\tfrac{\pi}{2}\l(\tfrac{d}{2}-i\nu_\sigma-2 i \nu_\phi\r)\G^2\l(\tfrac{d-4i\nu_\phi\pm2i\nu_\sigma}{4}\r)\G^2\l(\tfrac{d\pm2i\nu_\sigma}{4}\r)   {\cal F^{\{-\nu_\phi\}}_{\nu_\sigma}}(z,\bar z)  \\
& \hspace{1cm}- \sin \pi \l(\tfrac{d}{2}-2i \nu_\phi\r) \int^{+\infty}_{-\infty} \frac{d\nu'}{\nu'^2-\nu_\sigma^2}\frac{\G^2\l(\tfrac{d-4i\nu_\phi\pm2i\nu'}{4}\r)\G^2\l(\tfrac{d\pm2i\nu'}{4}\r)}{\G\l(\pm i\nu'\r)}{\cal F^{\{-\nu_\phi\}}_{\nu'}}(z,\bar z)\Bigg)~.
\end{split}
\end{align}

We see that the answer contains two terms, a discrete and a continuous contribution in the expansion in conformal partial waves (the first line and the second line in the parentheses, respectively). An expression of this form for the exchange diagram in dS was also derived in \cite{Sleight:2020obc}.\footnote{Our expression for this object, given in \eqref{treefull}, disagrees with that of \cite{Sleight:2020obc} in a relative coefficient between the discrete and continuous parts. We believe this is due a minor typo in this paper.} In the appendix \ref{app:general} we give the generalization of this result to the case of unequal external dimensions.

So far we have assumed $\sigma$ to be light, i.e. $i\nu_\sigma\in(0,\frac{d}{2})$. If we instead take $\sigma$ to be a heavy particle, i.e. real $\nu_\sigma$, the continuous part has a pole on the integration contour. This is similar to a resonant contribution to the scattering amplitude in flat space, in which case one needs to resum the bubble diagrams in order to incorporate the width of the resonance and shift the pole away from the physical kinematic region. In the next section we will see that a similar effect occurs in dS as well.

\subsection{Composite exchange: resumming bubble diagrams}\label{sec:compex}
\begin{figure}[t] 
	\centering
		\hspace{0cm}\includegraphics[width=0.7\linewidth,angle=0]{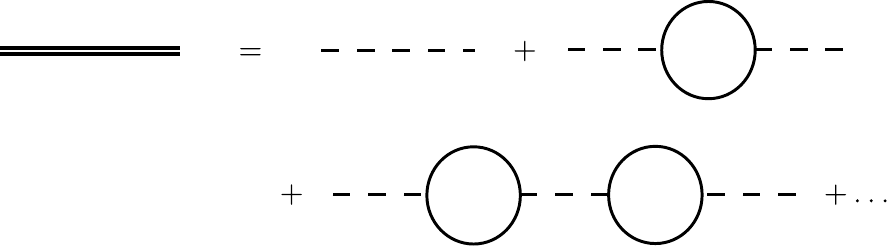}
	\caption{\small The ``full'' propagator of the $\sigma$ particle, obtained by resummation of the bubble diagrams. The resummation is needed to discuss the exchange diagram near the resonance $\nu=\nu_\sigma$. Note that in dS we need to also sum over assignments of the $r/l$ label to each vertex. In the appendices \ref{app:SR}-\ref{app:Bubble} we discuss methods that allow to avoid the resummation directly in dS, and to compute the spectral representation of the full propagator starting from EAdS or the sphere.}
	\label{fig:sigmaprop}
\end{figure}

As mentioned above, in order to obtain a physically sensible answer for all values of $\nu_\sigma$, including real values corresponding to a heavy $\sigma$ particle, one needs to resum the bubble diagrams, or in other words obtain the ``full'' propagator of the $\sigma$ particle, see figure \ref{fig:sigmaprop}. 
In order to resum the propagator of $\sigma$, we use the property that the spectral representation maps convolutions to products. This property can be derived using the analytic continuation from the sphere, or equivalently using the relation to the EAdS spectral representation, see the discussion in the appendix \ref{app:SR}. Either way, as a result we simply have to sum a geometric series involving products of the spectral representation of the propagator and of the bubble diagram. 

The answer can be expressed in the following form: 
\begin{align}
\label{sigma2pt}
 \l\langle\sigma(X^\alpha)\sigma(Y^\beta)\r\rangle & =\int_{-\infty-i(\ddt -\eps)}^{+\infty-i(\ddt -\eps)} d \nu \,\frac{\nu}{\pi i} f^{\rm 2 pt}(\nu)G_\nu^{\alpha\beta}(s)\,,\\
\label{f1loop}
 f^{\rm 2 pt}(\nu) & =\frac{1}{\nu^2-\nu_\sigma^2 - \frac{g^2}{2} \hat{B}_{\nu_\phi}(\nu)}\,,
\end{align}
where $ \hat{B}_{\nu_\phi}(\nu)$ is the ``bubble'' function given in \eqref{eq:Bfinal}. It was first computed in \cite{Marolf:2010zp}.  In $d=2$ the expression for the bubble simplifies and we get
\be
\hat{B}_{\nu_\phi}(\nu)\vert_{d=2}={\frac{i}{8 \pi \nu}}\l(\pi+\cot{\pi i \nu_\phi}\l(\psi\l(\frac{1}{2}+\frac{i\nu}{2}-i\nu_\phi\r)-\psi\l(\frac{1}{2}+\frac{i\nu}{2}+i\nu_\phi\r)\r)\r)\,.
\ee
The factor of $1/2$ in front of it in the denominator of \eqref{f1loop} is the symmetry coefficient of the bubble diagram. 

The function $\hat{B}_{\nu_\phi}(\nu)$ has the interpretation of the dS spectral density for the two-point function of the operator $\phi^2$ in a theory of a free scalar $\phi$, see the general definition of the spectral density in the appendix \ref{app:SR}. Note that the odd, or equivalently the imaginary part of $\hat{B}_{\nu_\phi}(\nu)$ is positive for real positive $\nu$ due to unitarity:
\be\label{eq:posB}
-i(\hat{B}_{\nu_\phi}(\nu)-\hat{B}_{\nu_\phi}(-\nu))>0~.
\ee
This is a check of the general statements in section \ref{sec:pos2pt}. In agreement with the arguments in the appendix \ref{app:SR}, $\hat{B}_{\nu_\phi}(\nu)$ is holomorphic in the lower-half $\nu$ plane with the possible exception of the interval $i\nu\in(0,\frac{d}{2})$, while it has poles in the upper half-plane at the location of ``double-trace operators'' in the mean field theory of the operator associated to the free field $\phi$. These are the following three families of poles:
\begin{equation}\label{eq:Bpoles}
\nu_{n++}=2\nu_\phi + i \left(\frac{d}{2}+2n\right)~,~~ \nu_{n+-}=i \left(\frac{d}{2}+2n\right)~,~~\nu_{n--}=-2\nu_\phi + i \left(\frac{d}{2}+2n\right)~,
\end{equation}
with $n$ a non-negative integer.

The resummed propagator $f^{\rm 2 pt}(\nu)$ inherits an interesting analytic structure from that of $\hat{B}_{\nu_\phi}$, that we discuss in more details in the next section. Being itself the spectral density of a two-point function, it is still holomorphic in the lower-half $\nu$ plane with the possible exception of poles for $i\nu\in(0,\frac{d}{2})$. In \eqref{sigma2pt} we chose to send the contour below all possible poles of $f^{\rm 2 pt}(\nu)$. Equivalently we could choose the real $\nu$ contour and explicitly include the contribution from all the residues of $f^{\rm 2 pt}(\nu)$ for $i\nu\in(0,\frac{d}{2})$. They could correspond to either single-$\sigma$-particle poles, or to two-$\phi$-particle poles, as long as $\sigma$ and $\phi$ are sufficiently light. Since we do not consider exactly massless particles, it suffices to take $\eps$ to be a small positive number. Note that when $\sigma$ is a heavy particle, the effect of the resummation is to shift the poles at $\nu =\pm \nu_\sigma$ from the real $\nu$ axis to the upper-half plane, as was noted previously in \cite{Marolf:2010zp,Krotov:2010ma}. This is thanks to the positivity condition \eqref{eq:posB}.  

\begin{figure}[t] 
	\centering
		\hspace{0cm}\includegraphics[width=1\linewidth,angle=0]{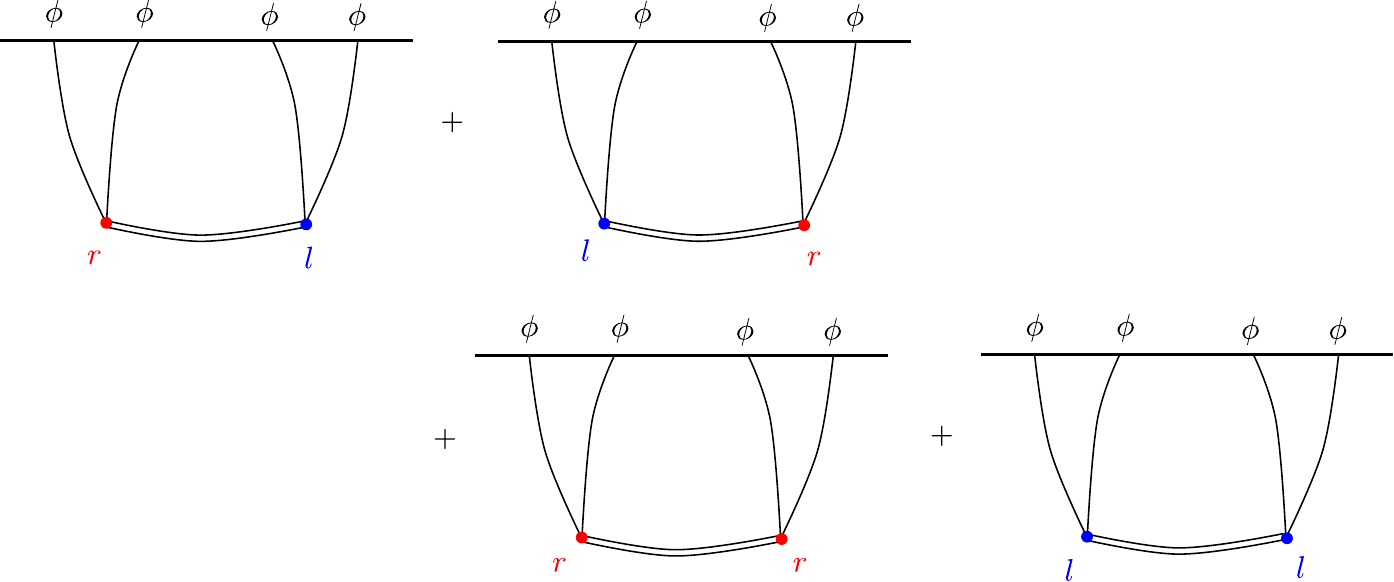}
	\caption{\small Exchange diagrams with the ``full'' propagator of the $\sigma$ particle, see fig. \ref{fig:sigmaprop}. The analytic structure of the four-point function is richer after the resummation, see section \ref{sec:OPEexch}, and going to the resonance at $\nu=\nu_\sigma$ we get a non-trivial check of the positivity constraint, see section \ref{sec:poscheck}.}
	\label{fig:compexch}
\end{figure}

Now we proceed to calculate the diagrams in figure \ref{fig:compexch}, using \eqref{sigma2pt} as a propagator for $\sigma$. The result is simply the tree level answer, given in \eqref{treefull}, but instead of a single particle with weight $\nu_\sigma$ we exchange a composite object characterized by the spectral density $f^{\rm 2 pt}(\nu)$:
\be
D^{s,{\rm loop}}=\int_{-\infty-i(\ddt-\epsilon)}^{+\infty-i(\ddt -\epsilon)}d\nu\frac{\nu}{\pi i}f^{\rm 2 pt}(\nu) D^{s,{\rm tree}}_{\nu}\,.
\ee
Due to our choice of contour, each value of $\nu$ corresponds to a ``particle'', which propagator behaves at large distances as that of a particle with mass of order $\eps$. As follows from \eqref{treefull}, there are two contributions in $D^{s,{\rm tree}}_{\nu}$, a continuum and a discrete one. Let us focus on the continuum part first, which reads:
\be
 \sim \int_{-\infty-i(\ddt-\eps)}^{+\infty-i(\ddt-\eps)} d \nu\frac{\nu}{\pi i}f^{\rm 2 pt}(\nu) \int_{-\infty}^{+\infty} \frac{d\nu'}{\nu'^2-\nu^2}\frac{\G^2\l(\frac{d-4i\nu_\phi\pm2i\nu'}{4}\r)\G^2\l(\frac{d\pm2i\nu'}{4}\r)}{\G\l(\pm i\nu'\r)}{\cal F^{\{-\nu_\phi\}}_{\nu'}}(z,\bar z)
\ee
We observe that, thanks to the holomorphicity of $f^{\rm 2 pt}(\nu)$ in the lower half-plane, there are no singularities that prevent us from pulling the $\nu$ contour all the way in the lower-half plane, and consequently this contribution vanishes. Thus the full answer comes from the discrete term in \eqref{treefull} and reads
\be
\label{1loopfull}
D^{s,{\rm loop}}= \l( \frac{\eta_c}{x_{12}}\r)^{d-2i\nu_\phi} \l(\frac{\eta_c}{x_{34}}\r)^{d-2i\nu_\phi} \int_{-\infty-i(\ddt-\eps)}^{+\infty-i(\ddt-\eps)} d \nu \, f^{\rm 4pt}(\nu) {\cal F^{\{-\nu_\phi\}}_\nu}(z,\bar z) \,,
\ee
where we defined
\begin{align}
\begin{split}
\label{rho1loop}
& f^{\rm 4pt}(\nu) = g^2 \, A(\nu) \, f^{\rm 2 pt}(\nu)~,\\
 A(\nu) & \equiv \frac{\G^4(i\nu_\phi)}{2^{10}\pi^{2d+5}}\sin^2\tfrac{\pi}{2}\l(\tfrac{d}{2}-i\nu-2 i \nu_\phi\r)\G^2\l(\tfrac{d-4i\nu_\phi\pm2i\nu}{4}\r)\G^2\l(\tfrac{d\pm2i\nu}{4}\r) \frac{\nu}{\pi i}\,.
\end{split}
\end{align}
Naturally, we can manipulate the contour in \eqref{1loopfull} and send it along the real $\nu$ axis, by picking explicit contributions from the poles of $f^{\rm 2 pt}(\nu) $ located in between. The shift of the poles at $\pm \nu_\sigma$, due to the positive imaginary part of $B_{\nu_\phi}(\nu)$, avoids the difficulty that one seemed to have  with pushing the $\nu$ contour to the real axis at the tree level. The function $f^{\rm 4pt}(\nu)$ is related to the spectral amplitude defined in secitions \ref{sec:Analyticity} and \ref{sec:Unitarity} by $f^{\rm 4pt}(\nu)=2\pi f(\tfrac{d}{2}-i\nu)$. Let us proceed with a more detailed discussion of the properties of $D^{s,{\rm loop}}$. 

\subsection{OPE decomposition of the four-point function}\label{sec:OPEexch} 
First we turn to the decomposition of the four-point function into conformal blocks. To do this we the relation between conformal partial waves and conformal blocks \eqref{eq:CPW}. Rewritten in terms of the $\nu$ variable, it takes the form
\beg\label{eq:FtoK}
{\cal F_{\nu}^{\{-\nu_\phi\}}}=C_\nu\,{\cal K}^{\{-\nu_\phi\}}_{\ddt+i\nu}\,+\,C_{-\nu}\,{\cal K}^{\{-\nu_\phi\}}_{\ddt-i\nu}\,,\\
C_\nu=\frac{\pi^\ddt \G(-i\nu)\G^2\l(\frac{d}{4}+\frac{i \nu}{2}\r)}{\G\l(\ddt+i\nu\r)\G^2\l(\frac{d}{4}-\frac{i \nu}{2}\r)}\,.
\eeg
This relation can be substituted into \eqref{1loopfull}, after which we can pull the $\nu$ contour to the lower-half plane ($\text{Re}\,i\nu > 0$) for the first term and to the upper-half plane ($\text{Re}\,i\nu < 0$) for the second term, thus generating an expansion into conformal blocks with positive real part of the dimension of the exchanged operator.

\begin{figure}[t] 
	\centering
		\hspace{0cm}\includegraphics[width=0.7\linewidth,angle=0]{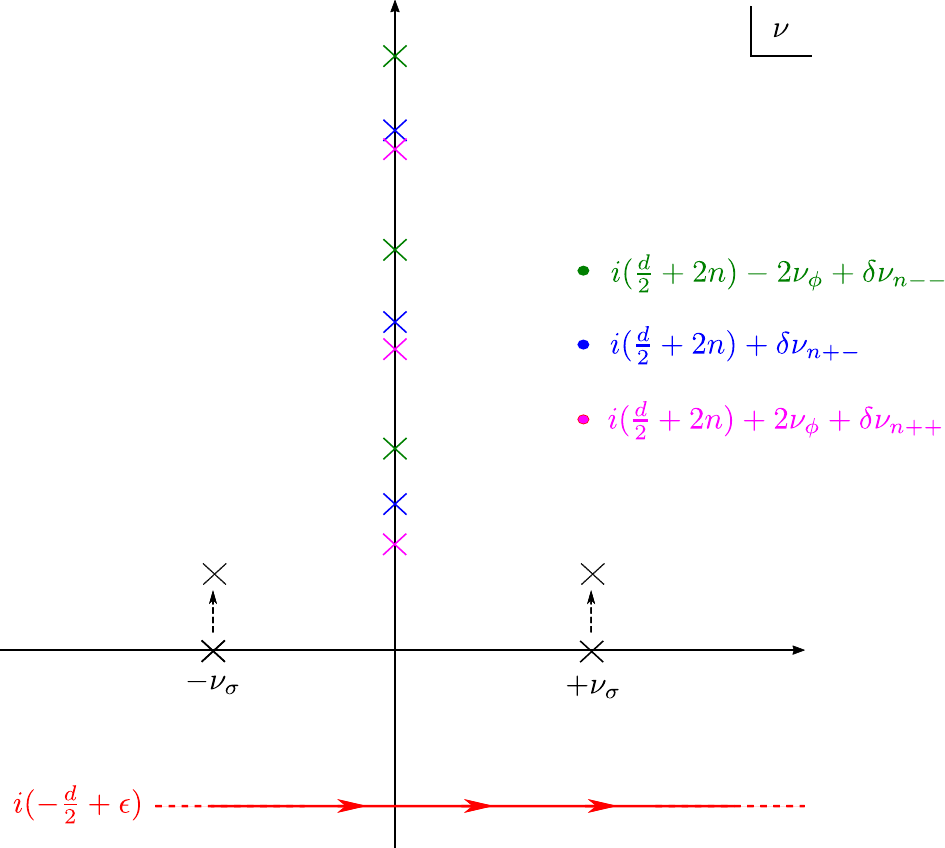}
	\caption{\small The analytic structure of the function $f^{\rm 4pt}(\nu)C_{-\nu}$ that encodes the conformal block expansion of the late-time four-point function. The function is holomorphic below the integration contour depicted by the red line. There are three sequences of poles on the imaginary $\nu$ axis that correspond to ``double-trace operators". Due to the interaction with the particle $\sigma$, their positions receive a shift $\delta\nu$ from the equally spaced values in the free theory. Before the resummation of bubble diagrams there are two poles on the real axis associated to particle $\sigma$. The dashed arrows denote the shift in the position of these poles induced by the interaction.}
	\label{fig:fOPE}
\end{figure}

We note that, while $f^{\rm 4pt}(\nu)$ is not holomorphic in $\nu$ in the half-plane ${\rm Im}\nu < -\frac{d}{2}$, due to the prefactor $A(\nu)$ in \eqref{rho1loop}, the product $f^{\rm 4pt}(\nu) \, C_{\nu}$ is. This means that in the part of the integral \eqref{1loopfull} involving $C_\nu\,{\cal K}^{\{-\nu_\phi\}}_{\ddt+i\nu}$ we can pull the contour all the way in the lower half-plane, and it does not contribute to the conformal block expansion. As a result the full conformal block expansion is obtained by pulling the contour for the term $C_{-\nu}\,{\cal K}^{\{-\nu_\phi\}}_{\ddt-i\nu}$ in the upper-half plane, and it is controlled by the poles of $f^{\rm 4pt}(\nu)C_{-\nu}$. We illustrate the analytic structure of $f^{\rm 4pt}(\nu)C_{-\nu}$ in figure \ref{fig:fOPE}. This provides an example of the analytic structure of the spectral amplitude discussed in section \ref{sec:Analyticity}.
We could have used the symmetry of the conformal partial wave to replace $f^{\rm 4pt}(\nu)$ inside the $\nu$ integral with its symmetrization under $\nu\to-\nu$, i.e. what we called $\rho(\nu)$ in section \ref{sec:Analyticity}. Here we see however that the symmetrization comes at the expense of loosing the analyticity of $f^{\rm 4pt}(\nu)$ in the lower-half plane. It is always possible to rewrite the integral representation involving $\rho(\nu)$ in terms of a density which is analytic in the lower half-plane. The two functions contain the same physical information, i.e. the conformal block expansion of the correlator. Moreover $\rho(\nu)$ is positive due to unitarity as explained in section \ref{sec:Unitarity}. It would be interesting to better understand more generally the physical interpretation of the analytic function $f^{\rm 4pt}(\nu)$. 

Let us focus on the poles of $f^{\rm 4pt}(\nu)C_{-\nu}$  when $d-4i\nu_\phi+2i\nu=-4 n$, where $n$ is a non-negative integer. Poles at this value of the spectral parameter correspond to ``double-trace operators" that contribute to this late-time four-point function even in the free theory of the field $\phi$, meaning that they are also present in the disconnected diagram. Note that even though the prefactor $A(\nu)$ in \eqref{rho1loop} has double poles at these locations, $f^{\rm 2 pt}(\nu)$ has a zero (due to the pole of $\hat{B}_{\nu_\phi}$), and, consequently, these are just single poles. In fact these poles should cancel with those in the disconnected contribution to the four-point function. The argument for this goes as follows. Since we have identical external particles, the diagram would be identical if instead we considered a theory with $N$ identical $\phi$-fields and took $N$ large. In this case, the exchange of $\sigma$ in the $t$- and $u$-channel would be suppressed compared to the $s$-channel and consequently the diagrams we consider would be the leading $1/N$ contribution, even at a finite coupling. Now we can use the EAdS version of our theory. For unitary theories in AdS such poles cancel with the disconnected contribution to the four-point function, and they get replaced by nearby poles, which can be thought of as the interacting version of the double-trace operators at the leading order in the $1/N$ expansion. In fact, as was shown in \cite{Carmi:2018qzm}, requirement of the poles to cancel is enough to bootstrap the AdS bubble function. Our theory is non-unitary from the AdS point of view, however this should not affect the pole cancellation.  Here we calculated the dS bubble function in a different way and we check the poles cancellation explicitly, as it serves as an important cross check of our computation.

The conformal partial waves decomposition of the disconnected contribution to the four-point function contains contributions with arbitrary even spin. Here we are only interested in the scalar contribution, which reads\footnote{We ignore the $s$-channel contraction which amounts to an exchange of the identity operator.} 
\beg
\l\langle\phi(x_1,\eta_c)\phi(x_3,\eta_c)\r\rangle\l\langle\phi(x_2,\eta_c)\phi(x_4,\eta_c)\r\rangle+ \l\langle\phi(x_1,\eta_c)\phi(x_4,\eta_c)\r\rangle\l\langle\phi(x_2,\eta_c)\phi(x_3,\eta_c)\r\rangle= \\
=\l( \frac{\eta_c}{x_{12}}\r)^{d-2i\nu_\phi} \l(\frac{\eta_c}{x_{34}}\r)^{d-2i\nu_\phi} \int_{-\infty}^{\infty} d\nu\,\rho^{\rm free}(\nu) {\cal F_{\nu}^{\{-\nu_\phi\}}}\,,
\eeg
where $\rho^{\rm free}(\nu)$ is the mean field theory expression \cite{Karateev:2018oml} multiplied by the normalization factor $\sqrt{N_{\nu}}$, defined in \eqref{KdS}, for each field: 
\beg
\label{rhofree}
\rho^{\rm free}(\nu)= \frac{\G\l(\ddt\r) \G^4\l(i\nu_\phi\r)}{64 \pi^{\frac{3 d}{2}+3}}
\frac{\G\l(\ddt\pm i\nu\r)\G\l(\frac{d-4i\nu_\phi\pm2i\nu}{4}\r)}{\G(\pm i\nu)\G\l(\frac{d+4i\nu_\phi\pm2i\nu}{4}\r)}\,.
\eeg
Note in passing that for $i\nu_\phi \in (0,\frac d2)$ we have $\rho^{\rm free}(\nu) \geq 0$ as required by unitarity, see section \ref{sec:Unitarity}. The cancellation of the double-trace poles is due to the following identity valid for any $n\in \mathbb{N}$
\begin{align}
\begin{split}\label{eq:freeres}
& {\rm Res}[2 \rho^{\rm free}(\nu) \, C_{-\nu} ]\vert_{\nu=2\nu_\phi + i \frac{d}{2} + 2 i n}=-{\rm Res}[f^{\rm 4pt}(\nu) \, C_{-\nu}]\vert_{\nu=2\nu_\phi + i \frac{d}{2} + 2 i n}\\
& =\frac{i  (-1)^{n+1} \Gamma \left(\frac{d}{2}\right) \Gamma (i \nu_\phi )^4 \Gamma (2 i \nu_\phi -2 n) \Gamma \left(\frac{d}{2}+n-i \nu_\phi \right)^2 \Gamma \left(\frac{d}{2}+n-2 i \nu_\phi \right)}{16 \pi^{d+3} n! \Gamma \left(\frac{d}{2}+n\right) \Gamma (i \nu_\phi -n)^2 \Gamma (2 i \nu_\phi -n) \Gamma \left(\frac{d}{2} +2 n-2 i \nu_\phi \right)}~.
\end{split}
\end{align}
Note that we have a factor of 2 in the free contribution, due to the fact that $\rho^{\rm free}(\nu)$ is shadow-symmetric -- i.e. $\rho^{\rm free}(-\nu)=\rho^{\rm free}(\nu)$ -- and therefore these double-trace poles receive equal contribution from the two terms in \eqref{eq:FtoK}. This is not the case for \eqref{1loopfull}, that as we explained above receives contributions only from the poles of $f^{\rm 4pt}(\nu) \, C_{-\nu}$ in the upper half-plane. It is interesting to note that even though the factors of gamma functions in $f^{\rm 4pt}(\nu)$ that give rise to the double pole are shadow-symmetric, the prefactor $A(\nu)$ contains also a sine$^2$ that precisely cancel the double-poles in the lower-half $\nu$ plane. 

The other poles of $f^{\rm 4pt}(\nu) \, C_{-\nu}$ give the OPE decomposition of the four-point function. We have three families of poles arising due to poles of the bubble function $\hat{B}_{\nu_\phi}(\nu)$ in \eqref{eq:Bpoles}. Let us first take $\nu = \nu_{n++} + \delta\nu_{n++}$ and expand $f^{\rm 4pt}(\nu) \, C_{-\nu}$ in the limit $\delta \nu_{n++}, g^2 \to 0$ with a fixed ratio $\delta\nu_{n++}/g^2$. We find that besides the pole at $\delta\nu_{n++} = 0$ --which cancels with the disconnected contribution as we discussed above-- $f^{\rm 4pt}(\nu) \, C_{-\nu}$ has a pole at
\begin{align}
\begin{split}
& \delta\nu_{n++} = -\frac{g^2 \Gamma \left(\frac{d}{2}+n\right)}{2^{d+3} \pi ^{\frac d2} n! \Gamma \left(\frac{d}{2}\right)} \\ & \times\frac{\sinh (\pi  \nu_{n++}) \text{csch}^2(\pi  \nu_\phi )  \Gamma \left(n-i \nu_\phi +\frac{1}{2}\right) \Gamma \left(\frac{d}{2}+n-i \nu_\phi \right) \Gamma \left(\frac{d}{2}+n-2 i \nu_\phi \right)}{  \left(\nu_{n++}^2-\nu_\sigma ^2\right) \Gamma (n-i \nu_\phi +1) \Gamma (n-2 i \nu_\phi +1) \Gamma \left(\frac{d}{2}+n-i \nu_\phi +\frac{1}{2}\right)}~.
\end{split}
\end{align}
This represents the shift in the scaling dimension of the free-theory double-trace operators induced by the coupling $g$. The residue at the shifted poles is the same as the one in the free theory in eq. \eqref{eq:freeres}.

Similarly expanding around $\nu_{n+-}$ we find a pole at $\nu =\nu_{n+-} + \delta\nu_{n+-}$ with
\begin{align}
\begin{split}
\delta\nu_{n+-} & = \frac{ig^2 \Gamma \left(n+\frac{1}{2}\right) \Gamma \left(\frac{d}{2}+n\right)}{2^{d+2} \pi ^{\frac{d}{2}} n! \Gamma \left(\frac{d}{2} \right)\Gamma \left(\frac{d+1}{2}+n\right)}  \frac{\sin \left(\frac{\pi  d}{2}\right) \text{csch}^2(\pi  \nu_\phi ) \Gamma \left(\frac{d}{2}+n\pm i \nu_\phi \right) }{\left(\nu_{n+-}^2-\nu_\sigma ^2\right) \Gamma (n+1\pm i \nu_\phi )}~,
\end{split}
\end{align}
and residue
\begin{align}
\begin{split}\label{eq:OPE2}
& {\rm Res}[f^{\rm 4pt}(\nu) \, C_{-\nu}]\vert_{\nu =\nu_{n+-}} = -\frac{g^2 \Gamma \left(-\frac{d}{2}-2 n+1\right) \Gamma \left(\frac{d}{2}+n\right)^3}{2^{d+2 n+9} \pi ^{\frac{3 d+11}{2}}  \Gamma \left(\frac{d+1}{2}+n\right)} \\ &~~~~~~~~~ \times \frac{\sin ^2\left(\pi  \left(\frac{d}{2}-i \nu_\phi \right)\right)\Gamma (i \nu_\phi )^4  \Gamma (-n-i \nu_\phi )^2 \Gamma \left(\frac{d}{2}+n-i \nu_\phi \right)^2}{\left(\nu_{n+-}^2-\nu_\sigma^2\right) }\delta\nu_{n+-} ~,
\end{split}
\end{align}
and expanding around $\nu_{n--}$ we find a pole at $\nu =\nu_{n--} + \delta\nu_{n--}$ with
\begin{align}
\begin{split}\label{eq:OPE3}
&\delta\nu_{n--}  = -\frac{g^2 \Gamma \left(n+\frac{1}{2}\right) \Gamma \left(\frac{d}{2}+n\right)}{2^{d+3} \pi ^{d/2} n! \Gamma \left(\frac{d}{2}\right) \Gamma \left(n+\frac{1}{2}\right)} \\ &\times \frac{ \sinh (\pi  \nu_{n--}) \text{csch}^2(\pi  \nu_\phi ) \Gamma \left(n+i \nu_\phi +\frac{1}{2}\right) \Gamma \left(\frac{d}{2}+n+i \nu_\phi \right) \Gamma \left(\frac{d}{2}+n+2 i \nu_\phi \right)}{ \left(\nu_{n--}^2-\nu_\sigma ^2\right) \Gamma (n+i \nu_\phi +1) \Gamma (n+2 i \nu_\phi +1) \Gamma \left(\frac{d}{2}+n+i \nu_\phi +\frac{1}{2}\right)}~,
\end{split}
\end{align}
and residue
\begin{align}
\begin{split}
& {\rm Res}[f^{\rm 4pt}(\nu) \, C_{-\nu}]\vert_{\nu =\nu_{n--}} = -\frac{g^2 \Gamma \left(\frac{d}{2}+n\right)^2}{2^{\frac{d}{2}+9} \pi ^{\frac{3 d+11}{2} }} \\ & \times\frac{\sin ^2\left(\frac{\pi  d}{2}\right) \Gamma (i \nu_\phi )^4 \Gamma (-n-2 i \nu_\phi )^2 \Gamma \left(\frac{d}{2}+n+i \nu_\phi \right)^3 \Gamma \left(-\frac{d}{2}-2 n-2 i \nu_\phi +1\right)}{ 2^{-i \nu_{n--}} \left(\nu_{n--}^2-\nu_\sigma ^2\right) \Gamma \left(\frac{d}{2}+n+i \nu_\phi +\frac{1}{2}\right)}\delta \nu_{n--}~.
\end{split}
\end{align}
We can interpret these residues, multiplied by $2\pi i$, as the square of the OPE coefficients between the two external operators and the exchanged operator with dimension fixed by the position of the pole. While in the free theory  these squared OPE coefficients are positive for any $n$ and any $i\nu_\phi \in (0,\frac{d}{2})$, and consequently they are also positive at the shifted poles in $\nu_{n++}+\delta\nu_{n++}$, for the other two families of poles we do not find a positive definite result. This is another signature of the lack of unitarity of the late-time CFT. We observe that some Gamma functions appearing in the expressions \eqref{eq:OPE2}-\eqref{eq:OPE3} give rise to poles for even (integer) $d$. In the case of \eqref{eq:OPE2}, this pole is precisely compensated by the zero of the factor of $\sin\left(\frac{d}{2}\right)$ in $\delta\nu_{+-}$, leading to a finite non-vanishing squared OPE coefficient for the double-trace operators of type $+-$. On the other hand, in the case of \eqref{eq:OPE3} we have an additional factor of $\sin\left(\frac{d}{2}\right)^2$, which leads to a vanishing OPE coefficient for the double-trace operators of type $--$ in even $d$.

We note that the leading asymptotic growth of these squared OPE coefficients is $\sim 2^{-{4n}}$ at large $n$, for all the three families. This is the same as the leading growth of the squared OPE coefficients in standard mean-field theory, see e.g. \cite{Fitzpatrick:2011dm}. By comparison, we conclude that the OPE expansion is convergent in this example. This is a notable result, because due to the lack of unitarity of the CFT, and more generally the lack of an operator-state correspondence (see the discussion in section \ref{sec:Analyticity}), we cannot use the Euclidean theorem about convergence of the OPE \cite{Pappadopulo:2012jk}. It is an important open question for the future to try to understand whether the OPE and its convergence can be established more generally for late-time dS correlators.

\subsection{Explicit check of unitarity and narrow resonances} \label{sec:poscheck}
As we discussed in section \ref{sec:Unitarity}, the spectral density of the four-point function of four identical operators corresponding to light external fields, i.e. with $\nu_\phi$ imaginary, has to be positive. Let us check that this is indeed the case for our one-loop result given by the sum of \eqref{rho1loop} and \eqref{rhofree}. Recall that the spectral density with positivity property is the symmetrization of $f^{\rm 4pt}(\nu)$ under $\nu\to-\nu$ (this symmetrization does not change the $\nu$ integral in \eqref{rho1loop} because the conformal partial wave is symmetric under $\nu\to-\nu$). Hence we wish to prove positivity of
\begin{align}\label{eq:resrho}
\begin{split}
\rho(\nu)  & = \rho^{\rm free}(\nu) + \frac{1}{2}(f^{\rm 4pt}(\nu) + f^{\rm 4pt}(-\nu)) \\
& = \rho^{\rm free}(\nu) +\frac{g^2}{2} \left(\frac{A(\nu)}{\nu^2 -\nu_\sigma^2 -\frac{g^2}{2} \hat{B}_{\nu_\phi}(\nu)}+ \frac{A(-\nu)}{\nu^2 -\nu_\sigma^2 -\frac{g^2}{2} \hat{B}_{\nu_\phi}(-\nu)}\right)~.
\end{split}
\end{align}
Away from the resonance at $\nu =\pm \nu_\sigma$ this condition is obviously satisfied, because the correction is subleading with respect to the positive free field contribution. However, for real $\nu_\sigma$ and for $\nu\approx \pm \nu_\sigma$ the resonant contributions is order one in terms of $g^2$ counting and it is not sign definite on its own. Thus, the positivity check becomes non-trivial. 

First of all, we separate $\hat{B}_{\nu_\phi}(\nu)$ in its real and imaginary part, which have the following parity properties for real $\nu$
\begin{align}
\begin{split}
{\rm Re} \hat{B}_{\nu_\phi}(\nu) & = {\rm Re} \hat{B}_{\nu_\phi}(-\nu)~, \\
{\rm Im} \hat{B}_{\nu_\phi}(\nu) & = -{\rm Im} \hat{B}_{\nu_\phi}(-\nu)~.
\end{split}
\end{align}
We note that the real part of $B_{\nu_\phi}(\nu)$ near the resonance amounts to order $g^2$ renormalization of the mass of the $\sigma$ particle, namely the poles at $\pm \nu_\sigma$ move to 
\begin{equation}
\pm\left(\nu_\sigma + \frac{g^2}{2}{\rm Re} \hat{B}_{\nu_\phi}(\nu_\sigma)\right)~.
\end{equation}
This shift can be absorbed in a redefinition of $\nu_\sigma$. This will be implied. 

\begin{figure}[t] 
	\centering
		\hspace{0cm}\includegraphics[width=0.7\linewidth,angle=0]{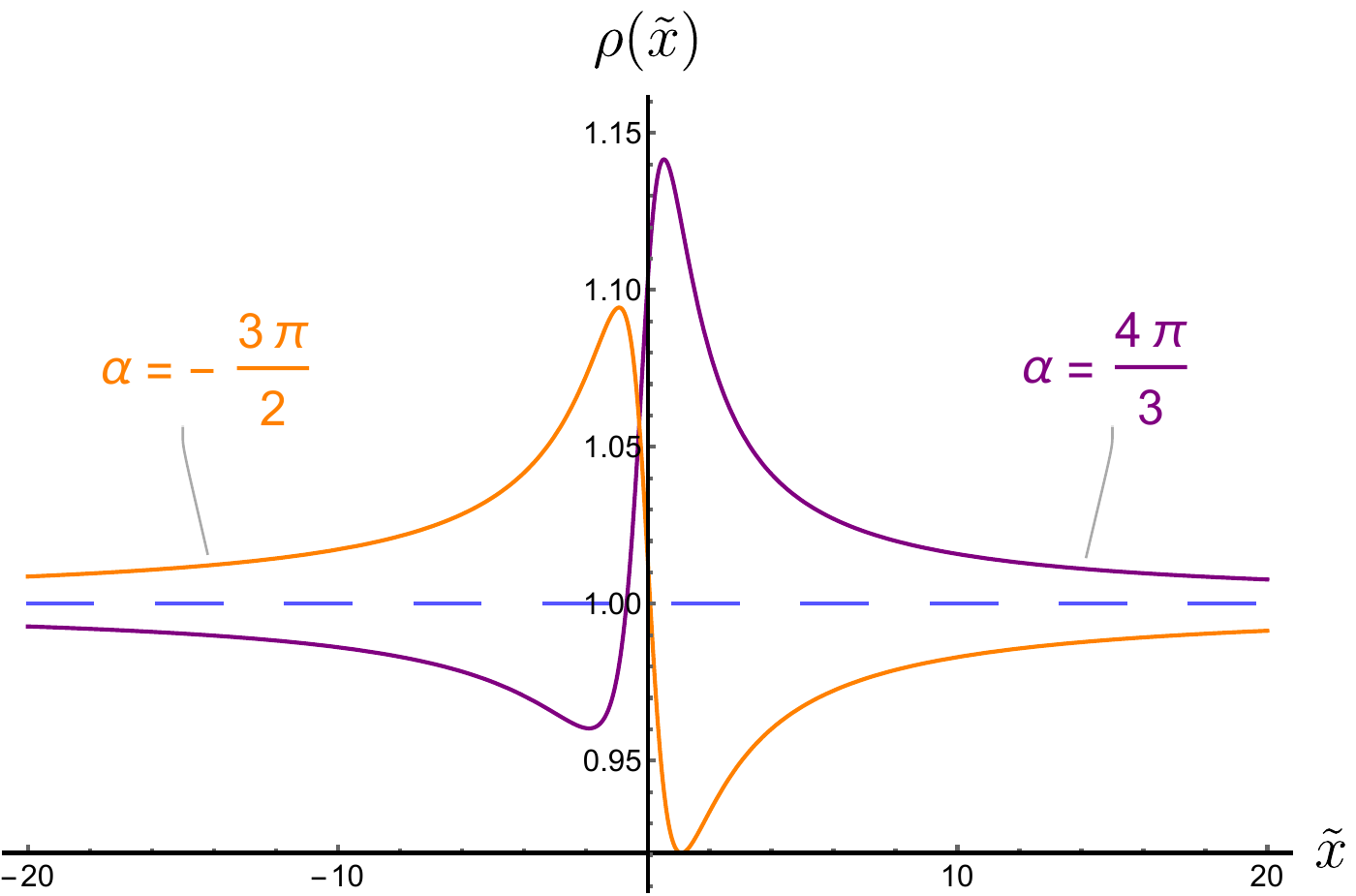}
	\caption{\small The dS version of the Breit-Wigner distribution associated to a resonance. Here we set the positive quantity $\rho^{\rm free}(\nu_\sigma)$ to 1, $\alpha=\pi\l(\ddt-2 i \nu_\phi\r)$ and $\tilde x= x/\gamma$ is the ratio between $x$, the displacement from the resonance $\nu=\nu_\sigma$ in units of the coupling $g^2$, and the width $\gamma$. The two curves are for $\nu_\sigma = 1$ and $\alpha =\frac{4\pi}{3}$ (purple)/$\alpha =-\frac{3\pi}{2}$ (orange). 
	For large positive and negative $\tilde{x}$, $\rho$ approaches the free value $\rho^{\rm free}(\nu_\sigma)$, indicated by the dashed blue line. The curve has two extrema at the positions \eqref{eq:extrema}.}
	\label{fig:resonance}
\end{figure}

We then plug $\nu = \nu_\sigma + \frac{g^2}{2}x$ and keep only the leading term in \eqref{eq:resrho} for small $g^2$ in order to zoom to the vicinity of the resonance. We obtain
\begin{align}
\begin{split}
\label{unit1}
\rho(\nu)=\rho^{\rm free}(\nu_\sigma)& \l[1+\frac{ 1}{32 \pi^{\frac{d}{2}+3}\G\l(\ddt\r)}
\frac{\G(\pm i\nu_\sigma)\G\l(\frac{d\pm4i\nu_\phi\pm2i\nu_\sigma}{4}\r)\G^2\l(\frac{d\pm2i\nu_\sigma}{4}\r)}{\G\l(\ddt\pm i\nu_\sigma\r)} \r.\\
& \times \l.\l(\l(1-\cos\pi\l(\tfrac{d}{2}-2 i \nu_\phi\r)\cosh\pi\nu_\sigma  \r)\frac{\gamma}{x^2 + \gamma^2} \r.\r.\\ & \l.\l. \phantom{\frac{\G(\pm i\nu_\sigma)}{\G\l(\ddt\pm i\nu_\sigma\r)}}-  \sin\pi\l(\tfrac{d}{2}-2 i \nu_\phi\r)\sinh \pi \nu_\sigma \frac{x}{x^2 + \gamma^2}\r )\r] +O(g^2)~,
\end{split}
\end{align}
where $g^2 \gamma$ is the width of the resonance with $\gamma$ given by
\be
\gamma=\frac{1}{2 \nu_\sigma}\text{Im}\hat{B}_{\nu_\phi}(\nu_\sigma)= \frac{  \Gamma\left(\frac{d}{4} \pm i \frac{\nu_\sigma}{2}\right)^2  \Gamma\left(\frac{d}{4} \pm i \frac{\nu_\sigma -2 \nu_\phi}{2}\right)  \Gamma\left(\frac{d}{4} \pm i \frac{\nu_\sigma +2 \nu_\phi}{2}\right) }{64 \pi ^{\frac{d}{2}+1}\Gamma\left(\frac{d}{2}\right)\,\nu_\sigma^2\, \G\l(\pm i \nu_\sigma\r) \Gamma\left(\frac{d}{2}\pm i \nu_\sigma\right)}\,.
\ee
Note the somewhat unusual shape of the resonance which contains both even and odd  terms with respect to $x=0$. This is illustrated in figure \ref{fig:resonance}.

To further simplify the expression, let us define $\tilde x= x/\gamma$ and $\alpha=\pi\l(\ddt-2 i \nu_\phi\r)$, as well as factor out $\gamma$ in the numerator and denominator. After these manipulations most of the lengthy overall factors cancel and we get
\be
\label{ressimple}
\rho(\nu)=\rho^{\rm free}(\nu_\sigma)\l(1+2\l(\frac{1-\cosh\pi\nu_\sigma\cos\alpha}{\sinh^2\pi\nu_\sigma}\frac{1}{\tilde x^2 + 1}-\frac{\sin\alpha}{\sinh\pi\nu_\sigma}\frac{\tilde x}{\tilde x^2 + 1}\r)\r)\,.
\ee
This expression approaches the positive value $\rho^{\rm free}(\nu_\sigma)$  for $\tilde{x}\to-\infty$, from above or below depending on the sign of $\alpha$, and it approach the same value for $\tilde{x}\to+\infty$, from the opposite direction. Moreover it is extremized at 
\be\label{eq:extrema}
\tilde x \in \l\{-\cot\l(\tfrac{\alpha}{2}\r)\tanh\l(\tfrac{\pi \nu_\sigma}{2}\r),\tan\l(\tfrac{\alpha}{2}\r)\coth\l(\tfrac{\pi \nu_\sigma}{2}\r)\r\}\,.
\ee
At these values of $\tilde x$ it takes the following values:
\be
\rho(\nu)=\rho^{\rm free}(\nu_\sigma)\l(\cosh\pi\nu_\sigma-\cos\alpha\r)\times \l\{\sinh^{-2}\l(\tfrac{\pi\nu_\sigma}{2}\r),\cosh^{-2}\l(\tfrac{\pi\nu_\sigma}{2}\r)\r\}\,,
\ee
which are manifestly positive, as long as $\nu_\sigma$ and $\alpha$ are real. Therefore $\rho$ as a function of $\tilde{x}$ behaves as sketched in figure \ref{fig:resonance}, for any real $\nu_\sigma$ and $\alpha$, and as a result it is always positive, as expected from the general arguments.

Equation \eqref{ressimple} gives us a simple expression for the behavior of the spectral density in the vicinity of the resonance. As it is usually the case, when the spectral parameter is ``on the resonance'', that is $|\tilde x | \approx 1$, or $|\nu-\nu_\sigma|\approx g^2\gamma$, the interacting part of the signal is of order one in the coupling constant, or, in other words comparable to the gaussian free theory part. This property of our representation provides a potentially useful way to look for the non-gaussian features associated to such resonances in experimental data. Of course, much work need to be done in order to make our approach applicable to the realistic cosmological scenarios. We sketch some of the steps in the conclusions. For now we can imagine that the qualitative shape of the resonance survives the necessary modifications. Then the shape of the resonance depends only on the parameter $\alpha$, which for the massless external particle in four-dimensional dS is equal to $-\frac{3 \pi}{2}$. The corresponding shape is drawn in orange in figure~\ref{fig:resonance}. As noted in the same context in \cite{Arkani-Hamed:2015bza}, the signal is suppressed by $e^{-\pi \nu_\sigma}$ when the particle is heavy. Of course the resonant search would be the most powerful if we somehow knew the mass of the exchanged particle in Hubble units.

\section{Summary and future directions}
\label{sec:Conclusions}
We conclude by summarizing our findings and listing possible generalizations and future directions. In this paper we studied several aspects of QFT on a dS background focusing on correlation functions of massive scalar fields located on the future boundary. We developed a formalism that allows to calculate such correlation functions with the help of an auxiliary EAdS theory with the field content doubled with respect to the original one. This technique simplifies the calculations dramatically, as well as it exhibits important analytic properties of the correlators. In particular, we expect that to all orders in perturbation theory dS correlators have the same meromorphic structure as AdS or, equivalently, CFT correlation functions. This is to some extent surprising because the volume of the spacial slice of dS is not bounded and the states are labeled by the continuum rather than by discrete parameters, thus one could have expected a more complicated analytic structure akin to that in flat space scattering amplitudes. And indeed, as we discussed, the OPE that correlators satisfy does not correspond to states in the dS Hilbert space. We also established the conditions that unitarity of the QFT in dS imposes on the correlators. Our most powerful result on this is the positivity of the spectral density in the CPW decomposition of the four-point function. The unitarity condition is independent of the EAdS technique and does not rely on perturbation theory in any way. 

The resulting structure of correlation functions is that of a real non-unitary CFT in the classification of \cite{Gorbenko:2018ncu}. As is completely generic for such theories \cite{Hogervorst:2015akt}, operators have complex dimensions and come in complex conjugate pairs. An additional feature of such cosmological CFT is the above-mentioned positivity condition.

We also calculated, using our new techniques, the exchange diagrams first at tree level and then at one-loop level by resumming the bubble diagrams for the propagator of the exchanged particle. This calculation gave us a check of both the analytic structure related to OPE as well as of the positivity conditions following from the unitarity. In particular, in the case of heavy exchanged particles the use of the full propagator is crucial for maintaining the correct structure in the vicinity of a narrow resonance.

We finally move on to the discussion of interesting open problems. While several of them are mentioned throughout the main text, we find it convenient to list them here, organized in several approximate categories.

\paragraph{Within QFT in dS:}

There are several directions to explore staying within the framework of massive QFT on a dS background. One of them is to establish the analytic structure uncovered in this paper at the non-perturbative level. It is tempting to conjecture that the meromorphic properties of spectral densities and the validity of the OPE holds beyond perturbation theory; however, at present we are lacking the tools to prove this. The reason is that our EAdS Lagrangian \eqref{LAdS} and hence corresponding Euclidean action is not bounded from below, thus one cannot use it in the usual path integral manipulations. Another related issue is the convergence of OPE, which is related to the asymptotic behaviors of the spectral densities involved in our calculations. At the moment, we are only confident in the convergence of OPE for the four-point function in our explicit examples. Understanding the physical meaning of ``operators'' in the bulk language could be very fruitful for this purpose. As we discussed in the subsection~\ref{subsec:QNM} the most promising direction is their relation to the QNMs of the static patch theory. This prompts to investigate the connections of our results and those of \cite{Jafferis:2013qia,Anninos:2020hfj}

Another question related to analyticity and to representations appearing as the intermediate states is the following. As briefly reviewed in Appendix~\ref{app:Gelfand}, the decomposition of the scalar dS Greens function into irreps of the isometry group involves some additional representations, as compared to the corresponding decomposition in EAdS. Nevertheless, these additional representations do not appear in our final answers. Even though there is no contradiction in these statements, it may be a hint to some properties of QFT in dS that we are missing so far. 

One more issue that is often brought up in the context of scalar fields in dS is that of IR divergences. Indeed, in this paper we assumed that perturbation theory is valid throughout the calculation, while it is known not to be the case once the fields become sufficiently light. The only known formalism for dealing with this problem is the approach of Starobinski \cite{Starobinsky:1984,Starobinsky:1994bd} promoted to a systematic framework in \cite{Gorbenko:2019rza}. In our approach the IR divergences of perturbation theory will manifest themself in the poles of the spectral density for $\nu\approx-i\frac{d}{2}$ whose residues are likely to be large and grow with the order of perturbation theory. One may wonder if our techniques can provide an alternative framework for dealing with such divergences.

A powerful tool for analyzing non-perturbative properties of QFTs is the use of a large-$N$ expansion. We can thus consider, for example, a large-$N$ $O(N)$ model on dS space at a finite coupling \cite{upcoming}. In the light physical mass regime and when the non-perturbativity is due to IR divergences both approaches should work. Then the eigenvalues of the Starobinski operator, computed to a subleading order in \cite{Gorbenko:2019rza} (see also \cite{Mirbabayi:2020vyt,Cohen:2021fzf}), should manifest themselves as poles in the spectral density, discussed in the appendix \ref{app:SR}, for $i \nu\in (0,\frac{d}{2})$. In \cite{upcoming} we will show that it is indeed the case, providing an interesting check of both techniques.

There are other famous tools for treating QFTs beyond perturbation theory: one is analytic and numerical bootstrap. During the last few years it was successfully applied in multiple settings to constrain flat space, AdS, and CFT observables. A very recent paper \cite{Hogervorst:2021uvp} initiated the bootstrap of cosmological observables. This paper uses the continuum set of states in dS and positivity of the spectral densities as an input for the bootstrap procedure. It is exciting to see the development of this program in the future. It could also be possible to use the discrete set of operators as a basis, in which case one would need the version of the CFT bootstrap applicable to non-unitary theories \cite{Gliozzi:2013ysa,El-Showk:2016mxr}.

The other is to use holography to study gravity duals of strongly-coupled large $N$ field theories in dS. This is an interesting setup on its own since the dual spacetime involves a FRW-like geometry with a singularity \cite{Maldacena:2010un,Hertog:2004rz,Hertog:2005hu,Turok:2007ry,Craps:2007ch,Barbon:2011ta,Barbon:2013nta,Kumar:2015jxa}. On this front, there are two natural questions to address. The first one is to compute the late-time correlators and QNMs in dS using the dual description and study their analytic structures. The second one is to understand a signature of the singularity in the bulk in the late-time correlators on the boundary. Both of these questions deserve detailed explorations.

\paragraph{Changing the state or the background:}
In this paper we tried to be succinct about the choice of state for the QFT calculations, even though that it is an issue that is often argued about. For our QFT calculations the choice of state is just a part of the definition of the theory. There are many ways to implement this definition, like $i\eps$ prescriptions in various coordinate systems, the no-boundary contour of \cite{Hartle:1983ai}, continuation of the calculations on the sphere, etc., and they all lead to the same results. The correlation functions respect dS invariance and reduce to the Minkowski ones for modes short compared to the Horizon scale. Thus we have a well-defined problem and a consistent framework to deal with it. There is a separate issue of what is the correct choice of state for our universe when we try to build a realistic model of cosmology, or when we decide to play with various toy models for cosmology to test our theoretical understanding. Ideally we would hope for a complete theory for the real-world universe that would also predict the initial state; however, we are nowhere close to having such a theory and for now all we have for the choice of state is our intuition, common sense, as well as the experimental data.

It would be interesting to see how our discussion gets modified when we consider other states, for example so-called alpha-states \cite{Mottola:1984ar,Allen:1985ux} that are also dS invariant, at least at the level of free theory. Of course the discussion of the initial state becomes really crucial when gravity is dynamical. In this case the closed analog of the state we considered is the Hartle-Hawking state. Recently, it was understood that other semiclassical contributions to the gravitational path integral, called bra-ket wormholes, may be relevant, at least in some lower-dimensional toy models of cosmology \cite{Chen:2020tes}, see also \cite{Penington:2019kki,Dong:2020uxp}. It is thus important to study QFT correlators on the bra-ket wormhole backgrounds using a generalization of our techniques.

Finally, let us mention a change of state that is definitely necessary to use our techniques for the inflationary phenomenology -- breaking of dS isometries by the clock field always present in all models of inflation \cite{Cheung:2007st}. There are different levels at which this breaking can be implemented. One important limit to consider is when only special conformal transformations are broken, while scale invariance, is preserved (see \cite{Baumann:2019ghk} for a ``boundary'' perspective on this). In this case the propagator has basically the same form with possibly a different speed of sound, consequently the procedure of section~\ref{sec:EAdS} can be readily generalized to this case. There is no problem with the continuation of the interaction vertices either. The discussion of sections ~\ref{sec:Analyticity} and~\ref{sec:Unitarity} more heavily relies on the representation theory of the conformal group, thus to extend the results of these sections to only scale invariant backgrounds would involve exploration of corresponding representation theory. On the other hand, in a scale invariant situation the functional form of the three-point functions is not fixed completely by symmetries, thus one may expect to find some non-trivial constraints from unitarity on them as well.

Eventually realistic inflationary backgrounds break also the scale, or in other words time-translation invariance by various slow-roll corrections. We believe that in many basic cases where the time-dependence of the inflationary background is smooth enough, our techniques have a good chance to work, in a way similar to \cite{Konstantinidis:2016nio} (see also \cite{McFadden:2011kk,Bzowski:2012ih}). On the other hand, models in which time-dependence involves so-called features \cite{Chluba_2015}  may require a special attention and case-by-case study.

\paragraph{Relation to the wave function approaches:}
Much of the recent literature on analytic properties of inflationary observables has focused on the properties of the wave function coefficients defined in section~\ref{sec:wavefunction} \cite{Arkani-Hamed:2017fdk,Hillman:2019wgh,Benincasa:2019vqr,Meltzer:2021zin,Pajer:2020wnj,Goodhew:2021oqg,Goodhew:2020hob,Jazayeri:2021fvk,Cespedes:2020xqq,McFadden:2011kk,Bzowski:2012ih,Baumann:2021fxj}. These studies were done in both dS invariant and boost-breaking settings. In particular, implications of unitarity on the wave function were derived in \cite{Goodhew:2020hob,Jazayeri:2021fvk}. It would be interesting to see how these developments are related to ours and especially how the two unitarity conditions are related. One way to proceed in this direction is through the equation \eqref{corrfrompsi} treated in perturbation theory, as done in some of the mentioned papers, or by using the conjectured relation \eqref{3cfts}. Another approach would be to apply the on-shell methods developed for the wave function calculations directly to the EAdS Lagrangian \eqref{LAdS}. The techniques of the recent paper \cite{Meltzer:2021zin} look particularly promising for this generalization.

\paragraph{Dynamical gravity:}
Needless to say, eventually we would like to study cosmological theories that have dynamical gravity. First, it is natural to ask which of the properties of QFTs in dS survive when coupled to a perturbative bulk gravity. As we discussed above, this already brings in several unresolved issues. For example, there is no obvious way to keep both locality and gauge invariance of correlation functions. The situation is slightly better in the inflationary backgrounds, where time-slicing can be fixed in an unambiguous way. Still spacial distances cannot be measured just in some coordinate system and, presumably, some sort of geodesic dressing is required. This dressing is intertwined with the subtle issue of the meaning of long, or soft, gravitational modes in inflationary context, which is subject of many ongoing discussions, {\it e.g.} \cite{Giddings:2011ze,Senatore:2012ya,Hinterbichler:2013dpa,Mirbabayi:2016xvc,Ferreira:2017ogo}. Inflation also inevitably breaks dS isometries at a level that cannot be arbitrary small \cite{Dubovsky:2011uy}, at least when certain energy conditions are preserved. There is a confidence, however, that in the inflationary setting and in perturbation theory correlators can be promoted to well-defined observables because eventually they are measured by a single experimental devise like Planck satellite. For perturbative inflationary correlators the idea of OPE was explored in \cite{Assassi:2012et,Kehagias:2012pd}. It is still an open question whether late-time correlators continue to be well-defined observables in a pure dS.

When gravity is treated at the non-perturbative level a whole new range of phenomena appear. First of all, the dS entropy becomes finite and is given by the horizon area in Planck units\cite{PhysRevD.15.2738}. What are the implications of this fact remains unclear at the moment, but it is expected that the perturbative semi classical description of correlation functions should break down at very late times, see for example \cite{Goheer:2002vf}. Motivated by the recent understanding of non-perturbative gravitational effects in the physics of black holes \cite{Almheiri:2020cfm}, one may conjecture that non-trivial saddle points become important also in the cosmological setting \cite{Susskind:2021omt,Chen:2020tes}. It would be interesting to see what are the effects of such contributions on the late-time correlation functions if they continue to be well-defined objects.

Another benefit of understanding dS QFTs coupled to gravity is due to the dS/dS correspondence \cite{Alishahiha:2004md,Alishahiha:2005dj}. In this approach a patch of dS is sliced by lower-dimensional dS geometries, which provides a holographic description by means of holographic CFTs on dS coupled to gravity, similar to the Randall-Sundrum models. In fact, the dS/dS correspondence has some qualitative similarity to the relation \eqref{3cfts}. To represent the dS/dS patch one needs two deformed CFTs coupled to gravity and to each other through some double-trace operators. It was also conjectured that in the particular case of dS$_3$/dS$_2$, the deformation may be of the $T\bar T$ type \cite{Gorbenko:2018oov}. 

\paragraph{Phenomenological applications:}
The next few years will bring us a great amount of new cosmological data from both Cosmic Microwave Background \cite{Ade_2019,CMB-S4:2020lpa} and Large Scale Structure \cite{Schlegel:2019eqc,2018AAS...23135423D} experiments. One of the goals of these experiments is the search for non-gaussian features in the primordial inflationary fluctuations. Translated in the language of our paper, this means distinguishing between an interacting and free theory that produces late time correlators. At the moment experimental data is consistent with a free theory \cite{Planck:2019kim}, even though at some level interactions must be present. In this paper we focused mostly on more formal aspects of QFT in cosmological background; however, it is interesting to think whether the progress we made here can have any applications in the analysis of experimental data. One natural place to start is to look for narrow resonances discussed in section~\ref{sec:poscheck}. As explained in \cite{Arkani-Hamed:2015bza}, one can extract an inflationary three-point function from a four-point function in QFT in dS by replacing one of the external legs with the inflaton background. In this paper the imprints of the resonance on the inflationary observables are studied in momentum space where it manifests itself as certain oscillations in the squeezed limit. Our spectral representation converts these oscillations into a resonant peak so that when the spectral parameter is close to the spectral parameter of the resonance the deviation from free theory is order one, no matter how weak the coupling is. In this sense our parametrization of the correlator by the spectral function is similar to the momentum representation of the flat space scattering amplitude, the spectral parameter $\nu$ being analogous to the center of mass energy. It is thus plausible that search for the resonances can be more efficient or reliable in the spectral representation. In practice, this would imply applying an analog of the CPW decomposition to the cosmological data, as opposed to the usual expansion in spherical harmonics. It remains to be seen how practical this procedure is, and, as we mentioned above, additional work needs to be done to incorporate the symmetries of realistic inflationary models. Even away from the resonance, the spectral density has to satisfy certain constraints that we discussed in details above. This can be useful in discriminating the primordial signal from various sources of background, whose spectral density isn't constrained in the same way. Reference \cite{Arkani-Hamed:2015bza} also studies properties of massive resonances with spin, adding such particles is another natural extension of our formalism. 

A different set of questions is whether analyticity and positivity properties of cosmological correlators that we uncovered in this paper can be useful for constraining inflationary models by applying dispersive arguments along the lines suggested in \cite{Baumann:2019ghk}. 

\paragraph{A microscopic theory?:}
One of the motivations for our project was to get any hints on properties of a possible UV-complete microscopic theory that would describe dS space, or a more general cosmology. Several strategies has been employed in attempts to produce such a theory differing by the type of observables it is supposed to include: dS/CFT \cite{Maldacena:2002vr,Anninos:2011ui}, dS/dS \cite{Alishahiha:2004md,Alishahiha:2005dj}, and static (causal) patch holography \cite{Anninos:2011af,Banks:2001px}. We do not have any additional suggestion that could help to construct such a microscopic theory; however, we established a set of constraints on cosmological correlators which have to be satisfied by any viable microscopic theory. For example, it should produce correlation functions with a positive spectral densities and the operator spectrum with certain properties, as explained in the main text. Of course for now we can only claim these properties only in the limit when we managed to prove them. In section~\ref{sec:alternate} we also conjectured a way in which the residual boundary gravity can be dealt with, which is relevant for some holographic ideas in cosmology.

It is interesting to fancy a thought whether one can hope to construct a theory, presumably a CFT coupled to some simple Euclidean gravity, that would directly describe late-time cosmological correlation functions. There are several curious properties that such a theory must have. First, if the bulk description is weakly coupled, the theory will contain operators with dimensions related by the shadow transform. Curiously, such operators exist in some well-studied critical theories like the long range Ising model \cite{Behan:2017dwr}. Moreover, the theory should have both a continuum spectrum, as well as a meaningful notion of a discrete set of operators, bearing some similarity to the Liouville-type theories, as briefly explained in section~\ref{sec:FAQ}.

\paragraph{Conclusion:}
To conclude, the program of studying cosmological observables at the fundamental level is at its very beginning. Our paper, together with several other recent publications, establishes some properties of late-time correlation functions in dS, which seems to be a necessary step in this program. What is truly exciting is the number of open problems, many of which seem accessible with the theoretical tools we have in our disposal.



%

\vspace{1cm}
\subsubsection*{Acknowledgments}
We thank Paolo Creminelli, Matthijs Hogervorst, Petr Kravchuk, Marco Meineri, Mehrdad Mirbabayi, Joao Penedones, Slava Rychkov, Eva Silverstein, Marko Simonovic, Sasha Zhiboedov and Kamran Vaziri for useful discussions. We are also grateful to Matthijs Hogervorst, Joao Penedones and Kamran Vaziri for sharing their work with us prior to publication. LD is partially supported by INFN Iniziativa Specifica ST\&FI. LD also acknowledges support by the program ``Rita Levi Montalcini'' for young researchers. VG is supported by the Simons Foundation Origins of the Universe program (Modern Inflationary Cosmology collaboration).

\appendix

\section{Split representation in dS}\label{app:Gelfand}
In this Appendix we briefly discuss the possibility of implementing an analog of a split representation directly in dS, without the continuation to EAdS. If possible, it would lead to a more direct way to factorization of Feynman diagrams for correlators. Let us first have another look at the split representation for the EAdS Greens function:
\be
\label{splitAdS}
G^{AdS}_\nu=\int_{-\infty}^{\infty} \frac{d\nu'}{\pi}\frac{\nu'^2}{\nu'^2-\nu^2} \int dP {N^{AdS}_{\nu'} N^{AdS}_{-\nu'}} K^{AdS}_{\nu'}(X_1,P)K^{AdS}_{-\nu'}(X_2,P)
\ee
This formula can be understood as a decomposition of the Greens function into plane waves, or Fourier harmonics, each carrying a fixed Casimir of the conformal group. That is, it is the closest analog of the Fourier decomposition of the flat space propagator:
\be
G^{R^{d+1}}_m=\int \frac{d^{d+1} p}{\l(2\pi\r)^{d+1}}\frac{e^{i p x} \cdot e^{-i p y}}{p^2+m^2}\,.
\ee
This is why \eqref{splitAdS} simplifies perturbative calculations -- for the same reason as momentum space calculations in flat space are much simpler. Bulk-to-boundary propagators play the role of plane waves labeled by $\nu$ and a $d$-dimensional vector $P$, which can be identified with a point on the boundary. One could wonder if an analog of \eqref{splitAdS} exists directly in dS, after all the symmetry group is identical so one may expect a similar formula to exist. This question is discussed in details in \cite{gelfand} for the case of dS$_3$ (with the expanding and contracting Poincar\'e patches identified, which projects away half of the harmonics). It is explained that Fourier decomposition of a generic function into Fourier series includes additional representations as compared to the EAdS case. 

Note that the relations between dS and AdS propagators discussed in section \ref{subsec:prop} do not allow to obtain a split representation for dS Green's functions from  \eqref{splitAdS}.  The reason is that after analytic continuation to dS the integral over $\nu$ in \eqref{splitAdS} would diverge for some values of the coordinates, as can be seen from \eqref{largenu}. This issue does not arise for the left-right propagator, since it is simply related to the AdS harmonic function which spectral representation does not involve an integral over $\nu$.  

To exhibit the full set of Fourier harmonics needed in dS we can study the decomposition of the delta-function, which is a problem closely related to decomposing a Green's function. Using embedding coordinates, we take
a bulk point $B$ and  a boundary point $P$ such that $B\cdot P=0$. Then the line $B+t P$, $t$ real, also belongs to dS space. Let us denote $\theta(X,B)=\arccos X\cdot B$, which is real, as long as $X\cdot P=0$. Then the delta function admits the following decomposition
\beg
\label{splitdS}
\delta(X,Y)= \frac{1}{16 \pi^3}\int d\nu\, \nu^2 \int dP \big |X\cdot P\big |^{-1-i\nu}  \big |Y \cdot P\big |^{-1+i\nu}+\\
\qquad\qquad\qquad\qquad+\frac{4}{\pi^2}\sum_n n \int dP e^{-2 i n \theta(X,B)}\delta\l(X\cdot P\r) \, e^{2 i n \theta(Y,B)}\delta\l(Y\cdot P\r)\,,
\eeg
where $B$ implicitly depends on $P$ because $B\cdot P=0$, but the expression is independent of the specific choice of $B$.

We refer the reader to \cite{gelfand} for the derivation and a discussion of the geometrical and group theoretic meaning of \eqref{splitdS}. Here we just make a few comments about this equation. The first, ``continuum'' term in \eqref{splitdS} is very similar to its AdS counter-part. The second, ``discrete'' term is new. To derive the split representation for the propagators one basically needs to divide the contribution of each harmonic by a corresponding Casimir, so the discrete contributions will remain in the expression. Nevertheless, in our final expressions for the correlation functions we see no trace of such discreet terms, as is obvious from the relation to EAdS space we used. It would be interesting to see a more direct physical reason for why it happens.

\section{Global coordinates}
\label{App:global}
\begin{figure}[t] 
	\centering
	   {\includegraphics[width=0.8\linewidth,angle=0]{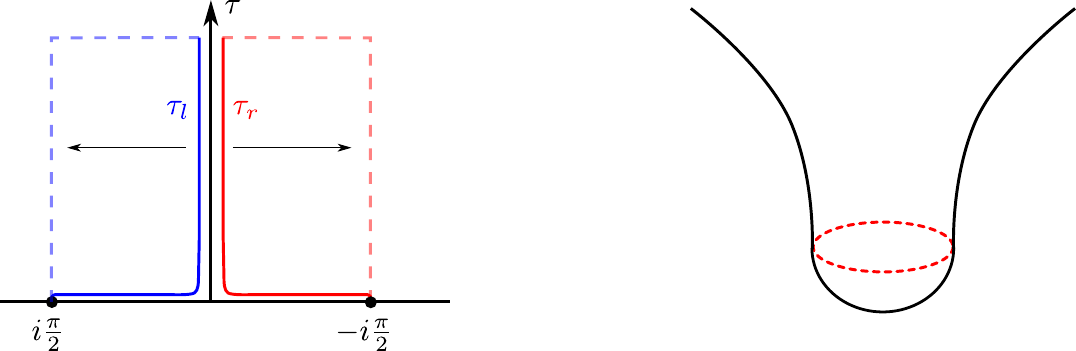}}
	\caption{\small \textbf{Left:} solid lines, the HH contours for $\tau^l$ and $\tau^r$, real part of $\tau$ grows upwards; dashed lines, contours after analytic continuation to EAdS. The horizontal part of dashed lines does not contribute at late times. \textbf{Right:} geometrical picture corresponding to the HH contour.}
	\label{fig:globalcont}
\end{figure}

In the main text, for the sake of simplicity, we mostly used Poincar\'e, or FRW, coordinates that cover an expanding Poincar\'e patch of dS space. However, our arguments do not rely on the choice of a coordinate system and any other parametrization of dS space can be used. In this appendix we show that all the steps can be repeated in global coordinates. These coordinates are also the most convenient to perform the analytic continuation to the sphere, discussed in Appendix \ref{app:SR}. In global coordinates the dS metric was given in \eqref{eq:metricGL}, which we copy here:
\be
ds^2=-dt^2+\cosh^2 \! \tau \,ds_{\Omega_{d}}^2\,.
\ee
The equal time slices are spheres; however, if we calculate correlation functions at $\tau\to\infty$ conformal invariance relates them to correlation functions on the plane at $\eta\to0$. In global coordinates the two-point invariant $s$, \eqref{sdef}, is given by 
\be
s(X^1,X^2)=-\sinh \tau_1 \sinh \tau_2+\cosh \tau_1 \cosh \tau_2 \,\,\vec y_1 \cdot \vec y_2\,,
\ee
where $\vec y_1$, $\vec y_2$ belong to the unit sphere. The $i\eps$ prescription in \eqref{dSprops} can now be implemented by choosing the Hartle-Hawking contour for integration over the global time $\tau$, see figure~\ref{fig:globalcont}. Infinitesimally, it corresponds to shifting $\tau^l\to\tau^l(1+i\eps)$ and $\tau^r\to\tau^r(1-i\eps)$, which, expanding to linear order in $\eps$, results into the necessary shift in $s$, as long as we are on the expanding part of the contour, $\tau>0$.\footnote{Local physics in the contracting part of the global dS space is not of interest for us here, but since expanding Poincar\'e patch covers a part of it, see figure~\ref{fig:Penrose}, it is clear that an analytic continuation for correlation functions to the entire global dS also exists.}


We can also proceed with our procedure of analytically continuing to EAdS in Global coordinates by deforming the contours for left and right time integrations as on figure~\ref{fig:globalcont}:
\be
\label{EAdSHH}
\tau^l\to\tau^l+i\frac{\pi}{2}\,,\qquad \tau^r\to\tau^r-i\frac{\pi}{2}\,.
\ee
This has exactly the same result on $s$ as the continuation \eqref{Wick}, thus, giving exactly the same EAdS theory\footnote{This continuation was used in \cite{Hertog:2011ky}. Formally, we produce the negative EAdS space. However, in our way description the negative sign in front of the metric is absorbed in various coupling constants in the EAdS Lagrangian.}. Thus the Poincar\'e patch and the no-boundary global geometry are two different analytic continuations of the EAdS space.

\section{Spectral representation: from the sphere to dS and to AdS}\label{app:SR}
In this appendix, we discuss the bulk two-point functions in dS.  We first review how its spectral representation can be obtained by the analytic continuation from the sphere. We then explain how to relate it to a spectral representation in EAdS by an analytic continuation. This second step provides another relation between dS and EAdS, similar but different from the one discussed in section \ref{sec:EAdS}, and is valid non-perturbatively. Finally, by rewriting the spectral representation in EAdS to a discrete sum and analytically continue back to dS, we obtain a representation of the two-point function in dS as a discrete sum, which is crucial for analyzing the two-point functions at large time-like separation discussed in subsection \ref{subsec:QNM}.
\paragraph{Step 1: sphere to dS.}
Consider the bulk two-point function of a scalar operator $O$ in dS
\begin{equation}
\langle O(X^\alpha) O(Y^\beta)\rangle = F^{\alpha\beta}(\zeta)~.
\end{equation}
This is a function of the chordal distance $\zeta$ between the two points which is defined in  \eqref{eq:chordal}. In this appendix we find it more convenient to use it instead of the two-point invariant $s$. Different choices for $\alpha,\beta = l,r$ correspond to different ways of regulating the singularity on the lightcone $\zeta = 0$ and implement different ordering prescriptions when $\zeta < 0$. 

As is the case with the correlation function in the Minkowski space, these correlation functions can be obtained from a unique two-point function in the Euclidean counterpart, namely $S^{d+1}$. 
The chordal distance on the sphere $\zeta$ is in the range $0<\zeta\leq 4$, and the only singularity is at $\zeta = 0$, i.e. at coincident points. We expand the sphere two-point function in a basis of spherical harmonics
\begin{equation}\label{eq:Sphere}
F^{S}(\zeta^S) =\sum_{l = 0}^\infty f^{S}(l) \, \Omega^S_l(\zeta)~,
\end{equation}
where
\begin{equation}
\Omega^S_l(\zeta) = \frac{(-1)^l(d+2l)\Gamma (d+l)}{(4 \pi )^{\frac{d+1}{2}} \Gamma(\frac{d+1}{2})\Gamma (l+1) }    \, _2 F_1\left(-l,d+l;\tfrac{d+1}{2};1-\tfrac{\zeta }{4}\right)\label{eq:Sharm}~.
\end{equation}
We normalized the harmonic function so that 
\begin{equation}
\sum_{l=0}^\infty \Omega^S_l(X,Y) = \delta^{d+1}(X,Y)~,
\end{equation}
where $\Omega^S_l(X,Y) = \Omega^S_l(\zeta(X,Y))$ and $\delta^{d+1}(X,Y)$ is the Dirac delta on the sphere with respect to the measure induced by the round metric with radius $L=1$. This is also the normalization in which
\begin{equation}
\Omega^S_l \star \Omega^S_m = \delta_{lm} \Omega^S_l~,
\end{equation}
where $\star$ denotes the convolution, again with the measure induced by the round metric.

The series expansion as written in eq. \eqref{eq:Sphere} is not very useful if continued term by term back to dS. This is because each term grows like $\zeta^l$ and the series is divergent if we go to a large space-like separation in dS, $\zeta\gg 1$. In order to obtain a more useful representation in such a regime, we apply the so-called Watson-Sommerfeld trick following \cite{Marolf:2010zp,Marolf:2010nz}. This amounts to replacing the sum with an integral in the complex plane of the variable $l$
\begin{equation}
\sum_{l = 0}^\infty (-1)^l f^{S}(l) \, \left[(-1)^l \Omega^S_l\right] = \int_\gamma \frac{dl}{2\pi i} \frac{\pi}{\sin(\pi l)} f^{S}(l)\, \left[(-1)^l \Omega^S_l\right]~,
\end{equation}
where the contour $\gamma$ surrounds all the poles of $\frac{\pi}{\sin(\pi l)}$ at non-negative integer values of $l$. Recalling the definition \eqref{eq:Sharm} we see that there is no issue in interpreting the quantity in square-brackets as an analytic function of the complex variable $l$. In fact we have that in their overlapping region of definition $0< \zeta\leq 4$
\begin{equation}
\left. \frac{\pi}{\sin(\pi l)}  \left[(-1)^l \Omega^S_l\right] \right\vert_{l=-\frac{d}{2}+i\nu} = - 2 i \nu \, G^{\alpha\beta}_\nu~,
\end{equation}
for any $\alpha,\beta = l,r$, since these choices are all equivalent in this range of $\zeta$.
We then deform the contour to run along the imaginary axis with ${\rm Re}(l) = -\frac{d}{2}$, because uniformly on this contour the function in square brackets has a power law decay at large distances. From the representation theory point of view, this line coincides precisely with the principal series representation of the dS isometry, which we reviewed in section \ref{sec:wavefunction}. 

\begin{figure}[t] 
	\centering
		\hspace{0cm}\includegraphics[width=0.5\linewidth,angle=0]{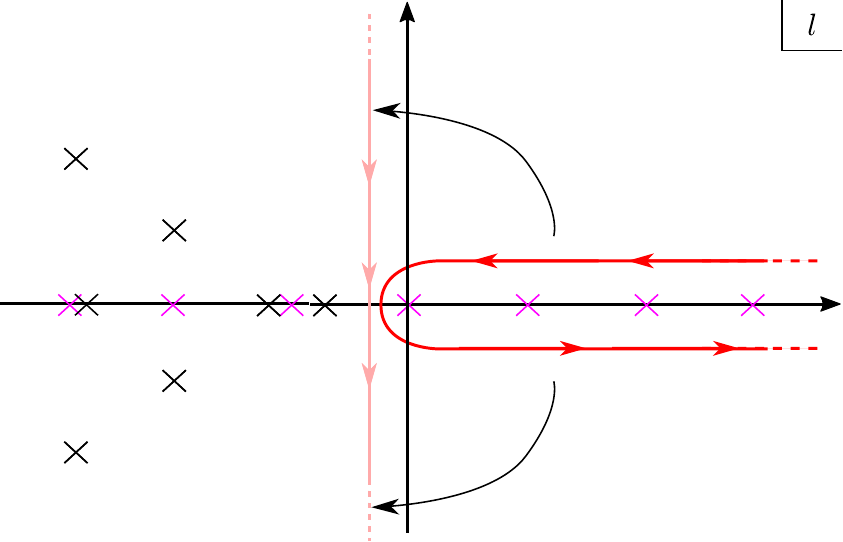}~~~~\raisebox{0.25cm}{\includegraphics[width=0.5\linewidth,angle=0]{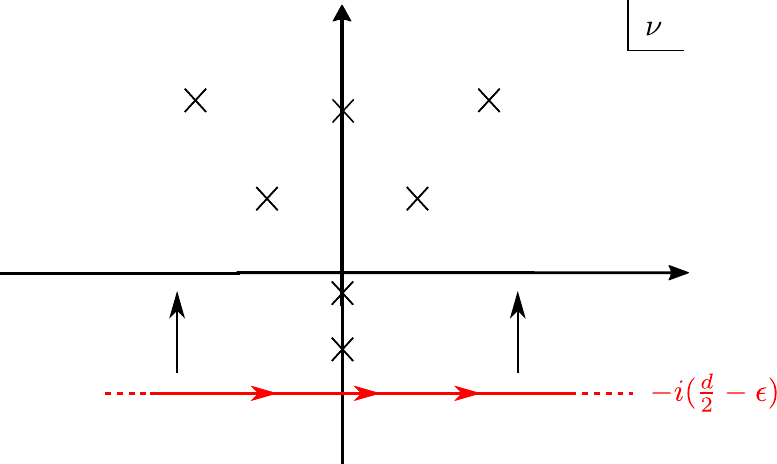}}
	\caption{\small On the left: using the Watson-Sommerfeld trick we rewrite the spectral representation of the two-point function on the sphere as a contour integral. We show the contour deformation in the complex $l$ plane and the singularities of $\frac{\pi}{\sin(\pi l)} f^{S}(l)\, \left[(-1)^l \Omega^S_l\right]$, which include poles at all the integers coming from the $\sin$ prefactor, denoted in purple. We assume no further singularity of $f^{S}(l)$ in the region where we perform the deformation. On the right: after changing variable to $\nu = -i(l + \frac{d}{2})$, we sketch of the analytic structure of the function $f(\nu)$ appearing in the integral representation \eqref{eq:dSintrep} of a two-point function in dS. The only singularities in the lower-half plane are in the interval $i\nu \in (0,\frac{d}{2})$. After the deformation showed on the left, the contour runs on the line ${\rm Im}\nu=-(\frac{d}{2}-\epsilon)$ for a small positive $\epsilon$. We can then deform it on the real $\nu$ axis, therefore obtaining a representation as an integral on the principal series, plus discrete contributions from the singularities in $i\nu \in (0,\frac{d}{2})$. Poles on the real $\nu$ axis, i.e. on the principal series, can be encountered in perturbation theory but are expected to get a positive imaginary part due to interactions and move to the upper half-plane. See the discussion in section \ref{sec:FAQ} and the example of this phenomenon in section \ref{sec:poscheck}.}
	\label{fig:analyticityf}
\end{figure}

When deforming the contour, we assume that the function $f^{S}(l)$ is holomorphic in the region where we are performing the contour deformation, except possibly for poles for $l$ in the interval $(-\frac{d}{2},0)$ of the real axis, which we discuss shortly. Renaming the integration variable in the deformed contour as $\nu = -i(l + \frac{d}{2})$ we obtain
\begin{align}\label{eq:dSintrep}
\begin{split}
F^{\alpha\beta}(\zeta) 
&= \int_{-\infty+ i(-\ddt+\epsilon)}^{+\infty+ i(-\ddt+\epsilon)} d\nu \, \frac{\nu}{\pi i}f(\nu)\, G^{\alpha\beta}_\nu(\zeta) ~,\\
&= \int_{-\infty}^{+\infty} d\nu \, \frac{\nu}{\pi i}f(\nu)\, G^{\alpha\beta}_\nu(\zeta) + \left(\text{poles for $i\nu \in (0,\tfrac{d}{2})$}\right)~,
\end{split}
\end{align}
where the two-point spectral amplitude in dS, $f(\nu)$, is defined by
\be\label{eq:Amp2ptdef}
 f(\nu) \equiv  - f^{S}(-\tfrac{d}{2}+i\nu)~.
\ee
This is the same object as the one we called $f^{\rm 2pt}(\nu)$ in section \ref{sec:pert}. Throughout this appendix, we drop the superscript since we only discuss the two-point functions.
The small parameter $\epsilon$ in the first line is such that the contour runs above the pole of $G^{\alpha\beta}_\nu$ at $\nu= - i\ddt$, and below any singularity of $f(\nu)$   in the interval $i\nu \in (0,\tfrac{d}{2})$. In the second line we deformed the contour to the real $\nu$ axis, picking the poles for $i\nu \in (0,\tfrac{d}{2})$. The motivation for the assumed analyticity property of $f^{S}(l)$ is that the poles for $i\nu \in (0,\tfrac{d}{2})$ give rise to contributions proportional to propagators of light scalar fields in dS, while other singularities would give propagators of scalar fields with unphysical values of the mass-squared. Stated in terms of $f(\nu)$, this is equivalent to saying that $f(\nu)$ is holomorphic everywhere in the lower-half $\nu$ plane except for possible light-particle singularities for $i\nu \in (0,\tfrac{d}{2})$. A sketch of the resulting analytic structure is shown in figure \ref{fig:analyticityf}.  It was shown in \cite{Higuchi:2010xt} that the analytic continuation from the sphere to dS is possible to all orders in perturbation theory. 

This integral representation was obtained in \cite{Marolf:2010zp,Marolf:2010nz} (see also \cite{Bros:1994dn, Bros:1995js, Bros:2003uw, Nguyen:2017ggc}). Thanks to the improved convergence property at large $|\zeta|$, the expression makes sense in the full range $\zeta \in \mathbb{R}$. The choice $\alpha,\beta$ for how to regulate the light-cone singularity on the left-hand side is fixed by the choice of propagator in the integrand on the right-hand side. However, this integral is still inappropriate for analyzing the large time-like separation, namely $\zeta\ll 0$. To see this, let us consider the convergence of the $\nu$ integral. Note that the dS propagator has the following large-$\nu$ asymptotics (recall that $\zeta = 2(1-s)$ is the chordal interval)
\begin{align}
\label{largenu}
\!\!\!\!\!\!\!\!\!\!G^{\alpha\beta}_\nu(\zeta) \!\!\underset{|\nu|\to\infty}{\sim}
\!\!\begin{cases}
\pi^{-\frac{d}{2}}|\nu|^{\frac{d}{2}-1}e^{-\pi |\nu|}\cos\left(\tfrac d 4 - |\nu| {\rm arccosh}(\frac{\zeta}{2}-1)\right)(\zeta(\zeta-4))^{-\frac{d}{4}}~,~{\text{for}}~\zeta>4~,\\
\frac12 \pi^{-\frac{d}{2}} |\nu|^{\frac{d}{2}-1}e^{-\pi |\nu|}e^{|\nu| {\rm arccos}(\frac{\zeta}{2}-1)}(\zeta(4-\zeta))^{-\frac{d}{4}}~,~~~~~~~~~~~~~~{\text{for}}~0<\zeta<4~,\\
\frac12 \pi^{-\frac{d}{2}} |\nu|^{\frac{d}{2}-1}e^{-i s^{\alpha \beta}\pi |\nu|{\rm arccosh(1-\frac{\zeta}{2})}}e^{-i s^{\alpha \beta}\pi\frac{d}{4}}(-\zeta(4-\zeta))^{-\frac{d}{4}}~,~~~~{\text{for}}~\zeta<0~.
\end{cases}
\end{align}
Here $s^{\alpha\beta} = \pm 1$ is the sign of the $i \epsilon$ prescription that is being used for $\zeta<0$, namely $\zeta\to\zeta + i s^{\alpha\beta} \epsilon$ with $\epsilon>0$. We see that the propagator decays exponentially like $e^{-\pi|\nu|}$ for $\zeta > 4$, for $0<\zeta<4$ it interpolates between an exponential decay and a power-law, and it behaves like a power-law with an oscillating prefactor for time-like interval $\zeta < 0$. Therefore if $f(\nu)$ only decays as a power-law for large $\nu$, as in the example that we will discuss in the next subsection, the convergence for a time-like separation $\zeta < 0$ is only in a distributional sense. 

\paragraph{Step 2: dS to EAdS.} Because of this distributional convergence, the integral representation \eqref{eq:dSintrep} is still not satisfactory for analyzing the two-point function at a large time-like separation which encodes the quasi-normal frequencies in the static patch. 
To overcome this problem, we first analytically continue  \eqref{eq:dSintrep} to EAdS. This is a generalization of what we discussed in section \ref{subsec:dStoAdS} for propagators and provides its non-perturbative generalization.  The result of the analytic continuation would look like
\begin{equation}\label{eq:fabcont}
F^{\alpha\beta}(\zeta) \,\,\to\,\, \int_{-\infty}^{+\infty} d\nu \, \rho^{\alpha\beta}(\nu)\,\Omega^{\rm AdS}_\nu(\zeta^{\rm AdS})~,
\end{equation}
where $\zeta^{\rm AdS}$ is defined as $\zeta^{\rm AdS} = -2(1+s^{\rm AdS})$, with $s^{\rm AdS}$ given in \eqref{sAdSdef}.
The spectral density $\rho^{\alpha\beta}(\nu)$ in this case depends on the choice of $\alpha,\beta=l,r$. 
In what follows, we derive the relation between $\rho^{\alpha\beta}$ and $f$, under the simplifying assumption that there are no contributions from light particles, and as a result $f$ is holomorphic in the lower-half $\nu$ plane. 

We first consider $\alpha=l$ and $\beta=r$. As can be seen in \eqref{splitW}, $G^{lr}_{\nu}$ is proportional to the harmonic function $\Omega^{\rm AdS}_\nu$ as a function of $\zeta$.  Using the symmetry of $\Omega^{\rm AdS}_{\nu}$ ($\Omega^{\rm AdS}_{\nu}=\Omega^{\rm AdS}_{-\nu}$), we then obtain 
\begin{equation}\label{eq:lrtodS}
\rho^{lr}(\nu) = - \frac{i\nu\Gamma(\pm i\nu)}{2\pi}(f(\nu)-f(-\nu))~.
\end{equation}
Note that $\rho^{rl}(\nu) = \rho^{lr}(\nu)$. If $O$ is a real operator $F^{lr}(\zeta)$ is a real function for space-like separation $\zeta > 0$. Therefore we see from the definition \eqref{eq:dSintrep} that $f(-\nu)=f(\nu)^*$ and $\rho^{lr}(\nu)$ is proportional to the imaginary part of $f(\nu)$:
\be
\rho^{lr}(\nu)= \frac{\nu\Gamma(\pm i\nu)}{\pi} {\rm Im}\, f(\nu)\period
\ee

Next, we consider the case $\alpha=\beta=l$. The analytic continuation to Euclidean AdS of the propagator $G^{ll}_{\nu}$ is given in \eqref{splitGreen}, so we obtain
\begin{align}\label{eq:llB}
\begin{split}
&F^{ll}(\zeta)  \\ &\to \int_{-\infty}^{+\infty} d\nu \frac{\nu}{\pi i}f(\nu) \, \frac{i \nu}{2 \pi} {\G(\pm i\nu)}\l(G^{\rm AdS}_{\nu}(\zeta^{\rm AdS})e^{ i \pi (\ddt+i\nu)}-G_{-\nu}^{\rm AdS}(\zeta^{\rm AdS})e^{ i \pi (\ddt-i\nu)}\r)~.
\end{split}
\end{align}
Looking at the expression \eqref{GOmegarel} for $G^{\rm AdS}$ in terms of the harmonic function, we see that there is a singularity when $\nu$ is real, which we regulate by giving it a small negative imaginary part. With this prescription we can substitute
\begin{equation}
G_{-\nu}^{\rm AdS} = G^{\rm AdS}_{\nu} + \frac{2\pi i}{\nu} \Omega^{\rm AdS}_{\nu}~,
\end{equation}
obtaining (here we omit the argument $\zeta^{\rm AdS}$)
\begin{equation}
\int_{-\infty}^{+\infty}d\nu\frac{\nu}{\pi i} f(\nu) \,  {\G(\pm i\nu)}\l(\frac{i \nu}{2 \pi}G^{\rm AdS}_{\nu}\l(e^{ i \pi (\ddt+i\nu)}-e^{ i \pi (\ddt-i\nu)}\r) + e^{ i \pi (\ddt-i\nu)} \Omega^{\rm AdS}_{\nu}\r)~.
\end{equation}
Both $G^{\rm AdS}_{\nu}$ and $f(\nu)$ are holomorphic in the lower-half $\nu$ plane, and the poles of $\Gamma(-i\nu)$ are canceled by the zeroes of $e^{ i \pi (\ddt+i\nu)}-e^{ i \pi (\ddt-i\nu)}$. Therefore in the term proportional to $G^{\rm AdS}_{\nu}$, we can close the contour in the lower-half $\nu$ plane and get a vanishing contribution. Hence as expected we are left with an integral representation only in terms of $\Omega^{\rm AdS}_{\nu}$, and upon symmetrizing the coefficient we find
\begin{equation}\label{eq:lltodS}
\rho^{ll}(\nu) = -\frac{i\nu\G(\pm i\nu) e^{i \pi\frac{d}{2}}}{2\pi}\left(  e^{\pi \nu} f(\nu) - e^{ -\pi\nu} f(-\nu)\right)~.
\end{equation}
Similarly one can derive
\begin{equation}\label{eq:rrtodS}
\rho^{rr}(\nu) = -\frac{i\nu\G(\pm i\nu) e^{-i \pi\frac{d}{2}}}{2\pi}\left(  e^{-\pi \nu} f(\nu) - e^{\pi\nu} f(-\nu)\right)~.
\end{equation}
Solving \eqref{eq:lrtodS}-\eqref{eq:lltodS} for $f(\nu)$, we obtain
\begin{equation}\label{eq:SRdSfromAdS}
f(\nu) = \frac{2\pi i}{\nu\G(\pm i \nu)(e^{2\pi\nu}-1)}\l(e^{-i\pi(\frac{d}{2}+i\nu)}\rho^{ll}(\nu)- \rho^{lr}(\nu)\r) = i (e^{-i\pi\frac{d}{2}}\rho^{ll}(\nu)- e^{-\pi \nu}\rho^{lr}(\nu))~.
\end{equation}
\paragraph{Step 3: discrete sum and back to dS.}As the next step, we rewrite the spectral representation in EAdS \eqref{eq:fabcont} into a discrete sum. This can be achieved by first decomposing $\Omega_{\nu}^{\rm AdS}$ into a sum of $G_{\nu}^{\rm AdS}$ and $G_{-\nu}^{\rm AdS}$ using \eqref{GOmegarel}, and performing the change of the variable $\nu\to -\nu$ to the first term. After this, we can close the contour onto the upper half plane, and replace the integral with a sum of contributions from poles. For instance, this gives the following expression for the analytically continued two-point function:
\be\label{eq:2ptdiscrete}
F^{\alpha\beta}(\zeta)\,\to\,\sum_{\substack{\nu_{\ast}:\text{ poles}\\{\rm Re}[i\nu^{\ast}]<0}}{\rm Res}\left[\nu\rho^{\alpha\beta}(\nu)\right]G_{-\nu^{\ast}}(\zeta^{\rm AdS})\period
\ee
Note that $G_{-\nu}(\zeta^{\rm AdS})$ behaves at $|\zeta^{\rm AdS}|\gg 1$ as
\be
G_{-\nu}(\zeta^{\rm AdS})\sim\frac{1}{(\zeta^{\rm AdS})^{\frac{d}{2}-i\nu}}\comma
\ee
and therefore the expansion is well-convergent in the long (AdS) distance for ${\rm Re}[i\nu]<0$.

The final step is to continue back to dS. This is basically undoing the analytic continuation we performed in section \ref{subsec:dStoAdS}. Written in terms of the chordal distance, it corresponds to the following replacement\footnote{Recall that the first one gives the anti-time-ordered two-point function while the second one is time-ordered.}:
\be
\zeta^{\rm AdS}\mapsto \begin{cases}e^{\pi i}\zeta\qquad &(\alpha,\beta)=(l,l)\\
e^{-\pi i}\zeta\qquad &(\alpha,\beta)=(r,r)\end{cases}\period
\ee
The discrete sum \eqref{eq:2ptdiscrete} remains well-convergent even after this replacement and provides an expansion of the two-point function at large time-like separation in the inverse powers of $\zeta$. Translating this to the static patch coordinates, one finds that each power corresponds to a distinct quasi-normal frequency. This establishes the relation between poles of $\rho(\nu)$ (and $f(\nu)$) and the quasi-normal modes in the static patch, which we discussed in section \ref{subsec:QNM}.
\section{Bubble function in dS}\label{app:Bubble}

In this appendix we obtain the integral representation defined in appendix \ref{app:SR} for the bubble function, defined as the product of two propagators. Since we can choose either the Wightman or the time-ordered propagators, there are different types of bubbles ($\alpha,\beta = l,r$)
\begin{equation}
B^{\alpha\beta}_\nu(X,Y) = G^{\alpha\beta}_\nu(X,Y)^2~.
\end{equation} 
For our purposes we can restrict to the case of two propagators of equal mass-squared. This function computes the bulk two-point function of the operator $\phi^2$ in a free scalar theory, namely $\langle\phi^2(X^\alpha)\phi^2(Y^\beta)\rangle = 2 B^{\alpha\beta}_\nu(X,Y)$. 

The integral representation \eqref{eq:dSintrep} is 
\begin{equation}\label{eq:dSintrepB}
B^{\alpha\beta}_\nu = \int_{-\infty}^{+\infty} d\nu' \,\frac{\nu'}{\pi i}\hat{B}_\nu(\nu') \, G^{\alpha\beta}_{\nu'}~.
\end{equation}
We are renaming the integration variable $\nu'$ because we are using $\nu$ to label the mass-squared of the propagators entering the bubble. We are also assuming that there are no ``light-particles'' poles, which we expect to be true for instance when the two propagators entering the bubbles are themselves propagators of ``heavy particles'', i.e. for $\nu\in\mathbb{R}$.  In this appendix we will obtain it using \eqref{eq:SRdSfromAdS} and computing the AdS spectral representations.

To compute $\hat{B}^{lr}_\nu(\nu')$ we can proceed in two ways: in the first approach we note that $B^{lr}_\nu(s)$ analytically continued to Euclidean AdS is proportional to the product of two AdS harmonic functions. We then use the following multiplication identity for the AdS harmonic function (see appendix D of \cite{Penedones:2010ue})
\begin{equation}
(\Omega^{\rm AdS}_\nu)^2 =\frac{\nu^2 \sinh ^2(\pi  \nu)}{32 \pi ^{\frac{d}{2}+4}\,\Gamma\left(\frac{d}{2}\right)} \int_{-\infty}^{+\infty} d\nu'\frac{  \Gamma\left(\frac{d}{4} \pm i \frac{\nu'}{2}\right)^2  \Gamma\left(\frac{d}{4} \pm i \frac{\nu' -2 \nu}{2}\right)  \Gamma\left(\frac{d}{4} \pm i \frac{\nu' +2 \nu}{2}\right) }{ \Gamma\left(\frac{d}{2}\pm i \nu'\right)}\Omega^{\rm AdS}_{\nu'}~.
\end{equation}
In this way we readily obtain
\begin{equation}\label{eq:BlrfromOmega}
\hat{B}^{lr}_\nu(\nu')  = \frac{  \Gamma\left(\frac{d}{4} \pm i \frac{\nu'}{2}\right)^2  \Gamma\left(\frac{d}{4} \pm i \frac{\nu' -2 \nu}{2}\right)  \Gamma\left(\frac{d}{4} \pm i \frac{\nu' +2 \nu}{2}\right) }{32 \pi ^{\frac{d}{2}+2}\,\Gamma\left(\frac{d}{2}\right) \Gamma\left(\frac{d}{2}\pm i \nu'\right)}~.
\end{equation}
In the second approach, we use eq. \eqref{splitW} to rewrite the analytic continuation of $G^{lr}_\nu(s)^2$ as the square of a linear combination of AdS propagators. Expanding the square we obtain a sum of terms, each proportional to a product of AdS propagators. We can then use the AdS bubble function
\begin{equation}\label{eq:GAdSsq}
G^{\rm AdS}_{\nu_1}G^{\rm AdS}_{\nu_2}  = \int_{-\infty}^{+\infty} d\nu' \,\hat{B}^{\rm AdS}_{\nu_1,\nu_2}(\nu')\,\Omega^{\rm AdS}_{\nu'}~,
\end{equation}
where\footnote{We thank C. Sleight and M. Taronna for finding a typo in the previous version of this formula \cite{Sleight:2021plv}.}
\begin{align}\label{eq:bubbleAdS}
\begin{split}
&\hat{B}^{\rm AdS}_{\nu_1,\nu_2}(\nu')  \\& \!\!\!\!\!\!\!\!\! =  \frac{\Gamma \left(\tfrac{d}{2}+i \nu_1\right) \Gamma \left(\tfrac{d}{2}+i \nu_2\right) \Gamma(1+i\nu_1+i\nu_2) \Gamma \left(\tfrac{d}{2}+i \nu_1+i\nu_2\right)}{\pi ^{\frac{d-1}{2}} 2^{d+2+i \nu_1+i\nu_2}} \left[\Gamma \left(\tfrac{d}{4} + i \tfrac{\nu_1+\nu_2}{2} +i\tfrac{\nu'}{2}\right) \phantom{\begin{matrix} a&b\\c&d
\end{matrix}} \right. \\  &\!\!\!\!\!\!\!\!\!\!\!\!\!\!\!\!\!\!\!\!\!\!\!\!\!\!\times
{}_7\tilde{F}_6\left(
\begin{matrix} 
\tfrac{d}{2}&\tfrac{1+i\nu_1+i\nu_2}{2}&\tfrac{2+i\nu_1+i\nu_2}{2}&\tfrac{d}{2}+i\nu_1+i\nu_2&\tfrac{d}{2}+i\nu_1&\tfrac{d}{2}+i\nu_2&\tfrac{d}{4} + i \tfrac{\nu_1+\nu_2}{2} +i\tfrac{\nu'}{2} \\
&\tfrac{d+i\nu_1+i\nu_2}{2}&\tfrac{d+1+i\nu_1+i\nu_2}{2} &1+i\nu_1+i\nu_2&1+i\nu_1&1+i\nu_2&\tfrac{d}{4}+1 + i \tfrac{\nu_1+\nu_2}{2} +i\tfrac{\nu'}{2}
\end{matrix}
;1\right) \\&\left. \phantom{\begin{matrix} a&b\\c&d
\end{matrix}} +\nu'\to-\nu'\right]~,
\end{split}
\end{align}
to represent each of these terms as an integral over the principal series. ${}_7\tilde{F}_6$ denotes the regularized hypergeometric function. Eq. \eqref{eq:bubbleAdS} is obtained by summing the expression of a product of AdS propagators as a series of ``double-trace propagators'', derived in the appendix C of \cite{Fitzpatrick:2011hu}. An alternative derivation in the case of equal external dimensions was given in \cite{Carmi:2018qzm}. For $d=2$, i.e. AdS$_3$, this expression simplifies to
\begin{equation}
\left.\hat{B}^{\rm AdS}_{\nu_1,\nu_2}(\nu')\right\vert_{d=2} = \frac{i }{8 \pi  \nu' }\left(\psi\left(\frac{-i \nu' +i \nu_1+i \nu_2+1}{2}\right)-\psi\left(\frac{i \nu' + i \nu_1+ i \nu_2+ 1}{2}\right)\right)~,
\end{equation}
where $\psi$ is the digamma function. Using this approach we obtain 
\begin{equation}\label{eq:BlrAdS}
\hat{B}^{lr}_\nu(\nu') = -\frac{\nu^2 \G(\pm i \nu)^2}{4 \pi^2}\left(\hat{B}^{\rm AdS}_{\nu,\nu}(\nu')+\hat{B}^{\rm AdS}_{-\nu,-\nu}(\nu')-2\hat{B}^{\rm AdS}_{\nu,-\nu}(\nu')\right)~.
\end{equation}
It can be checked that the two expressions \eqref{eq:BlrfromOmega} and \eqref{eq:BlrAdS} agree. It is remarkable that there is a large simplification in the particular linear combination \eqref{eq:BlrAdS} of AdS bubble functions, leading to the compact result \eqref{eq:BlrfromOmega}. In particular we observe that while the sum that defines the AdS bubble function is divergent for $d\geq 3$, reflecting a UV divergence that is expected for bulk dimension $\geq 4$, the $lr$ dS bubble is finite for any $d$. 

To compute $B^{ll}_\nu(\nu')$ we continue $(G^{ll}(s))^2$ using \eqref{splitGreen}, we expand the square and finally apply \eqref{eq:GAdSsq} to the resulting terms. We obtain 
\begin{equation}\label{eq:BllAdS}
\hat{B}^{ll}_\nu(\nu') = -\frac{\nu^2 \G(\pm i \nu)^2 e^{i\pi d}}{4 \pi^2}\left(e^{-2\pi\nu}\hat{B}^{\rm AdS}_{\nu,\nu}(\nu')+e^{2\pi\nu}\hat{B}^{\rm AdS}_{-\nu,-\nu}(\nu')- 2\hat{B}^{\rm AdS}_{\nu,-\nu}(\nu')\right)~.
\end{equation}
We can then substitute the expressions \eqref{eq:BlrAdS}-\eqref{eq:BllAdS} in \eqref{eq:SRdSfromAdS} to obtain
\begin{align}\label{eq:Bfinal}
\begin{split}
 \hat{B}_\nu(\nu') & = i \frac{\nu^2 \G(\pm i \nu)^2 }{4 \pi^2} e^{-\pi \nu'}\\  \times & \l[\l(1-e^{i\pi(\frac d2 -i \nu' +2 i \nu)}\r)\hat{B}^{\rm AdS}_{\nu,\nu}(\nu') +  \l(1-e^{i\pi(\frac d2 -i \nu' -2 i \nu)}\r)\hat{B}^{\rm AdS}_{-\nu,-\nu}(\nu') \right. \\ & \left. ~~~~~~~~~~~~~~~~~~~~~~~~~~~~~~~~~~~~~~~~~~~~~~~~~~-2\l(1-e^{i\pi(\frac d2 -i \nu' )}\r)\hat{B}^{\rm AdS}_{\nu,-\nu}(\nu')\r]~.
\end{split}
\end{align}
Each of the factors in round brackets has zeroes in the lower-half $\nu'$ plane precisely at the location of the poles of the AdS bubble function that they multiply, so that indeed $\hat{B}_\nu(\nu')$ is holomorphic in the lower-half $\nu'$ plane as expected from the arguments in the appendix \ref{app:SR}. The bubble function \eqref{eq:Bfinal} was computed by the analytic continuation from the sphere in \cite{Marolf:2010zp} although it seems our result is slightly different.

\section{Exchange diagram for general external dimensions}\label{app:general}

The calculation of the exchange diagram of section \ref{sec:tree} can be easily generalized to the case in which the four insertions at late times correspond to fields with different masses and correspondingly different values of the parameter $\nu$. In full generality we can take them to be all distinct: $\nu_i$, $i=1,\dots,4$, with the understanding that we always assume the insertions at late time to have a definite scaling dimension $\Delta_i = \frac{d}{2}-i\nu_i$. As a result when a certain $\nu_i$ is real, meaning that the associated field $\phi_i$ is heavy, the corresponding insertion will be a linear combination of $\phi_i$ and its time derivative, rather than simply $\phi_i$. We assume the exchange of the field $\sigma$ to take place between fields 12 and 34, i.e. the interactions are 
\begin{equation}
g_{12}\sigma \phi_1\phi_2+g_{34}\sigma \phi_3\phi_4~.
\end{equation}
The calculation follows exactly the same steps as in the case of equal $\nu_i$ described in details in section \ref{sec:tree}, so here we will only give the final result. We find
\begin{align}
\begin{split}
&D_{lr}+D_{rl}+D_{ll}+D_{rr} \\
&=  g_{12}g_{34}\left(\prod_{i=1}^4\l(-\eta_c\r)^{\l(\ddt-i\nu_i\r)}\frac{N_{\nu_i}}{N^{\rm AdS}_{-\nu_i}}\right)  \\
& \times\Bigg[ \Gamma(\pm i\nu_\sigma)4\sin\left(\tfrac{\pi}{2}\left(\tfrac{d}{2}-i\nu_\sigma -i \nu_1-i\nu_2\right)\right)\sin\left(\tfrac{\pi}{2}\left(\tfrac{d}{2}-i\nu_\sigma -i \nu_3-i\nu_4\right)\right)\,\mathcal{I}^{\{\nu_i\}}_\Omega
\\
&- 2\sin\left(\tfrac{\pi}{2}\left(d -i \nu_1-i\nu_2 -i \nu_3-i\nu_4\right)\right)\mathcal{I}^{\{\nu_i\}}_G \Bigg] ~,
\end{split}
\end{align}
where
\begin{align}
\begin{split}
& \left(\prod_{i=1}^4\frac{N_{\nu_i}}{N^{\rm AdS}_{-\nu_i}}\right)\mathcal{I}^{\{\nu_i\}}_\Omega\\ & = \frac{\prod_{i=1}^4\Gamma(i\nu_i)}{2^{12}\,\pi^{2d+5}\Gamma(\pm i\nu_\sigma)} \Gamma(\tfrac{d}{4}-i\tfrac{\nu_1+\nu_2}{2}\pm i\tfrac{\nu_\sigma}{2})\Gamma(\tfrac{d}{4}-i\tfrac{\nu_3+\nu_4}{2}\pm i\tfrac{\nu_\sigma}{2}) \\
& \hspace{3cm}\times\Gamma(\tfrac{d}{4}-i\tfrac{\nu_\sigma}{2}\pm i\tfrac{\nu_1-\nu_2}{2})\Gamma(\tfrac{d}{4}+i\tfrac{\nu_\sigma}{2}\pm i\tfrac{\nu_3-\nu_4}{2}) \widehat{\mathcal{F}}^{\{-\nu_i\}}_{\nu_\sigma}~, \\
& \left(\prod_{i=1}^4\frac{N_{\nu_i}}{N^{\rm AdS}_{-\nu_i}}\right)\mathcal{I}^{\{\nu_i\}}_G =\int_{-\infty}^{+\infty} d\nu \frac{1}{\nu^2-\nu_\sigma^2} \left.\left[\left(\prod_{i=1}^4\frac{N_{\nu_i}}{N^{\rm AdS}_{-\nu_i}}\right)\mathcal{I}^{\{\nu_i\}}_\Omega\right]\right\vert_{\nu_\sigma\to\nu}~.
\end{split}
\end{align}
Here $\widehat{\mathcal{F}}^{\{-\nu_i\}}_{\nu_\sigma}$ is the conformal partial wave with external dimensions $\Delta_i = \frac{d}{2}-i\nu_i$ and eigenvalue of the Casimir $\frac{d^2}{4}+\nu_\sigma^2$, normalized as in eq. \eqref{eq:CPW}. Starting from this result, one can easily repeat the steps performed in \ref{sec:compex} and obtain also the expression for the composite exchange in the case of general external dimensions.

 

\small
\bibliography{nBiblio}
\bibliographystyle{utphys}

\end{document}